**Recent Progress on Synthesis, Characterization, and Applications of Metal Halide Perovskites@Metal Oxide**

*Yanyan Duan, De-Yi Wang, and Rubén D. Costa**

Y.Y. Duan, Dr. D.Y. Wang
IMDEA Materials Institute, Calle Eric Kandel 2, 28906 Getafe, Spain.
Y.Y. Duan
Departamento de Ciencia de Materiales
Universidad Politécnica de Madrid
E.T.S. de Ingenieros de Caminos
Profesor Aranguren s/n
Madrid 28040, Spain
Prof. R. D. Costa
Chair of Biogenic Functional Materials
Technical University of Munich
Schulgasse 22, Straubing D-94315, Germany
E-mail: ruben.costa@tum.de

Metal halide perovskites (MHPs) have become a promising candidate in a myriad of applications, such as light-emitting diodes, solar cells, lasing, photodetectors, photocatalysis, transistors, *etc*. This is related to the synergy of their excellent features, including high photoluminescence quantum yields, narrow and tunable emission, long charge carrier lifetimes, broad absorption spectrum along with high extinction absorptions coefficients, among others. However, the main bottleneck is the poor stability of the MHPs under ambient conditions. This is imposing severe restrictions with respect to their industrialized applications and commercialization. In this context, metal oxide ($MO_x$) coatings have recently emerged as an efficient strategy towards overcoming the stabilities issues as well as retain the excellent properties of the MHPs, and therefore facilitate the development of the related devices' stabilities and performances. This review provides a summary of the recent progress on synthetic methods, enhanced features, the techniques to assess the MHPs/$MO_x$ composites, and applications of the MHPs@$MO_x$. Specially, novel approaches to fabricate the composites and new applications of the composites are also reported in this review for the first time. This is rounded by a critical outlook about the current MHPs' stability issues and the further direction to ensure a bright future of MHPs@$MO_x$.

## 1. Introduction

Perovskite, which is named after a geological Perovski, usually refers to a class of ceramic oxides whose general molecular formula is $ABO_3$. Recently, metal halide perovskites (MHPs) with a formula of $ABX_3$ have attracted remarkable attention – **Figure 1**.[1,2] Here, A is a monovalent cation, like organic cations-methylammonium (MA, $CH_3NH_3^+$), or formamidinium (FA, $HC(NH_2)_2^+$); or the inorganic alkali-metal cations-$Cs^+$, partial doping of $Li^+$, $Na^+$, $K^+$ and $Rb^+$.[3] The former constitutes the hybrid MHPs family, while the later constitutes the all-inorganic MHPs class. B is a divalent cation – *e.g.*, $Pb^{2+}$, $Sn^{2+}$. Pb-based MHPs usually show better stability and higher photoluminescence quantum yields (PLQYs), but the toxicity of the lead is still a big concern. Finally, X is the halide ion, $I^-$, $Br^-$, $Cl^-$.

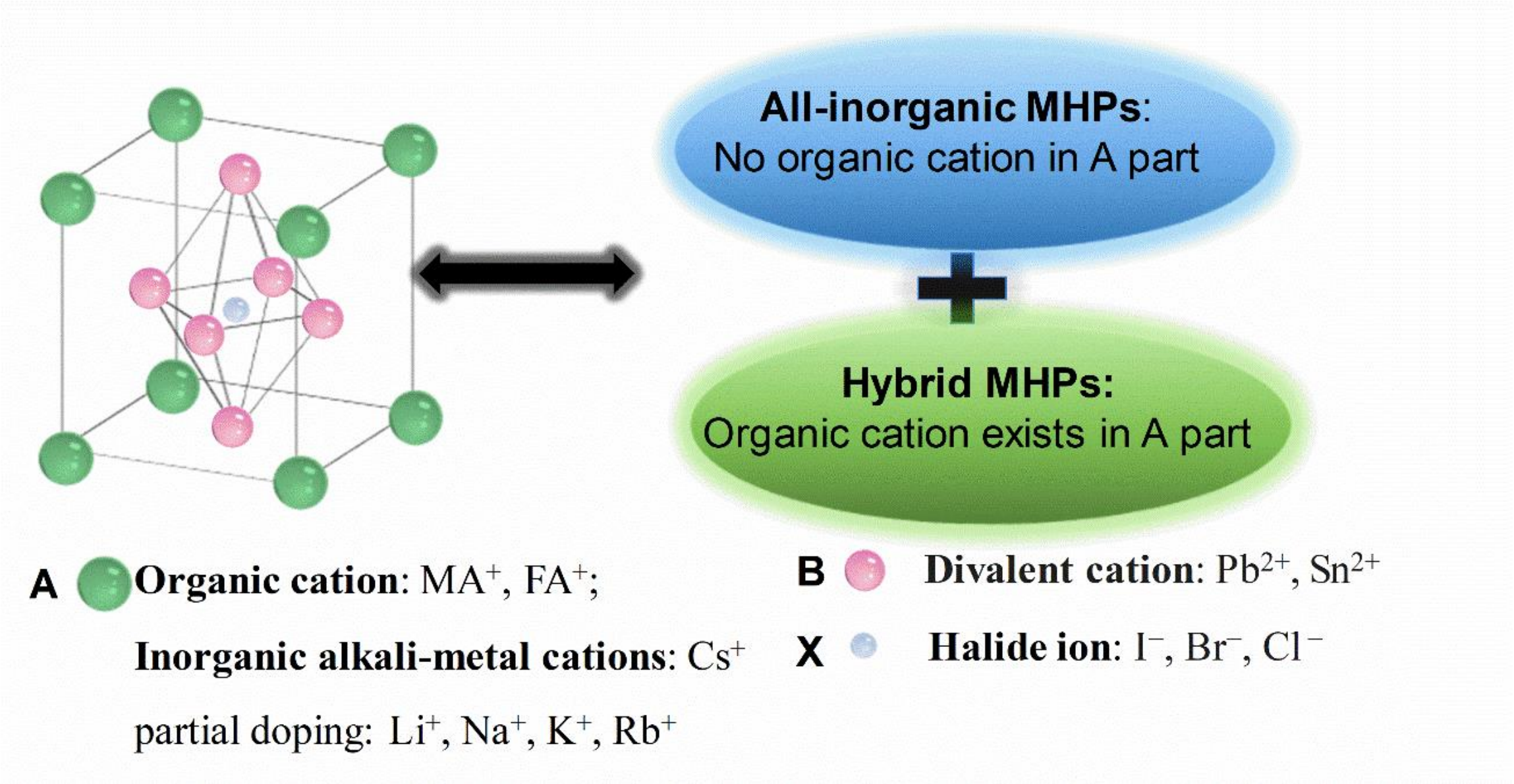

**Figure 1.** The crystal structure of the 3D cubic $ABX_3$ MHPs and the family classification based on the composition of A.

Nowadays, the MHPs show great promise in the optoelectronics, like lasing,[4] photodetectors,[5] photocatalysis,[6] transistors,[7] especially in light-emitting diodes (LEDs)[8-11] and solar cells.[12-14] In the pioneering work published in 2009, the hybrid $MAPbI_3$ was employed as the visible-light sensitizer in solar cells, reaching power conversion efficiencies (PCE) of 3.81%.[15] On 2021, the certified PCE has reached a milestone of ca. 25% and 29% for the single-junction and perovskite/silicon tandem solar cells, respectively.[16] Both, the easy film formation and the versatile chemical composition, have promoted the rapid development in terms of device performance.[17] Meanwhile, the Pb leakage issue has been significantly restrained,[18-20] while Snaith's team has introduced ionic liquids that simultaneously improve the efficiency and the long-term stability of solar cells.[21] In particular, these devices feature >1800 h at a continuous full-spectrum sunlight under 70 ~ 75 °C with an slight decrease of only about 5% in efficiency.

In parallel, the combination of high color purity[22]-corresponding to a full width at half maximum (FWHM) of about 20 nm – *cf.*, 40 nm for the organic compounds[23] and *ca*. 30 nm for quantum dots (QDs),[24] high PLQYs,[22,25] tunable emission[26,27] as well as excellent defect tolerance[28] have strongly encouraged the use of MHPs as alternative emission layers and color down-converting phosphors to achieve natural-color displays and highly performing flexible lighting devices.[29,30] In the case of electroluminescent (EL) perovskite thin-film lighting diodes (PeLEDs), the external quantum efficiencies (EQE) have increased from 0.1% in 2014[31] to 20.7% in 2018[32] for green LEDs and red LED devices, but low stabilities of 20 h. In phosphor-converted LEDs (pc-LEDs), the luminous efficiency has reached values of 109 lm $W^{-1}$[33] and the standard white emission with Commission internationale de l'éclairage (CIE) color coordinates of (0.33, 0.33) is also achieved.[34,35] In addition, the pc-LEDs which are fabricated based on the MHPs can cover up to 140% of the National Television System Committee (NTSC) color standard.[36] Likewise PeLEDs, the operational stabilities of *ca*. 1000 h are the main limitation.[37]

On the other hand, the MHPs have attracted profound attention in the photocatalysis area and have been recognized as a promising and intriguing alternative to the traditional semiconductor photocatalysts (like ZnO, $TiO_2$, CdS).[6] MHPs provide great potential to take the advantage of the rich visible light through extending the efficient light harvesting, benefiting from their excellent optoelectronic properties, like tunable bandgap.[38] In addition, other merits like high surface areas, high absorption coefficient, long photogenerated carrier lifetime, as well as low-cost facile processing also trigger the widespread applications of MHPs in photoelectrochemistry, ranging from the transformation of organic compounds[39-42] to the generation of solar fuels.[43-45] Specially, a turnover number (TON) of over 52,000 can be achieved for selectively visible photocatalyze carbon–carbon bond formation reactions using the as-prepared MHPs nanocrystals (NCs) as the photocatalyst.[39]

Besides, MHPs also show promising features as the optical gain media for lasers.[46] They have as high optical gain coefficients as the commercial laser materials (like GaAs);[47] and their tunable emission enables the tailorable lasing from visible to NIR region;[48-51] besides, they can be directly fabricated into optical microcavities or combined with industrial optical cavities to construct lasers because of their easy processability;[52] and their excellent charge carrier mobility also makes them feasible to be used in the electrically pumped laser devices.[53] Significant progresses have been achieved for improving the performances and expanding the capabilities since the first amplified spontaneous emission (ASE) based on the 2D MHPs films was reported in 1998.[54,55]

Despite the significant achievements in the above fields, the entrance into market in high performing products is, indeed, challenged by the low stability of MHPs under device fabrication, storage, and operation conditions.[28,56] Compared to their counterpart perovskite oxides – *i.e.*, $ABO_3$, MHPs are very sensitive to polar solvents, surface ligand exchange, and morphological changes upon exposure to organic solvents and ambient moisture.[33] In addition, density functional theory (DFT) calculations point out that the formation energy of MHPs is less favorable than $ABO_3$,[57] indicating that the MHPs structure is very fragile to different stress scenarios – *e.g.*, the easy formation of superoxide radicals under both $O_2$ and light.[58] Finally, the practical application of MHP-based devices implies a high thermal-stability is indispensable for working under outdoor conditions or facing device self-heating due to Joule effect.[59]

Therefore, many strategies have been developed to address the instability, like A-site cation engineering,[60] fabrication of the Ruddlesden-Popper (RP) phase perovskite,[61] surface engineering,[62,63] matrix encapsulation,[64-66] among others. Those are nicely summarized in several recent reviews.[57,59,65,67] Herein, we focus on the synthesis protocol of metal oxide ($MO_x$) coatings, through preparing MHPs@$MO_x$ towards enhancing the stability of the MHPs and MHPs based-applications. The benefits of using the $MO_x$ as the protection for the MHPs lie in several aspects. Firstly, inorganics, like $MO_x$, are more thermal stable and mechanically robust than the polymers.[65] Secondly, the confinement effect provided by the $MO_x$ will prevent the MHPs touching each other and therefore the enhanced stability.[68] Then, the $MO_x$ shell will prevent the permeation of the oxygen and water molecules from the surrounding environment and improve the stability.[69] Finally, the $MO_x$ shell helps to passivate the surface traps and enhance the radiative recombination.[70] At first, the reader will be introduced to the core-shell structure concepts and how to design and synthetize the MHPs@$MO_x$ structure including $SiO_2$, $TiO_2$, $ZrO_2$, $Al_2O_3$, $SnO_2$, $SiO_2/Al_2O_3$, $SiO_2/ZrO_2$ – section 2. This will be followed by a section devoted to the characterization of MHPs@MO$_x$ with respect to the morphological, structural, optical properties as well as stability studies under different stress scenarios – section 3. This will be discussed in concerned with a subsequent section highlighting the advances and remaining challenges in MHPs@MO$_x$ applications including LEDs, photoelectrochemistry, lasing, cell imaging, drug delivery, electrochemiluminescent (ECL) etc– section 4. Finally, the reader will find a critical outlook about the current limitations and the possible directions to tackle them – section 5. We aim to provide a comprehensive guideline for the researchers to fabricate MHPs@MO$_x$ structures assisted by a throughout analysis and end-use in energy. Overall, this review honors all the efforts of many groups working on MHPs@MO$_x$ and their applications to ensure a bright future of this approach. This will be useful for guiding researchers to find the easiest way to fabricate stable and highly efficient MHPs@MO$_x$ based devices as well as tracing the state-of-art in this field.

## 2. Design and Synthesis of MHPs@MO$_x$

MO$_x$ coatings is a promising approach for the protection of MHPs in a core-shell design – *i.e.*, MO$_x$ as the shell and MHPs as the core –, as they usually show a high barrier for oxygen, water, and other environment attacks.[57,65,71,72] In addition, the MO$_x$ shell can also passivate the surface trap of the MHPs, enabling high PLQYs.[65] In this section, we will present the different synthesis methods – *i.e.*, sol–gel, atomic layer deposition (ALD), template-assisted, physical approaches, successive ionic layer adsorption and reaction (SILAR) and other strategies, like the combined method and the low-temperature molten salts method– towards the preparation of the MHPs@MO$_x$ - **Table 1**.

### 2.1. Sol-gel Method

Sol-gel method is used to prepare an initial stable and transparent sol using the active components as the precursor, while a three-dimensional network structure of the gel will form through hydrolysis. In 1845, the French chemist J. J. Ebelmen mixed $SiCl_4$ with ethanol to form tetraethoxysilane (TEOS) and a gel was formed during hydrolyzing in wet air.[73] Currently, this method is mostly used to fabricate the shell coating with $SiO_2$,[74-86] $TiO_2$,[84, 87,88] $ZrO_2$,[70] $Al_2O_3/SiO_2$,[69] $SiO_2/ZrO_2$[89] around the surface of the MHPs core.

Starting with the well-known $SiO_2$-coating, (3-aminopropyl)triethoxysilane (APTES) has been widely used as the precursor for both, the $SiO_2$-coating and the capping agent for MHPs.[35] Specifically, the Cs-oleate was firstly prepared through degassing and heating the mixture of $Cs_2CO_3$, Oleic acid (OA), and octadecene (ODE), until getting a clear solution. At this point, the mixture of OA and APTES was slowly added to $PbX_2$-ODE solution under inert ($N_2$) atmosphere. The preheated Cs-oleate solution was quickly injected to the above solution, leading to the formation of $SiO_2$ coating through a hydrolytic process due to the trace water existing in the atmosphere – **Figure 2a**. Based on the results of Fourier transform infrared spectroscopy (FTIR) and X-ray diffraction (XRD), the $SiO_2$ shell is successfully formed and the formation process are depicted by the following reactions:

$$-SiOC_2H_5 + H_2O \rightarrow -SiOH + C_2H_5OH \quad (1)$$

$$-SiOH + -SiOC_2H_5 \rightarrow Si - O - Si + C_2H_5OH \quad (2)$$

$$-SiOH + -SiOH \rightarrow -SiOSi + H_2O \quad (3)$$

**Table 1.** Summary of the synthesis techniques, the features, and the applications of the MHPs/$MO_x$.

| MHPs/$MO_x$ | MHPs | | | $MO_x$ | | Synthesis techniques, the properties, and stability of the MHPs/$MO_x$ | | | | | | | | Application | Ref. |
|---|---|---|---|---|---|---|---|---|---|---|---|---|---|---|---|
| | Techniques | Synthesis conditions | Main regents and solvents | Synthesis conditions | Precursors for the $MO_x$ | Techniques for the formation of MHPs/$MO_x$ | PLQYs and emission peak | FWHM | Stability of the MHPs/$MO_x$: Water stability | Thermal stability | Photo stability | Air stability (storage stability) | Other stability | | |
| $CsPb(Br/I)_3/SiO_2$ | Hot-injection | Vacuum, $N_2$ protection, temperature (140 ºC, 100 ºC, 120 ºC), ice-bath | $PbBr_2$, $PbI_2$, OA, $Cs_2CO_3$ | Air, 60 ºC | APTES | Sol-gel | 88% (624 nm) | | | | | Only 5% loss in the PLQY after 3 months in air | | LEDs | [35] |
| $CsPbBr_3/SiO_2$ | Hot-injection | Vacuum, $N_2$ protection, temperature (140 ºC, 100 ºC, 120 ºC), ice-bath | ODE $PbBr_2$, OA, $Cs_2CO_3$ | Air, 20 ºC | APTES | Sol-gel | 85% (522 nm) | 22 nm | | | | Near no change in the PLQY after 3 months in air | | LEDs | [35] |
| $MAPbBr_3/SiO_2$ | LARP | Vacuum | MABr, $PbBr_2$, octylamine, OA, DMF, toluene | Ambient conditions | APTES | Sol-gel | 15–55% (452–524 nm) | | | | | | Almost 70% of the PL intensity remains in isopropanol after 2.5 h | | [63] |
| $MAPbBr_3/SiO_2$ | LARP | $N_2$ protection, freeze drying | MABr, $PbBr_2$, n-octylamine, OA, DMF, toluene | Humidity chamber (25 °C, RH of 60%) | TMOS | Sol-gel | 89% (505 nm) | ~ 32 nm | | | *$MAPbBr_3/SiO_2$ solution*: 61.03% of the PL intensity remains after illuminated under 450 nm LED light (175 mW $cm^{-2}$) for 49 h; *$MAPbBr_3/SiO_2$ powders*: 94.1% of the PL intensity remains after illuminated under 470 nm LED light (21 mW $cm^{-2}$) for 7 h *$MAPbBr_3/SiO_2/PMMA$ films*: 69.16% of the PL intensity remained after illuminated under 470 nm LED light (21 mW/$cm^2$) for 100 h | | | LEDs | [83] |
| $CsPbX_3$ (X=Cl, Br, I)/$SiO_2$ | Hot-injection | Vacuum, $N_2$ protection, temperature (150 ºC, 180 ºC, 120 ºC, 165 ºC), ice-bath | ODE, oleylamine, $PbX_2$, $Cs_2CO_3$, OA | Ambient conditions | $(NH_4)_2SiF_6$ | Sol-gel | 84% (~ 500 nm) for $CsPbBr_3/SiO_2$ | 20~25 nm for $CsPbBr_3/SiO_2$ | | *For $CsPbBr_3/SiO_2$:* 90% of its original emission intensity as the temperature rises to 353 K | *For $CsPbBr_3/SiO_2$:* 93% of initial PL can be reserved after 53 h under a 450 nm LED, 175 mW $cm^{-2}$ | | | LEDs | [77] |
| $CsPbX_3$ (X=Cl, Br, I)/$SiO_2$ Janus NCs | Hot-injection | Vacuum, $N_2$ protection, temperature (120 ºC, 150 ºC, 140 ºC), ice-bath | ODE, oleylamine, $PbX_2$, $Cs_2CO_3$, OA | Ambient conditions | TMOS | Sol-gel | 80% (517 nm) for $CsPbBr_3$/$SiO_2$ | 18 nm for $CsPbBr_3$/$SiO_2$ | *For $CsPbBr_3$/$SiO_2$ NCs:* around 80% of the PL stability remains after water treatment for 7 days | | *For $CsPbBr_3$/$SiO_2$ NCs/PMMA films:* ~ 2% PL drops after 10 h under continuous illumination with a 375 nm LED light (117 mW $cm^{-2}$) | *For $CsPbBr_3$/$SiO_2$ NCs/PMMA films:* high emission after 4 days in air | | LEDs | [76] |
| $CsPbBr_3$/$Ta_2O_5$ Janus NCs | Hot-injection | Vacuum, $N_2$ protection, temperature (120 ºC, 150 ºC, 140 ºC), ice-bath | ODE, oleylamine, $PbX_2$, $Cs_2CO_3$,OA | Ambient conditions | TTEO | Sol-gel | 85% (519 nm) | 19 nm | | | | | | | [76] |
| $CsPbX_3$ (X=Cl, Br, I)/$ZrO_2$ | Hot-injection | Vacuum, $N_2$ protection, temperature (120 ºC, 150 ºC, 140 ºC), ice-bath | ODE, oleylamine, $PbX_2$, $Cs_2CO_3$, OA | Ambient conditions | $Zr(OC_4H_9)_4$ | Sol-gel | 90% (514 nm) for $CsPbBr_3$/$SiO_2$; ~ 50% for $CsPbBr_2Cl_1$/$SiO_2$; ~ 70% for $CsPbBr_{2.5}I_{0.5}$/$SiO_2$; ~ 55% for $CsPbBr_{1.5}I_{1.5}$/$SiO_2$; ~ 40% for $CsPbBr0_{.5}I_{2.5}$/$SiO_2$; ~ 15% (~ 690 nm) for $CsPbI_3$/$SiO_2$ | 14 nm for $CsPbBr_3$ /$SiO_2$ | Around 80% of the PL stability can remain after water treatment for 8 days | | | | | LEDs | [70] |
| $CsPbBr_3$/$SiO_2$ | Hot-injection | Vacuum, $N_2$ protection, temperature (120 ºC, 140 ºC), ice-bath | ODE, $PbBr_2$, $Cs_2CO_3$, OA | 20 $^oC$, RH of 30% | APTES | Sol-gel | 80% (519 nm) | | | | | Almost no change in the PLQYs after 1 month in air; 77% of the initial PLQYs can be remained after 2 months | | | [85] |
| $CsPbBr_3$@$SiO_2$ | LARP | Room temperature | $PbBr_2$, CsBr, hydrobromic acid, DMF, BTPA-GA; toluene | 100 ºC in the oven | APTES | Sol-gel | 518 nm | 17 nm | | | 98% of the initial PL intensity remains under visible-light illumination and oxygen atmosphere for 1 h | | PL emission remains relatively unchanged over a period of 7 days in | Photocatalysis | [90] |
| | | | | | | | | | | | | | 1:1 alcohol/water solution | | |
| $CsPbBr_3/SiO_2$ | LARP | 90 ºC | CsBr, oleylamine, $PbBr_2$, OA, DMF, toluene, ammonia solution | Ambient conditions | TMOS | Sol-gel | 90% (501 nm) | 22 nm | 112% of the initial PL when the powders are dispersed in 2 mL of DI water and treated by ultrasonication | | | Characteristic peaks can be clearly identified in RH of 75%, 25 °C for 4 weeks | | | [86] |
| $CsPbBr_3@Cs_4PbBr_6/SiO_2$ | LARP | RT | CsBr, oleylamine, $PbBr_2$, OA, DMSO, toluene | Ambient conditions | APTES | Sol-gel | 48% (520 nm) | 22 nm | | No obvious degeneration upon heating−cooling for 10 cycles from 30 to 150 °C in RH of 50% under ambient condition | | A slight decrease of 9% after 2 months under ambient conditions with RH 50% ambient condition | | Anti-counterfeiting | [81] |
| $CsPbBr_3@SiO_2$ | LARP | RT | CsBr, oleylamine, $PbBr_2$, OA, DMF, ethanol, ultrapure water | Ambient conditions | TEOS | Sol-gel | 529 nm | 18 nm | | | | | 71% of the initial intensity remains after 168 h storage in ethanol | | [91] |
| $CsPbBr_3@SiO_2$ | LARP | RT | CsBr, oleylamine, $PbBr_2$, OA, DMF, toluene | Ambient conditions | APTES | Sol-gel | 71.6% (517 nm) | 25 nm | | 84% of the initial PL intensity remains after continuous heating at 60 °C for 80 min | | | | Lasing | [92] |
| $CsPbBr_3@SiO_2$ | Hot-injection | Vacuum, Ar protection, temperature (60 ºC, 90 ºC, 120 ºC, 150 ºC ), ice-bath | ODE, $PbBr_2$, $Cs_2CO_3$,OA | Ambient conditions | APTES | Sol-gel | 90% (519 nm) | 21 nm | | | | | Still emissive even after 3000 h in octane, ethanol isopropyl alcohol, n-butyl alcohol, | LEDs | [93] |
| | | | | | | | | | | | | | methyl acetate, and ethyl acetate | | |
| $CsPb(Br_{0.3}I_{0.7})_3$@ $SiO_2$ | Hot-injection | Vacuum, Ar protection, temperature (60 ºC, 90 ºC, 120 ºC, 150 ºC ), ice-bath | ODE, $PbBr_2$, $PbI_2$, $Cs_2CO_3$,OA | ambient conditions | APTES | Sol-gel | 663 nm | | | | | | | LEDs | [93] |
| $(CsPbBr_3/Fe_3O_4)$ @MPSs@$SiO_2$ | Hot-injection | Vacuum, $N_2$ protection, temperature (120 ºC, 150 ºC), ice-bath | ODE, $PbBr_2$, $Cs_2CO_3$, OA, oleylamine, n-hexane | Room temperatur e | TMOS | Sol-gel | 508 nm | 22 nm | | | | | | Capturing circulating tumor cells | [94] |
| $CsPbBr_3$ QDs/F$SiO_2$ | Hot-injection | Vacuum, $N_2$ protection, temperature (100 ºC, 120 ºC, 150 ºC, 160 ºC), ice-bath | ODE, $PbBr_2$, $Cs_2CO_3$, OA, oleylamine, toluene | 25 ºC | FPEOS | Sol-gel | 75% (519 nm) | 21 nm | | PL intensity remains 29.8% as temperature rises from 30 to 90 ºC | PL intensity remains above 90% after 50 h under continuous UV light irradiation (365 nm, 16 W, window area: 20 cm × 5 cm, irradiation distance of 20 cm) | 96.5% of PL intensity remains after storage for 90 days | | LEDs | [95] |
| $CsPbBr_3$ QDs/$SiO_2$ | Hot-injection | Vacuum, $N_2$ protection, temperature (120 ºC, 160 ºC), ice-bath | ODE, $PbBr_2$, $Cs_2CO_3$, OA, oleylamine, toluene, hexane | RH of 50%, room temperatur e | APTES | Sol-gel | 537 nm | | | | | PL intensity decreases to ca. 70 % after storage in air for 16 days | PL intensity is hardly reduced after integrated on blue LED chip (ca. 70 ºC) for 12 h | | [96] |
| $MAPbBr_3/ZrO_2$ | LARP | $N_2$ | MABr, PbBr, DMF, n-octylamine, OA, toluene | Room temperatur e | Zirconium n-propoxide | Sol-gel | 511 nm | | | | 45% of the initial PL value after 8.5 h under UV-light (365 nm, 35 mJ $cm^{-2}$) | | 61% of the initial PL value remains after 33.5 h in weak base; 68% of the initial PL value remains after 28 h in acid (pH=3.4) | | [97] |
| $CsPbBr_3/CsPb_2Br_5$@$SiO_2$ | Hot-injection | Glovebox, $N_2$ protection, temperature (120 ºC, 150 ºC), ice-bath | ODE, $PbBr_2$, $Cs_2CO_3$, OA, oleylamine, n-hexane | 50 ºC | TEOS | Sol-gel | 42% | | 2548.4% and 1796.1% of the initial PL intensity after 10 min and 30 min ultrasound in water, respectively | | | | | LEDs | [98] |
| $CsPbBr_3$@$SiO_2$ | Hot-injection | Vacuum, $N_2$ protection, temperature (100 ºC, | ODE, $PbBr_2$, $Cs_2CO_3$,OA, | Ambient conditions | APTES | Sol-gel | ~ 90% (~ 500 nm) | | | | | | | | [99] |

| | | | | | | | | | | | | | | |
|---|---|---|---|---|---|---|---|---|---|---|---|---|---|---|
| | | 120 ºC, 150 ºC ), ice-bath | oleylamine, n-butanol | | | | | | | | | | | |
| $CsPbBr_3/SiO_2$ | Hot-injection | 75 ºC | $Cs_2CO_3$, PbO, OA, ethanol, TOAB | Ambient conditions | APTES and TEOS | Sol-gel | 61.9% (514 nm) | 23 nm | | | | **Sol**: The PL intensity declines to 62.1% in 34 days **Films**: 82.8% of the initial PL intensity remains after 30 days | | [100] |
| $CsPbBr_3@SiO_2$ | LARP | RT and normal pressure | $PbBr_2$, CsBr, OA, oleylamine, DMF, toluene | Ambient conditions | TMOS | Sol-gel | 72% (520 nm) | | Nearly 95% of the original PL remains after introducing 20 µL of DI water to the QD toluene solution (1 mL) | ~82% of the PL intensity after 6 cycles heating-cooling from 20–120 °C | ~72% of the initial PL remains after irradiation for 5 h under UV light (365 nm, 20 W) | | LEDs | [101] |
| $CsPbBr_3$-coreactant@$SiO_2$ | Hot-injection | Vacuum, $N_2$ protection, temperature (100 ºC, 120 ºC, 150 ºC ), ice-bath | ODE, $PbBr_2$, $Cs_2CO_3$,OA, oleylamine, ethyl acetate | RH of 80% RT | TMOS | Sol-gel | 65% (518 nm) | 20 nm | 55% of ECL value remains after 48 h storage under 100% RH at RT | | | | ECL | [102] |
| $MAPbBr_3@SiO_2$ | LARP | Ambient conditions | MABr, $PbBr_2$, OA, DMF, ethanol, toluene | 25 ºC, 45 ºC | TEOS, APTES | Sol-gel | 60.3% (523 nm) | 25 nm | | 6.1% increase in PL intensity after continuous heating at 80 °C for 60 h | The emission remains nearly constant under 450 nm light (60 mW $cm^{-2}$) for ~ 350 h ($N_2$-protected) | | LEDs | [75] |
| $CsPbBr_3@SiO_2$ | Hot-injection | Vacuum, $N_2$ protection, temperature (120 ºC, 165 ºC ), ice-bath | ODE, $PbBr_2$, $Cs_2CO_3$,OA, oleylamine, 2-methoxyethanol | Ambient conditions | TEOS | Sol-gel | 533 nm | 18 nm | 73.8% of the initial value for the $CsPbBr_3/SiO_2$ film under 75% RH in air for 12 h | About 36.4% of the initial PL intensity remains after 15 h at 60 ºC | | The $CsPbBr_3/SiO_2$ solution maintains a bright green emission after 12 h in air | Lasing | [80] |
| CsPbX (X=Cl, Br, I)$_3$@$SiO_2$ | Hot-injection | Vacuum, $N_2$ protection, temperature (120 ºC, 180 ºC ), ice-bath | ODE, $PbX_2$, $Cs_2CO_3$,OA, oleylamine | Vacuum, 60 ºC | TEOS | Sol-gel | 11.2% for $CsPbCl_3/SiO_2$; 84% (523 nm) for $CsPbBr_3/SiO_2$; | | 40% of the emission remains after putting $CsPbX_3/SiO_2$ nanocomposit | | | PLQYs just decreases about 4.1%, 7%, 7.8% for $CsPbX_3/SiO_2$ composites (x = Cl, | LEDs | [79] |

| | | | | | | | | | | | | | | | |
|---|---|---|---|---|---|---|---|---|---|---|---|---|---|---|---|
| | | | | | | | 45% (634 nm) for $CsPb(Br_{0.3}I_{0.7})_3/SiO_2$ | | e in water for 4 h | | | Br, I), respectively; the PL of $CsPbI_3/SiO_2$ composites film remains unchanged in air after about 2 months | | | |
| $CsPbMnX_3@SiO_2$ | LARP | 80 ºC | $MnBr_2$, $PbBr_2$, $PbCl_2$, DMSO, toluene, CsBr, DMF, OA, n-octylamine | Rir, 25 °C, RH of 65% | TEOS | Sol-gel | 50.5% (446 nm and 607 nm) | | The PLQY of the QDs can maintain ≈90% of the initial value after being stored for 6 days in water | ≈90% of the initial intensity at 607 nm could maintain after being treated at 100 °C under vacuum for 1 h | | | | LEDs | [74] |
| $CsPbBr_3@SiO_2/Al_2O_3$ | Hot-injection | Vacuum, Ar protection, temperature (70 ºC, 120 ºC, 150 ºC, 180 ºC ), ice-bath | ODE, $PbBr_2$, $Cs_2CO_3$,OA, oleylamine, methyl acetate, toluene | 25 °C, 50%-60%, 50 °C, vacuum | DBATES | Sol-gel | 90% (519 nm) | 25 nm | | Ignorable change after 1 h treatment at 50 °C under vacuum | About 90% of the initial intensity remains after 300 h under a 470 nm LED light (21 mW $cm^{-2}$) | | *Storage stability under 80 °C and RH of 80%*: 90% of its initial intensity remains after 34 h | LEDs | [69] |
| $CsPbBr_3@SiO_2$ | Hot-injection | Vacuum, Ar protection, temperature (70 ºC, 120 ºC, 150 ºC 180 ºC ), ice-bath | ODE, $PbBr_2$, $Cs_2CO_3$,OA, oleylamine, methyl acetate, toluene | 25 °C, 50%-60%, 50 °C, vacuum | TMOS | Sol-gel | | | | | Ca. 25 of the initial intensity remains after 200 h under a 470 nm LED light (21 $mWcm^{-2}$) | | *Under 80 °C and RH of 80%*: ~30% of its initial intensity remains after 34 h | LEDs | [69] |
| DDAB-$CsPbBr_3@SiO_2$ | Hot-injection | Vacuum, $N_2$ protection, temperature (100 ºC, 120 ºC, 150 ºC 180 ºC ), ice-bath | ODE, $PbBr_2$, $Cs_2CO_3$,OA, oleylamine, DDAB, hexane | Room temperature | TMOS | Sol-gel | 80.45% (519 nm) | 19 nm | Bright green light in water after 40 min under ultrasonication | 70% of the initial PL intensity remains at 140 °C | | | | LEDs | [103] |
| $CsPbBr_3@Al_2O_3$ | Hot-injection | Vacuum, Ar protection, temperature (70 ºC, 120 ºC, 150 ºC 180 ºC ), ice-bath | ODE, $PbBr_2$, $Cs_2CO_3$,OA, oleylamine, methyl acetate, toluene | 25 °C, 50%-60%, 50 °C, vacuum | ASB | Sol-gel | | | | | | | *Under 80 °C and RH of 80%*: ~20% of its initial intensity | | [69] |
| | | | | | | | | | | | | | remains after 34 h | | |
| Mn:$CsPbCl_3$@$SiO_2$/$Al_2O_3$ | Hot-injection | Vacuum, $N_2$ protection, temperature (120 ºC, 160 ºC, 190 ºC), ice-bath | ODE, $PbCl_2$, $Cs_2CO_3$,OA, $MnCl_2\cdot(H_2O)_4$ oleylamine, hexane, acetone | 25 ºC, RH of 50%-60%, 50 ºC, vacuum | DBATES | Sol-gel | 56% (412 nm and 588 nm) | | | | | | 92% of the initial intensity after 7 days under 85 ºC, RH of 85% | LEDs | [104] |
| Mn:$CsPbCl_3$@$SiO_2$ | Hot-injection | Vacuum, $N_2$ protection, temperature (120 ºC, 160 ºC, 190 ºC), ice-bath | ODE, $PbCl_2$, $Cs_2CO_3$,OA, $MnCl_2\cdot(H_2O)_4$ oleylamine, hexane, acetone | 25 ºC, RH of 50%-60%, 50 ºC, vacuum | TMOS | Sol-gel | 49% (412 nm and 588 nm) | | | | | | ~30 % of the initial intensity remains after 5 days under 85 ºC, RH of 85% | | [104] |
| $CsPbBr_3$/a-$TiO_2$ | LARP | Ambient condition | $Cs_2CO_3$, propionic acid, $PbBr_2$, isovolumetric isopropanol/propionic acid/butyl amine mix solution, n-hexane, isopropanol | 120 °C, electronic oven | TBT | Sol-gel | | | | | | | | Photocatalytic $CO_2$ reduction | [87] |
| $CsPbBr_3$/$TiO_2$ | Hot-injection | Vacuum, $N_2$ protection, temperature (105 ºC, 120 ºC, 140 ºC, 150 ºC, 170 ºC), ice-bath | ODE, $PbBr_2$, $Cs_2CO_3$,OA, oleylamine, hexane, toluene | 25 °C, 30 % RH, 80 °C under vacuum, 300 °C under Ar | TBOT | Sol-gel | ~ 520 nm | | Ca.85% of initial emission intensity remains after 3 months immersion in water, along with no change in both peak position and FWHM | | 75% of the PL intensity maintains under UV light for 24 h | | The FWHM and PL position remains unchanged by mixing $CsPbBr_3$/ $TiO_2$ with $PbCl_2$/$PbI_2$ after 15 days | Photoelectrochemistry | [88] |
| PMMA-capped $CsPbBr_3$/$TiO_2$ | LARP | Vacuum, 120 ºC | $PbBr_2$, Cesium acetate, DA, DMF, DAm, hexane, methyl acrylate | Room temperature | TBOT | Sol-gel | 39% | | High PL emission after vigorous stirring with water for 6 h | 72% retention of PL at 70 ºC for 36 h | *Solution*: 66% of initial PL intensity remains after 12 h under 365 nm UV irradiation *Film*: 95% of initial PL intensity remains after 36 h under 365 nm UV irradiation | 68% of the initial PL intensity remains in the case of moisture exposure (RH: 70 ± 5%) after 12 h | 80% retention of PL intensity when mixing with isopropanol (10% by volume) after 36 h; 84% of PL intensity | LEDs | [105] |
| | | | | | | | | | | | | | remains after being exposed to 8% oleylamine solution for 36 h | | |
| $CsPbBr_3@SiO_2/ZrO_2$ | Hot-injection | Vacuum, $N_2$ protection, temperature (100 ºC, 120 ºC, 140 ºC, 150 ºC), ice-bath | ODE, $PbBr_2$, $Cs_2CO_3$,OA, oleylamine, hexane | Ambient conditions | APTES, ZTB | Sol-gel | 65% (524 nm) | 16 nm | 23% of the initial PLQY can be retained after 1 month in water | Neglectable emission changes after 24 h under 60 ºC | | The PLQY holds constant for 1 month in air | | LEDs | [89] |
| $CsPbBr_3@SiO_2$ | Hot-injection | Vacuum, Ar, $N_2$ protection, temperature (130 ºC, 150 ºC, 180 ºC), ice-bath | ODE, $PbBr_2$, $Cs_2CO_3$,OA, oleylamine, toluene and methyl acetate | Air, 45 °C | MPTMS | Sol-gel | 78% (527 nm) | 21 nm | 50% of the initial PL intensity remains after storing in water for over 20 days | | | | | Lasing | [106] |
| $CsPbBr_3@SiO_2$ | Hot-injection | Vacuum, Ar protection, temperature (120 ºC, 160 ºC, 180 ºC), ice-bath | ODE, $PbBr_2$, $Cs_2CO_3$, OA, oleylamine, hexane | 75 °C | PHPS | Sol-gel | 74% (510 nm) | 19 nm | Bright emission after 4 h in water | | | | | LEDs | [107] |
| $CsPbBr_{1.2}I_{1.8}@SiO_2$ | Hot-injection | Vacuum, Ar protection, temperature (120 ºC, 160 ºC, 200 ºC), ice-bath | ODE, $PbBr_2$, $PbI_2$ $Cs_2CO_3$, OA, oleylamine, hexane | 75 °C | PHPS | Sol-gel | 79 % (616 nm) | 31 nm | | | | | | LEDs | [107] |
| $CsPbBr_3/SiO_2$ | Hot-injection | Vacuum, $N_2$ protection, temperature (120 ºC, 150 ºC, 180 ºC), ice-bath | ODE, $PbBr_2$, $Cs_2CO_3$, OA, oleylamine, hexane | 70 °C | TEOS | Sol-gel | 64% | | | | 60% of the original PL intensity remains after 10 h under 365 nm light | | | PEC | [84] |
| $CsPbBr_3@SiO_2$ | Hot-injection | Vacuum, $N_2$ protection, temperature (120 ºC, 170 ºC), ice-bath | ODE, $PbBr_2$, $Cs_2CO_3$,OA, oleylamine, n-hexane | Ambient conditions | APTES | Sol-gel | 514 nm | | The PL intensity drops to 59%, 49.5%, 41.1% , 35.8% after 6 h, 12 h, 24 h, 48 h in water, respectively | | | | | Cell imaging | [108] |
| $CsPbI_3@SiO_2$ | Hot-injection | Vacuum, $N_2$ protection, temperature (120 ºC, 150 ºC, 210 ºC), ice-bath | ODE, $PbI_2$, HI, $Cs_2CO_3$,OA, oleylamine, hexane | Ambient conditions | APTES | Sol-gel | 84% | | Taking 1 h to lose the florescence in water | | | Ca. 92% of the original PLQY after storing in hexane for 28 days | | | [109] |
| $CsPbBr_3@SiO_2$ | LARP | 120 ºC | $PbBr_2$, CsBr, OA, DMF, oleylamine, ammonia solution, toluene | Ambient conditions | TMOS | Sol-gel | 10.2% (515 nm) | | 80% of the initial PL intensity remains after 24 h in water | | | | | Bioimaging and Drug Delivery | [110] |
| $Cs_4PbBr_6$:Sn@$SiO_2$ | LARP | Room temperature | $SnBr_2$, $PbBr_2$, CsBr, OA, DMF, oleylamine, toluene | 60 °C, vacuum | TMOS | Sol-gel | | | | | | | 59.6% of the initial PL intensity remains after 24 h in the mixture of toluene and water | | [111] |
| $CsPbBr_3@SiO_2$ | Hot-injection | Vacuum, $N_2$ protection, temperature (120 ºC, 165 ºC ), ice-bath | ODE, $PbBr_2$, $Cs_2CO_3$,OA, oleylamine, 2-methoxyethanol | Ambient conditions | TEOS | Sol-gel | 530 nm | 18 nm | | | | | | Lasing | [112] |
| $CsPbBr_3@SiO_2$ | Hot-injection | Vacuum, $N_2$ protection, temperature (120 ºC, 150 ºC,170 ºC ), ice-bath | ODE, $PbBr_2$, $Cs_2CO_3$,OA, oleylamine, TOPO | Ambient conditions | TEOS | Sol-gel | 519 nm (~ 87%) | 16 nm | | About 85% of the initial PL intensity is retained at 120 ºC | More than 80% of the initial intensity is preserved when exposed to UV irradiation for 168 h | ~90% of the initial PL intensity remains after a 30-day storage period | More than 70% of its initial fluorescence intensity remains after an 8-day storage in the mixture of the n-hexane and water (volume ratio: 1:1.5) | LEDs | [113] |
| $CsPbBr_3@SiO_2$ | Hot-injection | Vacuum, temperature (150 ºC) | ODE, $PbBr_2$, $Cs_2CO_3$,OA, oleylamine, tert-butyl alcohol | 20 kV; 100 µA | TEOS | Sol-gel | 525 nm | | 15% decrease in PL intensity after 96 h in water | 6.2% decrease in PL intensity after heating at 70 ºC for 24 h | | | | LEDs | [114] |
| $CsPbBr_3@SiO_2$ | Hot-injection | Vacuum, Ar protection, temperature (120 ºC, 140 ºC), ice-bath | ODE, $PbBr_2$, $Cs_2CO_3$, OA, APTES, ethyl acetate, toluene | 25 ºC and RH 75% | APTES, TEOS, and TMOS | Sol-gel | 82% (515 nm) | 18 nm | | PL peak remains unchanged after treatment at 60 ºC for 84 h | 81% of the initial PL intensity remains under UV-light (365 nm, 80 mW $cm^2$) for 6 h and FWHM remains stable for 57 h | 90% of the initial PL intensity after storage under RH of 75% and 25 ºC for 6 h | 85% of the initial PL intensity in ethanol：toluene (1:1 volume ratio) | | [115] |
| $CsPbBr_3@SiO_2$ | Hot-injection | Vacuum, inert protection, temperature (120 ºC, 140 ºC), ice-bath | ODE, $PbBr_2$, $Cs_2CO_3$, OA, hexane | Room temperature | APTES | Sol-gel | 98.56% (516 nm) | 19 nm | | | | | | Daytime radiative cooling | [116] |
| $CsPbBr_3@SiO_2$ | Hot-injection | Vacuum, Ar protection, temperature (120 ºC, 165 ºC,170 ºC), ice-bath | ODE, $PbBr_2$, $Cs_2CO_3$, OA, oleylamine, hexane | Room temperature | APTES and TMOS | Sol-gel | 530 nm | 20 nm | | | | No significant degradation after storage in air for more than 250 days | | LEDs | [117] |
| $CsPb_2Br_5@SiO_2$ | Hot-injection | Vacuum, $N_2$ protection, temperature (120 ºC, 150 ºC), ice-bath | ODE, $PbBr_2$, $Cs_2CO_3$, OA, oleylamine, toluene | Room temperature | TEOS | Sol-gel | ~ 5% (432 nm) | | Blue emission could still be observed clearly after 3 days in water | | | | | Biosensing | [118] |
| $Cs_2AgInCl_6/SiO_2$ | LARP | Temperature (349 K, 373 K) | CsCl, HCl, AgCl, $InCl_3$, ethanol | Room temperature | TEOS | Sol-gel | 505 nm and 580 nm | | | | | | 32.96% of the emission intensity remains after six air plasma bombardments | LEDs | [119] |
| $CsPbBr_3/TiO_2$ | Hot-injection | Vacuum, $N_2$ protection, temperature (120 ºC, 150 ºC, 180 ºC), ice-bath | ODE, $PbBr_2$, $Cs_2CO_3$,OA, oleylamine, hexane | 70 °C | TBOT | Sol-gel | 7% | | | | 82% of the original PL intensity remains after 10 h under 365 nm light | | | PEC | [84] |
| $CsPbBr_{3-x}Cl_x/SnO_2$ | Hot-injection | Vacuum, $N_2$ protection, temperature (120 ºC, 150 ºC, 180 ºC), ice-bath | ODE, $PbBr_2$, $PbCl_2$, $Cs_2CO_3$,OA, oleylamine, hexane trioctylphosphine | 70 °C | Tin(IV) isopropoxide | Sol-gel | 3% | | | | 94% of the original PL intensity remains after 10 h under 365 nm light | 70% of the original PL intensity remains after 6 days under humid air, RH of 70% | | PEC | [84] |
| Na: $CsPb(Br, I)_3@Al_2O_3$ | Hot-injection | Vacuum, $N_2$ protection, temperature (120 ºC, 150 ºC, 170 ºC), ice-bath | ODE, $PbBr_2$, $PbI_2$, $Cs_2CO_3$, OA, oleylamine, toluene | 50 ºC, vacuum, 25 ºC, RH 50~60% | ASB | Sol-gel | 82.21% | | Surviving in water for more 35 min under ultrasonication; | | 80% of the initial PL intensity remains after 16 h under UV lamp (365 nm, 10 W) at 3 cm distance | | | | [120] |
| $CsPbBr_3$ QDs@$SiO_2$/EVA | LARP | Aambient conditions | $PbBr_2$, CsBr, OA, DMF, oleylamine, toluene | 70 ºC in the drying oven | TEOS | Sol-gel | 521 nm | 24 nm | 99%, 93.9%, 81.1%, 63.6% of the initial PL intensity remains after 10, 20, 30, 40 h in water, respectively | | 85%, 80.4%, 72.6%, 72.1% of the initial PL intensity remains after 10, 20, 30, 40 h under UV light, respectively | 67.4%, 63.2%, 28.2% of the initial PL intensity remains after 30, 90, and 120 days in air, respectively | | | [121] |
| DDAB-$CsPbBr_3/SiO_2$ | LARP | Ambient conditions | $PbBr_2$, CsBr, OA, DMF, oleylamine, toluene, DDAB | Room temperature | TMOS | Sol-gel | 82% (~ 515 nm) | | | ~ 50% of the initial PL intensity remains at 100 ◦C; ~ 70% of the | | | 89% of initial intensity after 60 min in ethanol/toluene mixture; | LEDs and visible light wireless communication | [122] |
| | | | | | | | | | | initial PL intensity remains after heating at 60 ºC for 80 min | | | 50% of the initial PL remains after 210 min in the mixture of water/toluene (0.5 mL/3 mL) | | |
| $CsPbBr_3$@$AlO_x$ | Hot-injection | Vacuum, $N_2$ protection, temperature (100 ºC, 120 ºC, 150 ºC, 160 ºC), ice-bath | ODE, $PbBr_2$, $Cs_2CO_3$, OA, oleylamine, hexane, acetone, octane | 50 ºC, 0.15 Torr. | TMA | ALD | ~55% (~ 522 nm) | | PL properties do not change after soaking in water for 1 h | No additional XRD peaks appear after annealing in air at 200 ºC | Minimal PL quenching after 8 h and no energy shifts after being irradiated under simulated solar spectrum (10 mWcm$^{-2}$) in air for 8 h | No apparent change in the PL properties is observed after 45 days in air | | | [123] |
| $CsPbBr_3$@$AlO_x$ | Hot-injection | Vacuum, $N_2$ protection, temperature (100 ºC, 120 ºC, 165 ºC), ice-bath | ODE, $PbBr_2$, $Cs_2CO_3$, OA, oleylamine, hexane, ethyl acetate, octane | $N_2$ flow and $O_2$, 35 °C | TMA | c-ALD | 90.9% | | Solutions can keep stable in water for more than 1 week | | | | | | [124] |
| $CsPbBr_3$@/$Al_2O_3$ | CVD | 40 mTorr, Ar, 550 °C | $PbBr_2$ and CsBr | 200 °C, $N_2$ | TMA | ALD | | | Keep stable in water for 31 days | | | | | Lasing | [4] |
| $CsPbBr_3$@$SiO_2$/$AlO_x$ | Hot-injection | Vacuum, $N_2$ protection, temperature (80 ºC, 120 ºC, 170 ºC), ice-bath | ODE, $PbBr_2$, cesium stearate, OA, oleylamine, ethyl acetate, toluene | 25 °C, RH of 45%, 50 ºC, $N_2$ | TMOS and TMA | Sol-gel and ALD | 65% (519 nm) | 20 nm | The PL intensity keep the same after immersion in water for 2 h and still have strong emission after 20 days in water | 8% loss in the PL intensity after treatment at 100 ºC under $N_2$ | 80% of the initial PL intensity remains after 40 h under 450 nm blue light (200 mW cm$^{-2}$) | | | | [125] |
| $CsPbBr_3$@$SiO_2$ | Impregnation | 150 °C, vacuum | CsBr, $PbBr_2$, DMSO | Temperature (80 °C, 100 °C, 550 °C), muffle Oven, vacuum | TEOS | Template-assisted method | 32.5% (524 nm) | 23 nm | | | | | | Anti-counterfeiting | [126] |
| $MAPbBr_xI_{3-x}$@$SiO_2$ | Impregnation | 95 °C | MABr, MAI, $PbBr_2$ | Temperature (35° ~ 80 °C , 500 °C, 150 | TEOS, TMOS, TPOS | Template-assisted method | 5.5 ± 1.1% (520 nm) for $MAPbBr_3$ @$SiO_2$ | | | | | | | | [68] |
| | | | | °C), muffle Oven, vacuum | | | | | | | | | | | |
| A (A= MA, FA, Cs)PbX (X= Cl, Br, I)$_3$@$SiO_2$ | Impregnation | Temperature: 120 °C (for MA and FA salts) or 150°C (for Cs salts), 80 °C, vacuum oven | FABr, MABr, CsBr, MAI, CsI, MAI, PbI, $PbBr_2$ | Dried at 150 °C under vacuum before using | Commercial MS | Template-assisted method | Average PLQY: 48 ± 24% (530 nm) for $CsPbBr_3$ NCs@$SiO_2$; | 20~22 nm for $CsPbB_3$ NCs@ $SiO_2$ | | | Retaining the same high QY after 6 days of photo-annealing under the UV lamp (365 nm, 400 μW/cm$^2$). | | | | [127] |
| $CsPbBr_3$/$Al_2O_3$ | Impregnation | Temperature: 100 °C, 160°C in a dry oven | CsBr, $PbBr_2$, DMSO | 500 °C | Commercial α-$Al_2O_3$ | Template-assisted method | 522 nm (8%–15%) | 17 nm | | Less than 1 nm blue shift in PL peak and no drop for PLQY at 150 °C for 20 days | | No change in the PL peak and PLQY over five months under ambient conditions | | LEDs | [128] |
| $CsPbBr_3$@$SiO_2$ | Impregnation | | CsBr, $PbBr_2$, water | Muffle furnace, temperature: 400 °C, 500 °C, 600 °C, 700 °C, 800 °C, and 900 °C | All-silicon molecular sieves (MCM-41) | Template-assisted method | 71% (~ 520 nm) | 20 nm | 105% of the initial PLQYs remains after immersion in water for 50 days | | | | *Water and photo stability*: no change in PLQY even testing in water under strong illumination (450 nm LED, 175 mW cm$^{-2}$) for 50 days; *Chemical stability*: no obvious change in PLQY after immersed in 1 M HCl for 50 days. | LEDs | [37] |
| $MAPbBr_3$@HSNSs/PVDF | Impregnation | Vacuum, 60 ºC, 110 ºC | MABr, $PbBr_2$, MFA | Vacuum-drying at 60 °C | TEOS and DMDMS | Template-assisted method | 85.5% (520 nm) | 27 nm | 55% of the initial PL intensity remains after 2 h | | 88.1 % of the initial PL intensity remains after 50 h under illuminating (365 nm, 6 W) | 97% of the initial PL intensity remains after 2 months | | LEDs | [129] |

| | | | | | | | | | | | | | | |
|---|---|---|---|---|---|---|---|---|---|---|---|---|---|---|
| | | | | | | | | | in water | | | storage in open air | | |
| $CsPbBr_3/SiO_2$ | Hot-injection | Vacuum, $N_2$ protection, temperature (90 ºC, 120 ºC, 150 ºC), ice-bath | ODE, $PbBr_2$, $Cs_2CO_3$, OA, oleylamine, toluene | 100 ºC | TEOS | Template-assisted method | 515 nm | 20 nm | | | | | LEDs | [130] |
| $Cs(Pb_{0.66}/Mn_{0.34})Cl_3@SiO_2$ | Impregnation | Vacuum, 50 ºC, 120 ºC | $Cs_2CO_3$, $PbCl_2$, $MnCl_2$, DMSO | 80 °C 100 °C, 150 °C vacuum | TEOS | Template-assisted method | 32% (588 nm) | | | 3.6% lose in the PL intensity after the heating-cooling process from room temperature to 180 °C and then room temperature | 80% and 50% of the initial PL intensity remains under continuous UV irradiation (365 nm, 0.5 W $cm^{-2}$) for 24 h, 72 h, respectively; | 90% of the initial intensity remains after 12 h under RH of 40% | LEDs | [131] |
| $CsPbBr_3/Al_2O_3$ | Impregnation | Vacuum, 60 ºC oven drying | CsBr, $PbBr_2$, DMF | 60 °C in a hot-air oven and calcinated at 500 °C | ASB | Template-assisted method | 42% (507 nm) | 24 nm | | | | | | [132] |
| $CsPbBr_3/MS@SiO_2$ | Impregnation | 80 ºC | CsBr, $PbBr_2$, DI water | 80 °C; 40 °C and RH 80%; 60 °C under vacuum; 600 °C oven | TMOS and commercial MS (SBA-15) | Template-assisted method and sol-gel | 93% (513 nm) | 18 nm | 93% of the initial intensity remains after dispersed into water for 120 h | 94% of the initial PL intensity remains after 12 h under continuous heating at 150 °C | 90% of the initial stability remains after continuous illumination with a 450 nm LED (350 mW $cm^{-2}$) for 240 h | ~ 100% of the initial PL intensity after storage (25 °C, RH of 60%) for 60 days | LEDs | [133] |
| $CsPbBr_3@SiO_2$ | Hot-injection | Vacuum, $N_2$ protection, temperature (10 ºC, 120 ºC, 150 ºC, 190 ºC), ice-bath | ODE, $PbBr_2$, $Cs_2CO_3$, OA, oleylamine, hexane | | Commercial SBA-15 | Physical method | 519 nm | 20 nm | | Nearly the same PL intensity after the heating-cooling testing from 25 °C to 100 °C, and 100 °C to 25 °C | About 80% of the initial PL intensity remains after 96 h under continuous UV-light (365 nm, 6 W) irradiation | | LEDs | [134] |
| $CsPbBr_3@SiO_2$ | Hot-injection | Vacuum, $N_2$ protection, temperature (100 ºC, 120 ºC, 150 ºC, 180 ºC), ice-bath | ODE, $PbBr_2$, $Cs_2CO_3$, OA, oleylamine, hexane | | Commercial mesoporous silica | Physical method | 46.2% (521 nm) | 25 nm | Can survive more than 3 h in aqueous solution | 35% of the PL intensity remains as the temperature increases from 298 K to 398 K | 80% of the initial PL intensity remains under 365 nm light after 120 h | PL emission and shape do not change after 1 month under ambient condition | LEDs | [135] |
| $CsPbBr_3@SiO_2$ | Hot-injection | Vacuum, $N_2$ protection, temperature (120 ºC, 150 ºC), ice-bath | ODE, $PbBr_2$, $Cs_2CO_3$, 2-hexyldecanoic acid, oleylamine, hexane, n-butanol | | Commercial $SiO_2$ spheres | Physical method | 96% (517 nm) | 17.4 nm | | 32% of the initial PL intensity remains after annealed at 70 °C for 10 h | | No significant reduction in PLQY after 70 days | Lasing | [136] |
| $CsPbBr_{1.5}I_{1.5}/TiO_2$ inverse opal electrode | Hot-injection | Vacuum, $N_2$ protection, temperature (120 ºC, 165 ºC, 200 ºC), ice-bath | ODE, $PbBr_2$, $PbI_2$ $Cs_2CO_3$, OA, oleylamine, hexane, acetone | 500 °C | Butyltitanate | Physical method | 608 nm | | | | | | PEC sensing | [137] |
| $CsPbBr_3@SiO_2$ (disperse in different solvents, toluene-T, hexane-H, acetone/octane mixture-AO) | Hot-injection | Vacuum, Ar protection, temperature (150 ºC) | ODE, $PbBr_2$, $Cs_2CO_3$, OA, oleylamine, toluene, acetone | 100 °C, 550 °C | TEOS | Physical method | *$CsPbBr_3@SiO_2$-T*: 5.3% (519±2 nm); *$CsPbBr_3@SiO_2$-H*: 4.9% (520±2 nm); *$CsPbBr_3@SiO_2$-AO*: 7.8% (517±2 nm) | *$CsPbBr_3@SiO_2$-T*: 19±2 nm; *$CsPbBr_3@SiO_2$-H*: 19±2 nm; *$CsPbBr_3@SiO_2$-AO*: 19±2 nm | | | *$CsPbBr_3@SiO_2$-AO*: brightly luminescent green spots can still be observed in the film under intense UV light (365 nm, pulse laser of 200 μJ $cm^{-2}$) for 1 h. | *$CsPbBr_3@SiO_2$-T*: 85% and 31% of the original PL intensity remains after 7 days and 110 days in air, respectively; *$CsPbBr_3@SiO_2$-H*: 11% of the original PL intensity remains after 110 days in air | | [138] |
| $CsPbBr_3@SiO_2$ | Hot-injection | Vacuum, $N_2$ protection, temperature (120 ºC, 180 ºC), ice-bath | ODE, $PbBr_2$, $Cs_2CO_3$, OA, oleylamine, acetone, toluene | 95 ºC, 35 ºC | TEOS | SILAR | 68% for powder and 83% for solution (519.3 nm) | | Can survive more than 30 min in water under violent ultrasonication | 67% of the initial emission can remain under 373 K | About.80% of the original PL intensity remains after being exposed to the UV-light (365 nm) for 120 h | | LEDs | [139] |
| $CsPbBr_3@SiO_2$ | Hot-injection | Vacuum, $N_2$ protection, temperature (100 ºC, 120 ºC, 150 ºC, 170 ºC), ice-bath | ODE, $PbBr_2$, $Cs_2CO_3$, OA, oleylamine, hexane | 60 ºC, 75 ºC, 110 ºC drying oven | APTES/TEOS | SILAR | 89% (~ 514 nm) | | | 65% of the initial PL intensity remains after being treated under 100˚C heating | 89% of the initial PL intensity remains after being continuously exposed to UV light for 72 h | | LEDs | [140] |
| $SiO_2@CsPbBr_3@SiO_2$ | Hot-injection | Vacuum, $N_2$ protection, temperature (120 ºC, 150 ºC), ice-bath | ODE, $PbBr_2$, $Cs_2CO_3$, OA | 60 ºC under vacuum | APTES and TEOS | SILAR and sol-gel | >82% (521 nm) | 24 nm | 20% decrease in PL intensity after1 h in DI water | 40% decrease in PL intensity when being heated to 110 °C | 40% decrease in PL intensity after 100 h of UV irradiation | 15% decrease in PL intensity after 60 days in air | Determination of $Fe^{3+}$ in Water | [141] |

| | | | | | | | | | | | | | | | |
|---|---|---|---|---|---|---|---|---|---|---|---|---|---|---|---|
| $CsPbBr_3$@h-$Al_2O_3$ | Hot-injection | Vacuum, $N_2$ protection, temperature (120 ºC, 170 ºC), ice-bath | ODE, $PbBr_2$, $Cs_2CO_3$, OA, oleylamine, methyl acetate | 120 ºC, 180 ºC, 700 ºC drying oven | $Al_2(SO_4)_3$·18 $H_2O$ | SILAR | 80% (522 nm) | 27 nm | High stability against destruction in water for more than one month | The PL intensity undergoes an ignorable change at 100 °C, and shows 100% PL recovery after the thermal process | | | | | [142] |
| $CsPbBr_3$/mesoporous-$SiO_2$ | Low-temperature molten salts | 350 ºC in a furnace, 40 ºC vacuum oven | CsBr, $PbBr_2$, DMSO or DMF, inorganic salts ($KNO_3$−$NaNO_3$−KBr) | | Commercial MCM-41 mesoporous-$SiO_2$ | Low-temperature molten salts | 90% | | ~95% PL remains after 30 days in air | ~90% PL retention after 3h at 180 °C | Retaining the initial luminescence after 240 h under high irradiation (450 nm, 200 mW $cm^{-2}$) | | ~55% PL remains after 30 days in in aqua regia; PLQY is 83% after 24 h of incubation at RT in an mixture aqueous solution of NaCl, $CaCl_2$, $MgCl_2$, $Na_2SO_4$, and $NaHCO_3$ | LEDs | [143] |

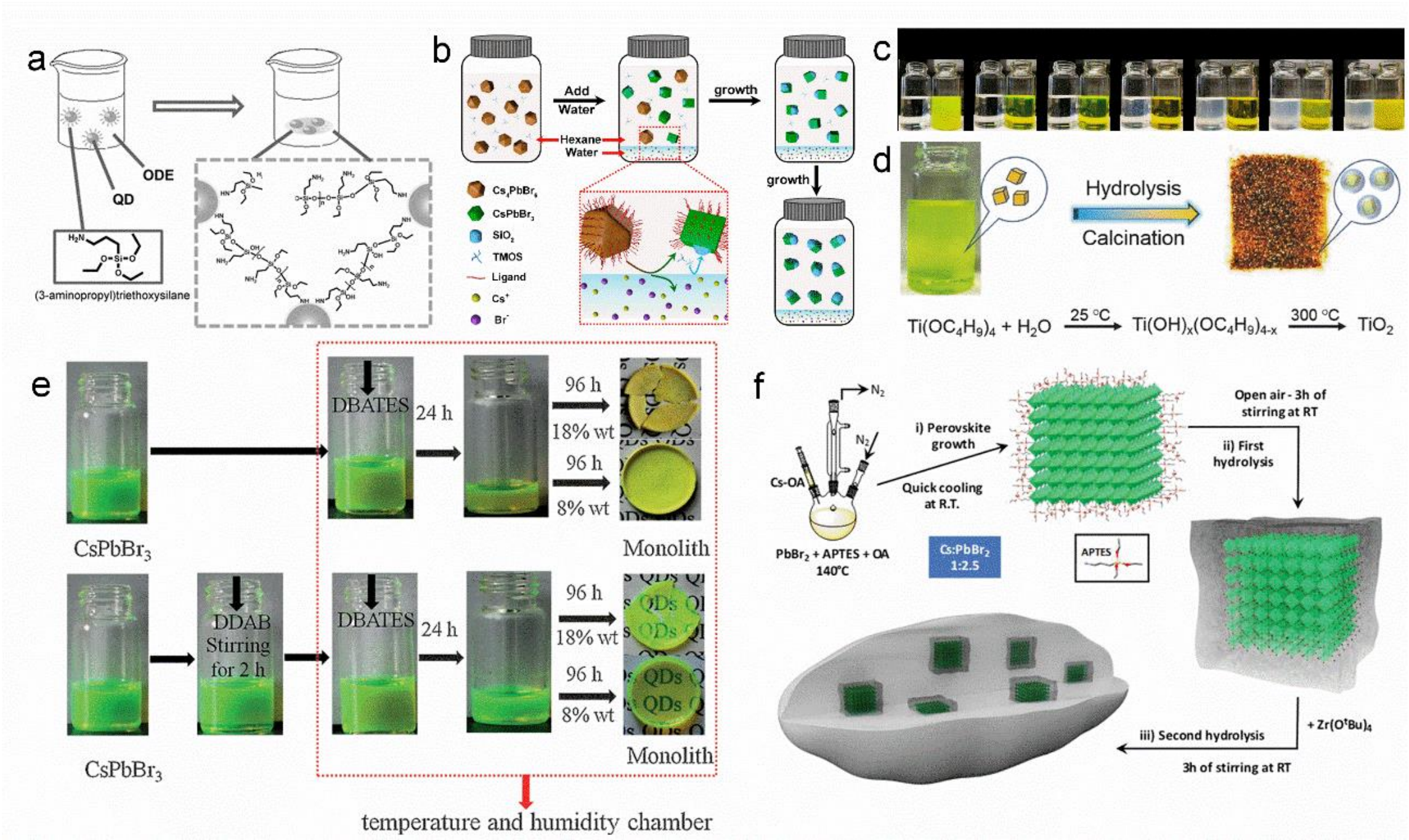

**Figure 2.** a) The sol-gel process for the formation of the MHPs@$SiO_2$ structure. Reproduced with permission.[7] 35 Copyright 2016 ~~2017~~, Wiley-VCH. b) The water-triggered transformation process and sol-gel process for the formation of $CsPbBr_3$@$SiO_2$ Janus NCs. Reproduced with permission.[76] Copyright 2018, American Chemical Society. c) The formation process for $CsPbBr_3$@$TiO_x$ using TBOT as the precursor: TBOT in toluene without $CsPbBr_3$ NCs (left) and with $CsPbBr_3$ NCs (right); d) The fabrication process for the $CsPbBr_3$@$TiO_2$ NCs. c,d) Reproduced with permission.[88] Copyright 2018, Wiley-VCH. e) The sol-gel formation of $CsPbBr_3$@ $SiO_2$/$Al_2O_3$ without (up part) and with (down part) DDAB treatment. Reproduced with permission.[69] Copyright 2017, Wiley-VCH. f) Schematic of the $CsPbBr_3$@$SiO_2$/$ZrO_2$ prepared through the sol–gel reaction. Reproduced with permission.[89] Copyright 2020, Wiley-VCH.

However, the above method is suitable for the preparation of the MHPs@$MO_x$ composite at the multiple-particle level. This will be very harmful in some applications, for example, single nanoparticle is highly desired in the bio-related fields as well as the LED films. Therefore, it is essential to develop a new method to prepare the MHPs@$MO_x$ composite at a single particle level. For instance, monodisperse $CsPbX_3$@$SiO_2$ and $CsPbBr_3$@$Ta_2O_5$ Janus NCs were prepared through a water-triggered transformation process and a sol-gel process – **Figure 2b**.[76] Herein, $Cs_4PbX_6$ NCs were firstly prepared through the hot-injection process that was developed in 2015[27] and then dispersed in hexane to form a solution. Then, tetramethoxysilane (TMOS) was added to the above solution, and deionized (DI) water was quickly injected into the mixture under vigorous oscillation for 5 min. During this process, the CsBr surface will peel from the hexane/water surface as CsBr is highly soluble in water. And the hydrophobic capping ligands around the CsBr, like OA and oleylamine, will be removed at the same time, followed by the adhesion of the Si-precursors. After being kept undisturbed for 12 h, the $CsPbX_3@SiO_2$ architecture is successfully formed. In this system, the oscillation (using a vortex reactor) is the key for the formation of the $CsPbBr_3@SiO_2$ Janus structure. Indeed, free $SiO_2$ particles are observed if strong stirring is applied. The authors highlighted that the selection of Si-precursors is important, as only $CsPbBr_3$ NCs can be obtained after a standing time of 12 h when TEOS (lower hydrolysis rate) is used. Likewise, $CsPbBr_3@Ta_2O_5$ Janus structures were prepared through using tantalum(V) ethoxide (TTEO) as the precursor for $Ta_2O_5$. In a followed up work, the same group successfully synthesized the $CsPbBr_3@ZrO_2$, using $Zr(OC_4H_9)_4$ as the precursor.[70] Here, the Zr-precursor concentration has a significant effect on the spectroscopic features like excited state lifetime and PLQY values. The increasing amount of the Zr precursor favors the formation of more $ZrO_2$ to protect the $CsPbBr_3$ NCs. However, free $ZrO_2$ and undesired hydrolytic byproducts will form when excess amount of Zr precursor is introduced, which will be harmful for both the structure and optical properties of $CsPbBr_3$.

Similar to MHPs/$SiO_2$, others like $CsPbBr_3@TiO_2$[84,87,88] and $CsPbBr_3@SnO_2$,[84] $CsPbBr_3/ZrO_2$,[70] Na: $CsPb(Br, I)_3@Al_2O_3$ nanocomposites[120] were also prepared *via* the sol-gel method. Specifically, the temperature is key for the synthesis of $CsPbBr_3@TiO_2$. As the temperature has an important effect on the crystallinity of the obtained $TiO_2$ shell. For example, amorphous shell will form under low-temperature hydrolysis, while high temperature favors the formation of compact anatase phase $TiO_2$ shell and therefore increases the stability as well as the charge carrier transfer. As reported in previous sol-gel prepared $TiO_2$ nanoparticle works, titanium butoxide (TBOT) toluene solution (as the Ti-precursor) was introduced to the $CsPbBr_3$ toluene solution under stirring and a constant relative humidity (RH) of 30% at room temperature (RT).[88] As the time increases, the intermediate product $Ti(OH)_x(OC_4H_9)_{4-x}$ formed and gradually deposited on the surface of $CsPbBr_3$– **Figure 2c**. After that, when the obtained precipitates were only treated under vacuum and RT, the amorphous $TiO_x$ matrix forms as proved by the TEM image that gray shadow area is surrounding the $CsPbBr_3$ NCs. However, the $CsPbBr_3@TiO_2$ structure with a shell of 5 ± 3 nm were realized when a post calcination process with a high temperature of 300 ℃ was applied – **Figure 2d**. As expected, the $CsPbBr_3@TiO_2$ shows better water stability than those for the $CsPbBr_3@TiO_x$ – this will be discussed in section 3. Similar results were obtained in $CsPbBr_3$@amorphous-$TiO_2$ formed when the hydrolytic product was dried at a low temperature of 120 ℃ using tetrabutyl titanate (TBT) hexane solutions as precursor.[87]

Finally, the sol-gel method also allows to fabricate binary $MO_x$ coated MHPs. Li's group selected a single molecular di-sec-butoxyaluminoxytriethoxysilane (DBATES, $(sec\text{-}BuO)_2$-Al-O-$Si(OEt)_3$) as the precursors for both Al and Si.[69] When DBATES was directly added into the $CsPbBr_3$ toluene solution, the color changed from green to yellow along with the quenching of the fluorescence and the obtained product was not transparent. And a low PLQY of ca. 30 % was obtained. During this process, the byproducts, like water and alcohols, produced from the hydrolysis and condensation process will disrupt the surface of $CsPbBr_3$ QDs, leading to the loss of the ligands and the aggregation of the QDs. Therefore, didodecyl dimethyl ammonium bromide (DDAB) was firstly introduced to achieve ligand exchange, since most of OA and oleylamine ligands were replaced by DDAB. Herein, DDAB has excellent steric hindrance, can provide better passivation for $CsPbBr_3$ QDs than OA or oleylamine. Finally, transparent, green, and high emissive $CsPbBr_3@SiO_2/Al_2O_3$ with a high PLQY of 90% was obtained after a 96 h sol-gel reaction – **Figure 2e**. As expected, the as-prepared $CsPbBr_3@SiO_2/Al_2O_3$ exhibits better photo-stability and thermal stability than that of the pristine $CsPbBr_3$, $CsPbBr_3@SiO_2$ and $CsPbBr_3@Al_2O_3$ because of the compact $SiO_2/Al_2O_3$ shell proving high barrier for both oxygen and moisture.

Inspired by Li's work,[69] our recently published paper reported the first dual metal oxide‐coated $CsPbBr_3@SiO_2/ZrO_2$ composite which was prepared through a one‐pot synthesis via the kinetic control of the sol–gel reaction (**Figure 2f**).[89] In our work, $CsPbBr_3@SiO_2$ was firstly prepared following the method reported by Yu's group.[35] Then, the zirconia source—i.e., zirconium(IV) tert‐butoxide (ZTB)—was added dropwise to the $CsPbBr_3@SiO_2$ system. Taking advantage of the presence of trace amount of water in the reaction medium, the $CsPbBr_3@SiO_2/ZrO_2$ composite can be obtained through controlling the molar ratio of $CsPbBr_3$:APTES:ZTB at 1:28.5:8.5. Finally, the $CsPbBr_3@SiO_2/ZrO_2$ composite powders can be achieved through centrifugation and drying under ambient environment overnight. This simple and easily scalable method enables the excellent stability of the $CsPbBr_3@SiO_2/ZrO_2$ composite against surrounding environmental as well as outstanding optical properties. For example, about 100% of the initial PLQY can be retained after 1 month exposed in air, and a narrow FWHM of 16 nm. Other examples about using the sol-gel method to fabricate the MHPs@$MO_x$ composites through sol-gel method are shown in **Table 1**.

All-in-all, the successful fabrication of MHPs@$MO_x$ through the sol-gel method requires a fine control of the i) the type precursor to control the reaction rate, ii) the concentration of the precursor to achieve ultrathin and single particle level core/shell structure, iii) the temperature to tune the crystalline state of the shell, and iv) the ligand exchange to fulfill transparent and highly emissive features.

### 2.2. ALD

ALD technique allows nm-controlled deposition of metal oxides layers on patterned surfaces and materials.[144-147] The nature of the coatings can be controlled by the precursors, while the thickness of the coatings can be tuned by varying the ALD cycles.[144,148,149] For instance, ALD is an effective strategy for the deposition of $MO_x$, such as $TiO_2$,[150] $Al_2O_3$,[4,123,125] $SiO_2$,[151] ZnO[152] to stabilize QDs.[123,148,153,154] Typically, previous works have proved that the $Al_2O_3$ coating that is deposited on the surface of PbSe through ALD could improve the oxidative and photothermal stability, and impedes the movement of the internal atomic and molecular.[148]

The first work dealing with ALD to fabricate the $CsPbBr_3@MO_x$ was reported in 2017.[123] Here, the prepared $CsPbBr_3$ NCs were firstly spin-coated on substrates, like glass or p-doped silicon, to form a uniform layer. The deposition of the amorphous $AlO_x$ was realized using trimethylaluminum (TMA) and ultrapure $H_2O$ for the Al-precursor and $O_2$ source, respectively. Each deposition cycle consists of 4 stages: TMA pulse, purging, $H_2O$ pulse, and purging. As the number of cycles increases, two different growth regimes are proposed: the infilling (<75 cycles) and the overcoating processes – **Figure 3a** and **b**. For the former step, a large surface area is exposed, providing abundant sites for the nucleation of alumina and the TMA can react with the partially oxidized QD surface. This leads to a higher alumina growth per cycle (GPC) of 1.06 ng $cm^{-2}$ $cycle^{-1}$. During the overcoting process, a steady state linear regime with a lower GPC of 0.21 ng $cm^{-2}$ $cycle^{-1}$ is reached.[144] The filling and the overcoating processes can also be proved by the cross-sectional scanning TEM (STEM) electron energy loss spectroscopy (EELS) image of the $CsPbBr_3@AlO_x$ – **Figure 3c**. Here, Al disperses onto the $CsPbBr_3$ layer and the bottom substrate layer to form the desired core/shell structure. In addition, the parameters during the deposition process have an important effect on the PL emission and the stability of the $CsPbBr_3@AlO_x$ film. The PLQY of the obtained $CsPbBr_3@AlO_x$ increases when minor amount of TMA (can be tuned by the TMA exposure time) is added, and a reversed trend is observed when increases the concentration of TMA. The first increase is related to the passivation of the surface trap states, while the decrease is caused by the ligand exchange of the QDs. Besides, the working cycles of the ALD also has an important effect on the stability of the layer. The $CsPbBr_3@AlO_x$ layer show better stability against water as the increasing ALD cycles (**Figure 3d**). Meanwhile, the temperatures (50 ºC, 60 ºC, 70 ºC) also significantly affect the PL emission of the composite, lower temperature (50 ºC) favors higher emission because sintering and changes in structure will happen under higher temperatures (60 ºC, 70 ºC). [155]

Inspired by this work, Chen's group firstly dispersed the $CsPbBr_3$ NCs on the $SiO_2$ spheres (9.8 nm) using the sol-gel technique.[125] Then, $Al_2O_3$ was deposited on the surface of this material using ALD. The advantage of this process is the lack of negative impact related to break of the surface ligands of $CsPbBr_3$ NCs as well as will not affect the optical features with respect to emission performance. As depicted in **Figure 3e**, the TMA molecules are chemisorbed on the unsaturated sites of both, the $CsPbBr_3$ NCs surface and the $SiO_2$ surface, and the unsaturated sites are gradually protected as the growth of the alumina. The FTIR spectra show a strong peak located at 1085 $cm^{-1}$, which is in according with the stretching vibration of the Si−O−Si bond **– Figure 3f**. After the ALD deposition of $Al_2O_3$, no significant change is noted for the Si−O−Si bond. This means that the ALD does not cause the change of the $CsPbBr_3$ NCs structure. And the PLQYs increase from 45% to 65% after the ALD process instead of PL quenching. They also investigated the effect of the ALD deposition temperature, the ALD working cycles on the PLQYs and the stability of the obtained composite. Similarly, the PLQY increases as the temperature rises from 20 to 50 °C, but the PLQY decreases q uickly under higher temperatures (from 50 °C to 100 °C). In Addition, better water stability is realized for the $CsPbBr_3$ QDs embedded in Silica and Alumina coated Luminescent Sphere (QDs-SALS) composite as the ALD working cycles increase from 10 to 50.

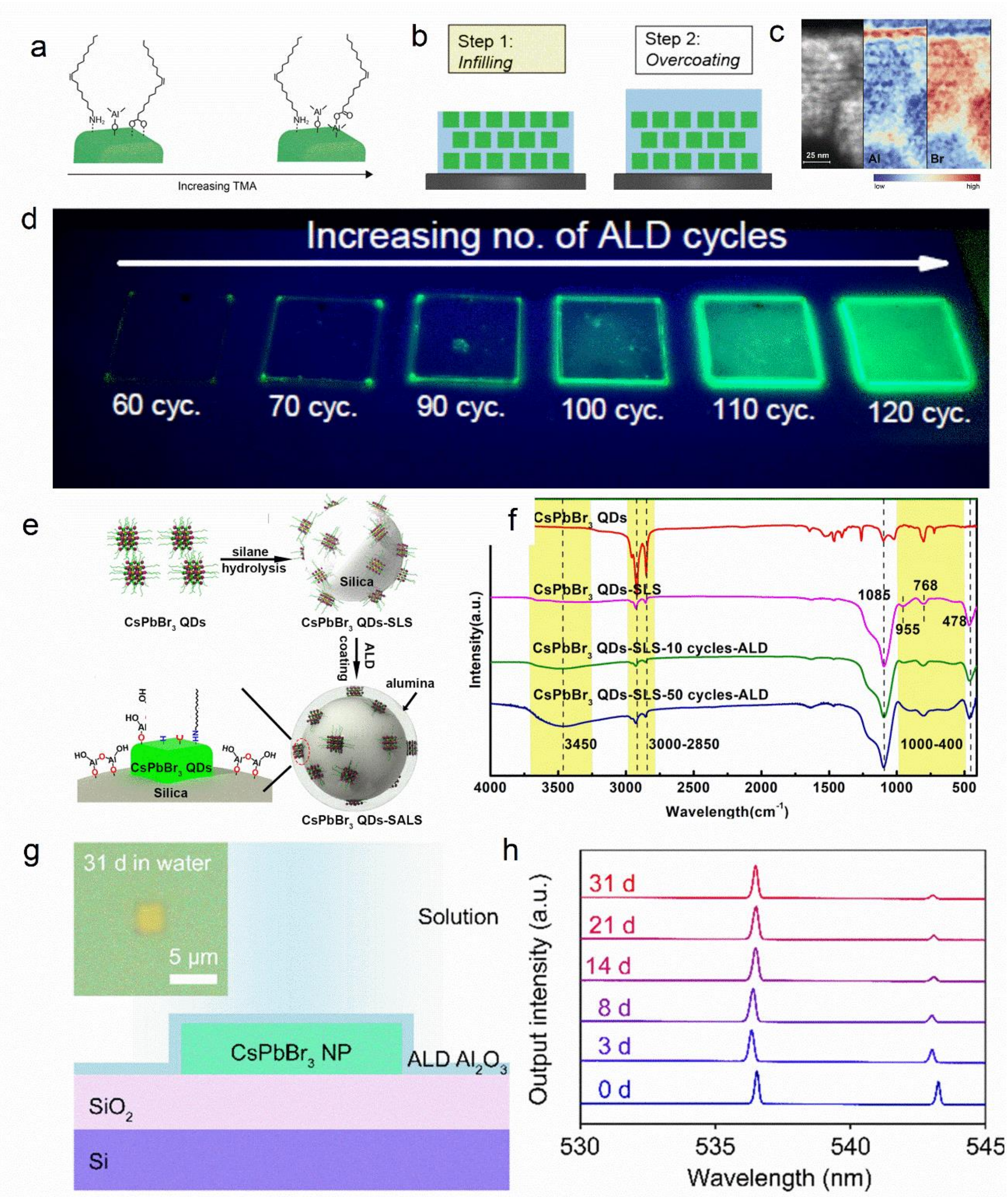

**Figure 3.** a) The proposed nucleation process of $AlO_x$ during the ALD deposition process; b) The schematic diagram for the infilling and over-coating processes; c) STEM-EELS elemental dispersion maps for Al and Br; d) The degradation behavior of the $CsPbBr_3$ QD/$AlO_x$ nanocomposite films prepared by different ALD cycles. a,b,c,d)Reproduced with permission.[123] Copyright 2017, Wiley-VCH. e) The fabrication process for the $CsPbBr_3$ QDs SALS; f) The corresponding FTIR spectra. e,f) Reproduced with permission.[125] Copyright 2018, American

Chemical Society. g) Schematic of the $Al_2O_3$ coated $CsPbBr_3$ nanoplate laser on a Si/$SiO_2$ substrate; h) The normalized lasing spectra collected for the devices with a 50 nm $Al_2O_3$ layer collecting day by day. g,h) Reproduced with permission.[4] Copyright 2020, American Chemical Society.

Furthermore, a waterproof $CsPbBr_3$ nanoplate laser was realized through a large-scale ALD prepared $Al_2O_3$ layer (**Figure 3g**).[4] Herein, the $CsPbBr_3$ nanoplate was firstly prepared by the chemical vapor deposition (CVD) method. Then, $Al_2O_3$ was coated on the $CsPbBr_3$ nanoplate layer through ALD system, using TMA and DI water as the precursors. The temperature for ALD was set at 200 °C and $N_2$ was used to carry these two precursors' vapors to a sealed reaction chamber. The thickness of the $Al_2O_3$ layer is decided by the ALD working cycles (each cycle contributes to a thickness of 1.06 Å). During the ALD process, a minor amount of water vapor will quickly react with TMA to form the $Al_2O_3$ layer. The $Al_2O_3$ layer will protect $CsPbBr_3$ nanoplate from the following ALD process. The stability of the laser devices with a 50 nm $Al_2O_3$ layer was investigated through immersion in water while collecting the laser spectra day by day. The device can still lase without obvious lasing peak shifts after days (**Figure 3h**). More importantly, the topography of $CsPbBr_3$ nanoplate does not change after the measurement (inset **Figure 3g**). Similar as the above reports, the thickness of $Al_2O_3$ layer also plays an important role in the water stability of the laser devices. When immersed in water, the devices with different thicknesses of $Al_2O_3$ layer, like 30, 20, 10 nm, can not lase after 16 days, 72 h, and a few minutes, respectively.

The ALD method must be considered an effective way to control the thickness and morphology of the coatings, but it is difficult to predict how the temperature, the precursor concentration, and the thickness of the coatings rule the final photophysical features. In addition, the number of works are still small and only focusing on $AlO_x$ coatings currently. Finally, the ALD deposition requires a dedicated instrumentation that is not widely spread and it is not straightforward upscale.

### 2.3. Template-Assisted Method

Template-assisted method is an effective strategy for preparing nanomaterials, such as oxides[156-158] and carbons.[159] In this method, all the reactions are artificially limited in a space for both, liquid phase or gas phase reactions. The size, shape, structure, and properties of the nanomaterials can be precisely controlled by using the template as the carrier. Moreover, these templates are usually commercially available materials with regular structure – *i.e.*, pores and layers, like anodized aluminum oxide (AAO),[160] metal organic framework (MOF),[43,161-165] mesoporous silica (MS),[68,127] zeolites.[166] In addition, these pore structures act as an individual room to isolate these nanomaterials for the direct *in-situ* growth, and therefore the aggregation that always causes the instability of nanomaterials is highly restricted.

Yamauchi's group adopted highly ordered MS as the template for the $MAPbBr_xI_{3-x}$.[68] The MS matrices exhibit different pore sizes of about 7.1, 6.2, 4.2, 3.7, and 3.3 nm. The MS template powders were treated under vacuum to reach reduced pressure before using. Firstly, the MHPs precursors were slowly added to the MS matrices, followed by vortex mixing. The capillary forces drive the precursor solution to fill the channels of the template because of the low interfacial tension between N,N-Dimethylformamide (DMF) and the MS surface. The $MAPbBr_xI_{3-x}$ NCs form gradually through self-organized process within the MS as the evaporation of the solvent. Importantly, there are no side-products forming in this process, as no other ligands are introduced. Final products are obtained through drying under vacuum at 95 °C to ensure the complete removal of the solvent from the pores. These $MAPbBr_xI_{3-x}$ NCs are closely dispersed along the channels and no obvious damage or doping has been noted for the MS templates. Additionally, the colors of the obtained powders are closely related with the pore size of the MS – **Figure 4a**, changing form light color to dark color as the pore size of the template increases because of the quantum confinement effect. For the $MAPbX_3$, the exciton Bohr radius is about 20 Å, and therefore the quantum confinement can only be observed when the radius of the crystals is less than 5 nm.[167] This means that the crystal size is consistent with the distribution of the pore size. Moreover, the authors also described that the MS templates can improve the resistance to the UV irradiation. They compared the lifetime of the $MAPbBr_3$@MS composite before UV irradiation, after UV irradiation as well as after an overnight recovery. The lifetime of the composite can be recovered or even longer than before, while limited recover happens for the bulk $MAPbBr_3$. However, the PLQYs of the $MAPbBr_3$@MS composite are very low (<5.5%) because of the lack of the ligands.

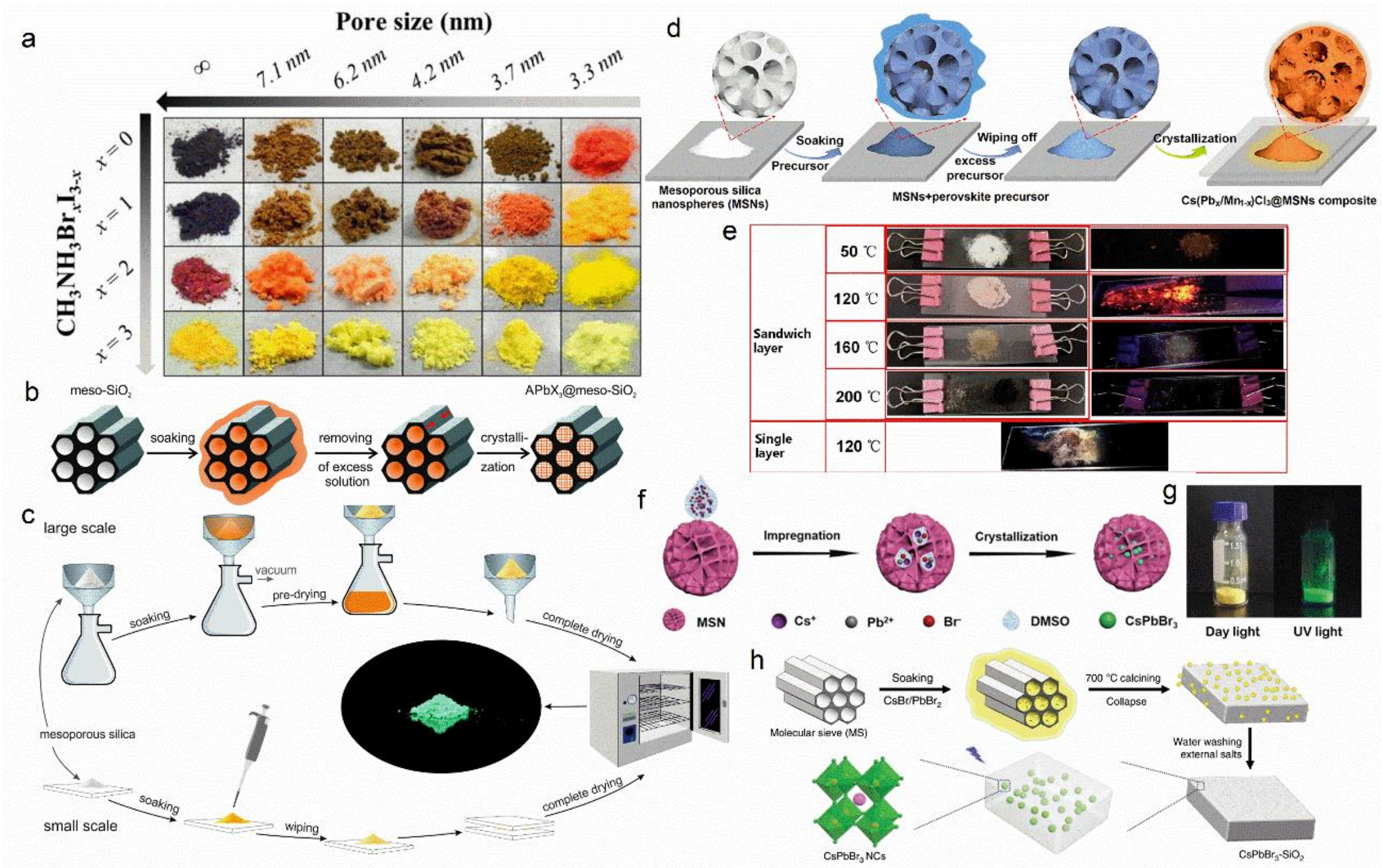

**Figure 4.** a) Photographs of the obtained $MAPbBr_xI_{3-x}$/MS powders with different pore sizes and compositions. Reproduced with permission.[68] Copyright 2016, American Chemical Society. b) The process for the template-assisted synthesis of $APbX_3$ NCs using MS; c) The procedure for large-and small scale synthesis of $APbX_3$/MS. b,c) Reproduced with permission.[127] Copyright 2016, American Chemical Society. d) The preparation process of the $Cs(Pb_x/Mn_{1-x})Cl_3$@MSNs; e) Optical photographs of the obtained products processed at different conditions. d,e) Reproduced with permission.[131] Copyright 2020, American Chemical Society. f) The schematic diagram for the preparation of the $CsPbBr_3$ NCs@MSNs using MSNs as the template; g) The obtained $CsPbBr_3$ NCs@MSNs powders under day light (left) and UV light (right). f,g) Reproduced with permission.[126] Copyright 2020, American Chemical Society. h) The schematic diagram for the fabrication of the ceramic-like highly stable $CsPbBr_3@SiO_2$ composite. Reproduced with permission.[37] Copyright 2020, Nature Publishing Group.

Kovalenko's group also adopted the same method to fabricate the A (A= MA, FA, Cs) $PbX_3$ (X= Cl, Br, I)/MS composite – **Figure 4b**,[127] but removed the extra solvent through damping with filter paper and vacuuming on a glass filter for the small- (<50 mg of MS powders) and large- (>50 mg of MS powders) scale syntheses before drying thoroughly in the vacuum oven, respectively – **Figure 4c**. This work shows higher PLQYs for $CsPbBr_3$ NCs/MS, with the average and the highest values of 48 ± 24% and ∼ 90%, respectively; this is comparable to the $CsPbBr_3$ NCs prepared using excess ligands.

Importantly, the temperature and the way to stack the powders during the drying process are very important to get highly emissive powders for the MHPs/$MO_x$ that fabricated through the template-assisted method. Yellow emission $Cs(Pb_x/Mn_{1-x})Cl_3$@mesoporous silica nanoparticles (MSNs) were fabricated following the above works (**Figure 4d**).[131] Specifically, MSNs were dispersed in the mixture of $Cs_2CO_3$, $PbCl_2$, $MnCl_2$, DMSO, using filter paper to remove the excess solution. Then, the powders were dried at 120 ºC through being sandwiched between 2 glass sides (**Figure 4e**). Only weak yellow emission can be achieved when the saturated MSNs are stacked on one piece of glass, maybe because that the heating on the surface is not homogeneous. Higher emission will be observed when the saturated MSNs are sandwiched between 2 glass sides. In addition, too low (50 ºC, $Cs(Pb_x/Mn_{1-x})Cl_3$ can not form) or too high temperature (more than 120 ºC, the organic solvents will be carbonized and the $Cs(Pb_x/Mn_{1-x})Cl_3$ cannot grow homogeneously) is also harmful for the emission.

Similarly, $CsPbBr_3$ NCs@MSNs were also prepared through using the MSNs as the template (**Figure 4f**).[126] CsBr and $PbBr_2$ were well dissolved in dimethyl sulfoxide (DMSO)and then MSNs were impregnated in the precursor solution. Subsequently, the CsBr/$PbBr_2$ solution would permeate into the pore structure of MSNs. Finally, the yellowish $CsPbBr_3$ NCs@MSNs powders were obtained through drying the mixture in the oven and green emission can be observed under UV light (**Figure 4g**). However, the powders also show a low PLQY of 32.5% because the lack of the ligands.

The main drawback of the MS template is the lack of a full protection or encapsulation of the MHPs from the surrounding stresses, as many pores are easily exposed to the moisture and oxygen. For instance, the green emission $CsPbBr_3$ NCs@MSNs would loss quickly when treated with moisture.[126] This is unsuitable for the application of LEDs. Therefore, Li's group developed a facile template-assisted strategy to synthesize ceramic-like highly stable $CsPbBr_3$@$SiO_2$ composite.[37] In this work, ultrapure water is the solvent for CsBr and $PbBr_2$, and the all-silicon molecular sieves (MCM-41) is the template. After treating the obtained products at high temperatures, the molecular sieve structure collapsed (as proved by the decrease in the surface area) at 600 ~ 900 °C. A ceramic -like and dense $SiO_2$ coating formed around the $CsPbBr_3$ – **Figure 4h**. This is an all solid-state reaction, without using organic solvents or organic ligands. The collapsed molecular sieves will limit the growth and the interaction of $CsPbBr_3$ NCs. This $CsPbBr_3@SiO_2$ features a high PLQY of ~ 70% and a narrow FWHM of *ca*. 20 nm is achieved as well. Remarkably, the water stability of the $CsPbBr_3@SiO_2$ is even better than that of the commercial ceramic $Sr_2SiO_4:Eu^{2+}$ under the same conditions – 105% and 88% of the initial PLQYs for $CsPbBr_3@SiO_2$ and $Sr_2SiO_4:Eu^{2+}$ after immersion in water for 50 days, respectively. Finally, this material can also bear harsh conditions, like surviving in 1 M HCl for 50 days, owing to the dense $SiO_2$ coating. This results in the highest reported stability of >1000 h in on-chip LEDs operating at 20 mA, 2.7 V. Meanwhile, a ceramic and dense $Al_2O_3$ coating was also achieved for the $CsPbBr_3/Al_2O_3$ composite through a similar strategy, using the α-$Al_2O_3$ powder as the template and then calcined under a high temperature, like 500 °C. [128]

### 2.4. Physical Method

Physical method consists in the dispersion of as-prepared nanomaterials in a solvent and then mixing the mixture with templates or carriers through mechanical stirring – **Figure 5**. Herein, the typical templates and carriers are always MS powders, polystyrene (PS),[168] or MOFs.[169,170] During the stirring process, the solvent with the nanomaterials will fill the pores of the carriers. It is a very facile and rapid process for the efficient separation of the nanomaterials, preventing their aggregation.[171-173]

Hung-Chia *et al.* mixed pre-prepared green $CsPbBr_3$ NCs with commercial MS powders with a pore size of *ca*. 15 nm. During this process, non-polar solvents, such as hexane, were selected to disperse the $CsPbBr_3$ NCs and to prevent side effects caused by the solvent.[134] The final mesoporous silica green PQD nanocomposite (MP-G-PQDs) can be obtained after purification and drying – **Figure 5a**. The MS template not only improves the thermal stability and the photo stability, but also can efficiently prevent the undesired ion exchange processes. For example, nearly the same PL intensity can be reserved after the heating-cooling testing from 25 °C to 100 °C, and 100 °C to 25 °C for $CsPbBr_3@MS$, while the intensity decreases to 60% for the pristine $CsPbBr_3$ under the same testing condition. Moreover, about 80%, and 40% of the initial PL intensity can be remained for $CsPbBr_3@MS$ and $CsPbBr_3$ after 96 h under continuous UV-light (365 nm, 6 W) irradiation. More importantly, red and green peak severely move to yellow when $CsPbBr_3$ was mixed with $CsPb(Br_{0.4}I_{0.6})_3$ PQDs in silicone resin ( **Figure 5b**). However, the spectra shift can be stopped and white LEDs (WLEDs) will be obtained MP-G-PQDs and $CsPb(Br_{0.4}I_{0.6})_3$ PQDs (R-PQDs) are mixed in silicone resin ( **Figure 5c**). In addition, the same strategy was also applied in Baranov's work[138] and Xiang's work,[135] through embedding the $CsPbBr_3$ QDs in the prepared porous $SiO_2$ microspheres and the commercial MS, respectively.

In addition, MHPs with desirable photonic and electronic properties are good candidates for photoelectrochemical (PEC) sensing.[174] However, they are suffering from degradation in aqueous solution and the widely used encapsulation matrices, like $SiO_2$ and polymers, are insulative. A PEC sensor consists of a $TiO_2/CsPbBr_{1.5}I_{1.5}$/Nafion electrode was fabricated through directly mixing the ahead prepared three-dimensional (3D) $TiO_2$ inverse opal photonic crystals (IOPCs) with $CsPbBr_{1.5}I_{1.5}$ QDs (**Figure 5d**).[137] In detail, a glass substrate with 3D $TiO_2$ IOPCs was vertically put into the $CsPbBr_{1.5}I_{1.5}$ QDs cyclohexane solution. Taking advantage of the capillary force, the QDs will loosely self-organized into the free space of the 3D $TiO_2$ IOPCs as the evaporation of the solvent. In this structure, the 3D $TiO_2$ IOPCs structure with high specific surface area provides rich sites to immobilize the $CsPbBr_{1.5}I_{1.5}$ QDs as well as good encapsulation to protect the QDs against water ( **Figure 5e**). And Nafion was deposited on the surface of the mixture to further fix the QDs. Under visible light, the $CsPbBr_{1.5}I_{1.5}$ will be activated and the energy band alignment of the $TiO_2/CsPbBr_{1.5}I_{1.5}$ favors the separation photo-generated carriers (**Figure 5d**).

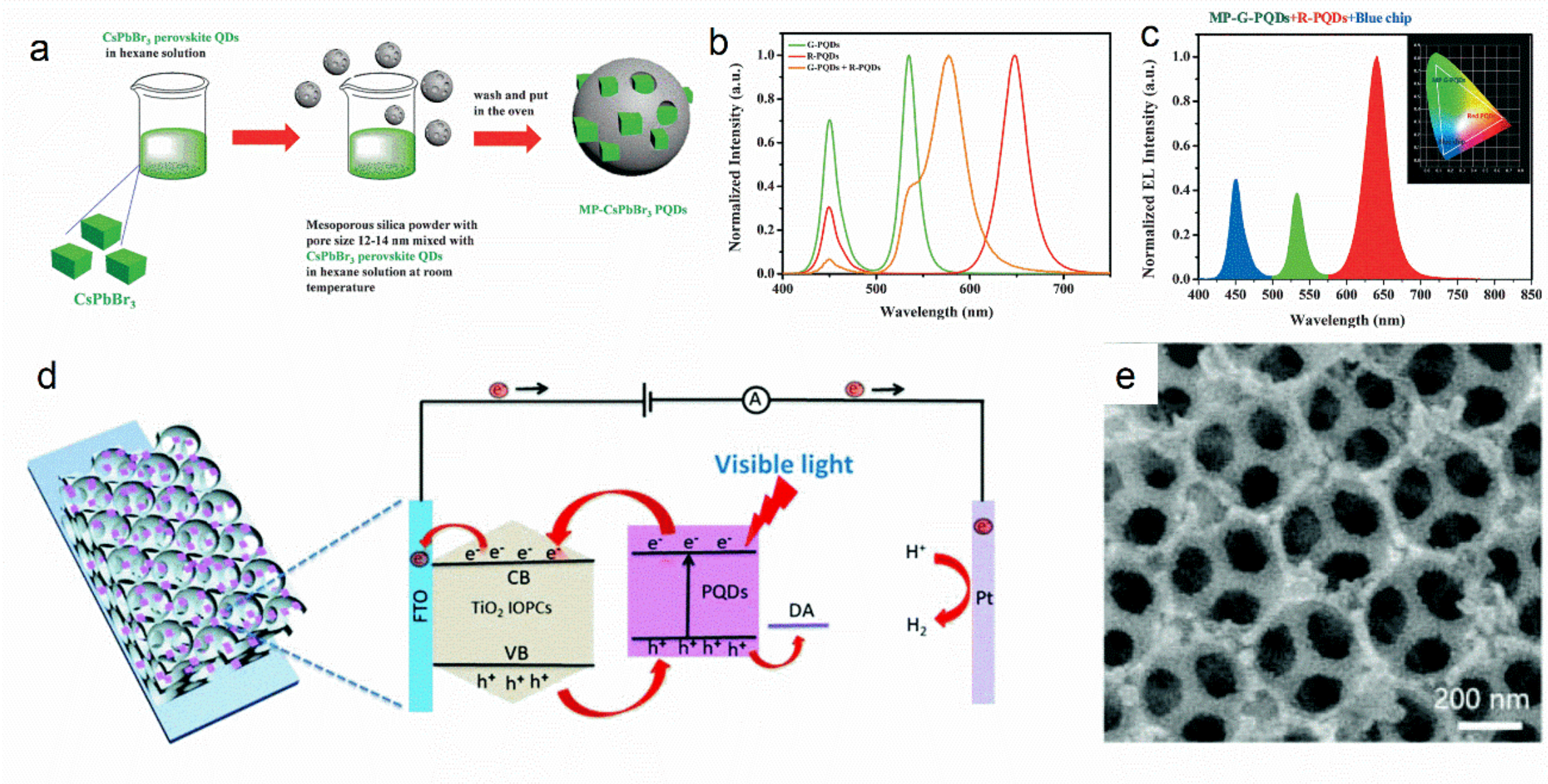

**Figure 5.** a) The physical method for the fabrication of the MP-G-PQDs composite; b) Spectra of the separate G-PQDs, R-PQD, and mixed PQDs under 450 nm excitation; c) Spectra of mixed MP-G-PQDs and R-PQDs under 450 nm excitation. The inset is the color gamut of the WLEDs. a,b,c) Reproduced with permission.[134] Copyright 2016, Wiley-VCH. d) The schematic illustration and the working mechanism of the $CsPbBr_{1.5}I_{1.5}/TiO_2$/Nafion electrode; e) The SEM image of the $CsPbBr_{1.5}I_{1.5}/TiO_2$composite. d,e) Reproduced with permission.[137] Copyright 2018, Royal Society of Chemistry.

### 2.5. SILAR Method

In the above MHPs/$MO_x$ obtained from the physical method and template-assisted method, the adhesion between MHPs and the templates is very week and the PLQYs are very low without using the ligands. Therefore, a new strategy of combining the hot-injection method (for the formation of $CsPbBr_3$ QDs) and the MS as the template was proposed through the *in situ* growth of $CsPbBr_3$ QDs in MS (**Figure 6a**).[139] The method is based on the hot-injection technique, the difference is adding the MS into the solvent for the preparation $CsPbBr_3$ QDs. In detail, MS was firstly well-dispersed in the mixture of oleylamine, OA, ODE. Here, both the inner and outer surface of MS will be modified by oleylamine and OA. Then, $PbBr_2$ was added to the solution and the Pb and Br ions will permeate into both the pores and channels of MS. This is followed by the hot-injection of the Cs-oleate solution. And $Cs^+$ will in situ react with $Pb^{2+}$ and $Br^-$ to form $CsPbBr_3$ inside the channels of the MS. And the TEM-energy dispersive spectroscopy (TEM-EDS) images (**Figure 6b**) indicate that Cs, Pb and Br are well dispersed in the $SiO_2$ matrix. And the pore structure will also help restrict the growth of the $CsPbBr_3$. Moreover, a stronger adhesion is formed between the surface of MS and MHPs. The obtained $CsPbBr_3$@MS has an excellent PLQY of 83 % (for solution, and 68% for powder).

Importantly, the morphology and the precursor (mainly the active amino groups) have an important effect on the optical properties of the obtained MHPs/$MO_x$. The $CsPbBr_3$ QDs were also encapsulated to the dual-shell hollow $SiO_2$ spheres through the SILAR method(**Figure 6c**).[140] Firstly, the dual-shell hollow $SiO_2$ spheres were prepared though tuning the ratio of the APTES and TEOS. The morphologies of obtained $SiO_2$ evolve from regularly spheres to cracked spheres as the increasing ratio of the APTES and TEOS. Then, both $PbBr_2$ and $SiO_2$ spheres were well dispersed in ODE to ensure the dissolution of $PbBr_2$ and the maximal absorption of $Pb^{2+}$ into the channel of $SiO_2$ spheres. During this process, the amino groups favor the adsorption of $Pb^{2+}$ ions on the inner shell. This is followed by the addition of the OA, Cs-Oleate, oleylamine to the mixture. The $CsPbBr_3$ QDs will form at this stage. The PLQYs of the $CsPbBr_3$@$SiO_2$ changes with the morphology of the $SiO_2$. And a high PLQY value of 89% was achieved when the ratio of the APTES/(APTES+TEOS) is 20% (**Figure 6d**). For the $CsPbBr_3$@$SiO_2$ that fabricated through only using TEOS as the precursor, $CsPbBr_3$ QDs and $SiO_2$ spheres are isolated from each other because of the absence of the binding sites. As the increasing APTES ratio, the amino groups will help $CsPbBr_3$ QDs to bind with $SiO_2$ and therefore improve the PLQYs. Further increasing the amount of APTES will induce the adhesion between $CsPbBr_3$ and the outer surface of $SiO_2$. And $CsPbBr_3$ will escape from the cracked spheres and thus the decreased PLQYs. Benefited from the excellent protection of the dual-shell hollow $SiO_2$ spheres (APTES ratio is 20%), excellent thermal and photo stability can be achieved for the $CsPbBr_3$@$SiO_2$ composite. For example, 89% and 65% of the initial PL intensity remain after being continuously exposed to UV light for 72 h and 100 ˚C heat treatment, respectively.

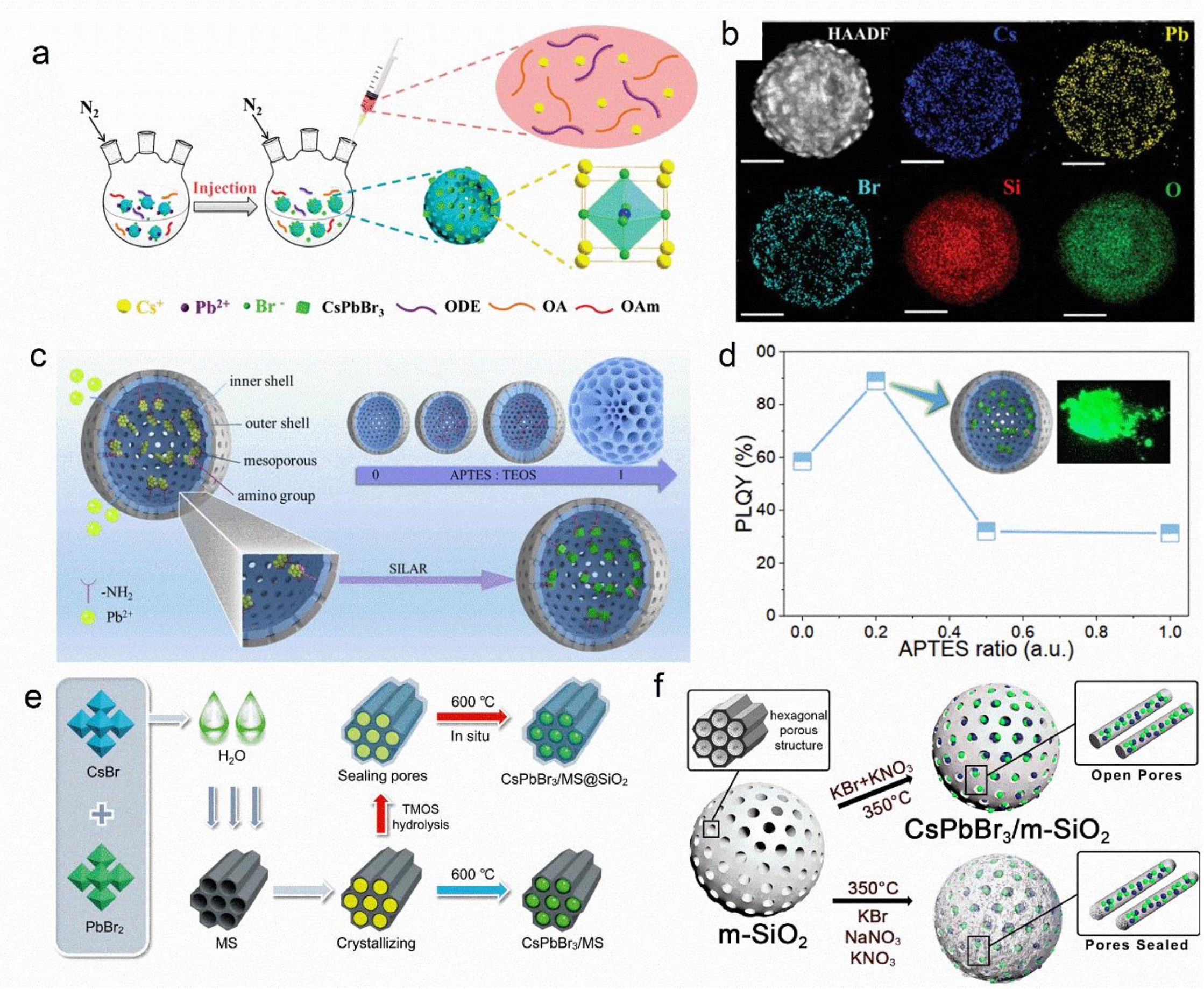

**Figure 6.** a) Schematic of the *in situ* growth of $CsPbBr_3$ QDs in the MS template; b) The TEM-EDS images of the $CsPbBr_3$@MS. a,b) Reproduced with permission.[139] Copyright 2019, Royal Society of Chemistry. c) The preparation process for the $CsPbBr_3$@$SiO_2$ dual-shell hollow nanospheres through SILAR method; d) The PLQYs of the $CsPbBr_3$@$SiO_2$ obtained from different APTES/(APTES+TEOS) ratio. c,d) Reproduced with permission.[140] Copyright 2020, Royal

Society of Chemistry. e) The process for the preparation of the $CsPbBr_3$/MS@$SiO_2$ through the template-assitated and sol-gel methods. Reproduced with permission.[133] Copyright 2021, Wiley-VCH. f) The synthesis of the $CsPbBr_3$/mesoporous-$SiO_2$ composites through the molten salts approach, combining different kinds of salts. Reproduced with permission.[143] Copyright 2021, American Chemical Society.

### 2.6. Other Synthesis Techniques

In addition to the above synthesis techniques, some groups also combine two or more techniques to control the sizes and size distribution as well as get dense protection for the MHP@$MO_x$ composites.[133,141] In Li's work,[133] the $CsPbBr_3$/MS was firstly prepared using the template-assitated method, through putting a certain amount of MS templates to the mixture of CsBr and $PbBr_2$. Then, the intermediate product was dried and TMOS was introduced to the $CsPbBr_3$/MS-toluene solution. After the complete hydrolysis of the TMOS, the final $CsPbBr_3$/MS@$SiO_2$ can be obtained through grinding and calcining the dried powders at 600 °C ( **Figure 6e**). Herein, the uniform pores in the MS template favor the precisely confined growth of $CsPbBr_3$ and the $SiO_2$ formed through the pre-hydrolysis of the TMOS helps to seal the pores of the MS template. Therefore, exceptional stability against water, light, and heat, as well as excellent optical properties can be achieved (See **Table 1** for the details).

On the other hand, a new solvent-free strategy, that is the molten-salts-based approach was also proposed by Manna's group to prepare the $CsPbBr_3$/mesoporous-$SiO_2$ composites.[143] In the work, the mixture of CsBr, $PbBr_2$, inorganic molten salts ($KNO_3$−$NaNO_3$−KBr) salts, and commercial MCM-41 mesoporous-$SiO_2$ was firstly ground and then calcined at 350 °C in the furnace. The excess inorganic salts and the unprotected $CsPbBr_3$ particles that disperse outside of the mesoporous-$SiO_2$ were removed by DMF or DMSO washing (**Figure 6f**). As a result, the composites show high PLQYs (89 ± 10%) and excellent resistance to heat, water, and even to aqua regia (See **Table 1** for the details). In addition, the organic salts play an essential role in achieving the excellent performances. Among them, $NaNO_3$ attributes to the excellent stability (sealing the pores), KBr and $KNO_3$ are responsible for the good optical properties.

### 2.7. Take Home Message Considering MHP@$MO_x$ Synthesis Methods

After having provided an overview of the MHP@$MO_x$ synthesis methods (**Table 1** and **Table 2**), the sol-gel method is the most widely used strategy for the fabrication of the MHPs/$MO_x$ structure. And high PLQYs and high stability are usually achieved through the sol-gel method. But the shell thickness is not easily controlled as well as the presence of byproducts (like water and alcohols) during the hydrolysis process.[69] Using the solvent-free strategy, like the molten-salts-based approach can perfectly avoid the above byproducts. Meanwhile, ALD can be used to precisely tune the thickness and the composition of the composite, but it has been restricted to $Al_2O_3$ and the instrumentation requirements are not standard in all the labs.

The template-assisted method is a good way to achieve the large-scale synthesis and the outstanding stability, but the obtained PLQY is usually cannot reach as high as that for the sol-gel method and the ALD method because of the absence of ligands. The preparation process of the MHPs/$MO_x$ using the physical method is almost the same as the template-assisted method, the only difference is that the MHPs are ahead prepared. No byproducts form during the fabrication process of MHPs/$MO_x$ for both template-assisted and physical methods because the $MO_x$ are ahead prepared. The size of the MHPs can be tuned through using templates with different pore sizes. However, the size of MHPs should be controlled to fit the pore or channel size of the $MO_x$ during the physical method. The SILAR method can be recognized as the combination of the benefits of the hot-injection method for the formation of $CsPbBr_3$ QDs and the handleability of the template-assisted method. Other advantages for this method are that the MHPs can grow uniformly inside the $MO_x$ and no byproducts formation during the process. In addition, the composites obtained from the template-assisted, the physical method, and the SILAR method usually show low stability because the lack of a full protection or encapsulation of the MHPs from the surrounding stress. However, the combination with a last high temperature sintering or two or more techniques has led to fully dry and highly emissive $CsPbBr_3$@$SiO_2$ composites featuring recorded thermal- and photo-stabilities under severe scenarios that are, in addition, noted in devices for lighting applications.

**Table 2**. The comparison of the advantages and disadvantages of the fabrication processes for MHPs/$MO_x$ composites.

| Fabrication method | Advantages | Disadvantages |
|---|---|---|
| Sol-gel method | a. Easy accessibility | a. By-products |
| | b. Most widely used | b. Aggregation |
| | c. Applied to all kinds of $MO_x$ | |
| | d. Gentle reaction condition | |
| ALD | a. Controllable thickness | a. Limited to $AlO_x$ currently |
| | b. Homogenous deposition | b. Expensive instruments |
| | c. Large-scaled preparation | c. Harsh reaction condition<br>d. Difficult to operate<br>e. By-products<br>f. Water necessary |
| Template-assisted method | a. In-situ growth<br>b. Tunable size<br>c. Large-scaled preparation<br>d. No by-products<br>e. Time-saving | a. Low PLQYs<br>b. Non-dense coating |
| Physical method | a. Direct, facile<br>b. No by-products<br>c. Large-scaled preparation | a. Non-dense coating<br>b. Diffusion resistance<br>c. Difficult to predict the distribution of $MO_x$ (outside or inside the template)<br>d. Size mismatch between MHPs and $MO_x$ |
| SILAR | a. In-situ growth<br>b. Tunable size<br>c. No by-products<br>d. High PLQYs | a. Non-dense coating<br>b. Limited to $SiO_2$ template currently |
| Combined synthesis | a. In-situ growth<br>b. Tunable size<br>c. Confined growth<br>d. Dense coating | a. Energy consuming<br>b. Tedious process |
| Low-temperature molten salts | a. Solvent free<br>b. Dense coating<br>c. No aggregation | a. High energy consuming (~ 350 °C)<br>b. High requirements for inorganic salts |

All in all, every method has both advantages and disadvantages. Methods should be chosen based on the application purpose and the available instruments of the lab. MHP@$MO_x$ prepared via the sol-gel method and the SILAR method or the combination of both and/or other techniques favors the utilization in the LEDs because of the high PLQYs and excellent stability. ALD fabricated MHP@$MO_x$ is suitable to use in the situation where has a high requirement for the film quality, like lasers. Template-assisted method is desired for the moisture related anti-counterfeiting, where reversible transition between luminescent $CsPbBr_3$ and non-luminescent $CsPb_2Br_5$ can realize owe to the lack of the ligands.

## 3. Characterization of the MHPs@$MO_x$

Standard characterization protocols of the MHPs@$MO_x$ composites with respect to the size, shell thickness, the composition, the optical properties, and the environmental stability towards moisture, heat, irradiation, and $O_2$ are key to ensure a quick advance in the field. In this context, several widely used characterization techniques, such as TEM, Scanning electron microscope (SEM), X-ray photoelectron spectroscopy (XPS), XRD, PL, Ultraviolet–visible (UV-Vis) spectroscopy, and time-resolved photoluminescence (TRPL) spectra are discussed in this section to better understand the best synthesis method.

### 3.1. Microscopic Analysis

TEM is the most common microscopic technique for the characterization of the MHPs@$MO_x$ structure. TEM has a higher resolution than that of the SEM – *i.e.*, 0.2 nm. TEM can also provide a direct information of the MHPs@$MO_x$ structure, confirming the boundaries between the $MO_x$ shell and the MHP core, and the size of both, the $MO_x$ coating and the MHPs. Finally, the distribution of MHPs inside the $MO_x$ coating can also be clearly observed.[37,74,78,79,82,83,102,175]

As shown in **Figure 7a**, $CsPbMnX_3$ NCs are all in *quasi* square shaped with a uniform size of *ca*. 11 nm. And the crystal lattices of 5.69 Å (as labeled in **Figure 7b**) is also well matched with the $CsPbMnX_3$ size. In a core-shell $CsPbMnX_3$@$SiO_2$ structure, every $SiO_2$ shell covers an individual $CsPbMnX_3$ core – **Figure 7c**. And a clear boundary can be observed in the high resolution TEM (HRTEM) graph in **Figure 7d**. Specially, the measured thickness of $SiO_2$ shell is about 2 nm. The distinct lattice fringes in all the $CsPbMnX_3$ NCs mean the high crystallinity. The core part remains intact without breakage means the sol-gel process does not affect the properties of the core.[74]

Besides, the high-angle annular dark-field STEM (HAADF-STEM) and energy-dispersive X-ray spectroscopy (EDS) are usually used to provide all important information concerning the composition and the elemental distribution of the MHPs@$MO_x$ structure. **Figure 7e** and **f** display the TEM and HRTEM images of the $CsPbBr_3$@$SiO_2$ Janus NCs prepared using the sol-gel method.[76] Obvious contrast can be observed in these two parts, the light part belongs to the amorphous $SiO_2$ formed through the sol-gel process, while the darker part is the $CsPbBr_3$ NCs. From **Figure 7g**, the elements Cs, Pb, and Br that from the $CsPbBr_3$ NCs are mainly dispersed in the darker part, while Si elements are mainly distributed in the lighter part. At the same time, some Si elements are observed in the $CsPbBr_3$ part caused by the permeation of $SiO_2$.

On the other hand, specific morphological features exist in the above MHPs@$MO_x$ composites that prepared through different methods. For example, the composites that obtained through the sol-gel method usually tend to show some aggregations and there is no clear boundary between each $MO_x$ particle (**Figure 8a**).[35] However, the boundary between $MO_x$ particle can be seen clearly for the composite prepared by the SILAR method (**Figure 8b**),[139] template-assisted method (**Figure 8c, d, e**).[126] In addition, the MSNs can still keep monodisperse after incorporation with the MHPs (the SEM image in **Figure 8c**).[126] Normally, the $MO_x$ structure in the composites from the template related methods, like SILAR method, template-assisted method, physical method, is always highly regular, as shown in both SEM and TEM images of the composites (**Figure 8b, c, e, f, g**).[68,126,137] And the TEM is also the best way to show the evolution of the sizes of the MHP particles as the pore size of the template (**Figure 8e, f**). Moreover, the cross-section SEM is a good tool to detect the interface between $MO_x$ and MHPs for the composite prepared by the ALD method. For example, **Figure 8h** illustrates that a clear interface between $Al_2O_3$ and $CsPbBr_3$, means that the ALD process does not affect the $CsPbBr_3$.[4] However, the unsharp interface in **Figure 8i** suggests that $Al_2O_3$ fill in the interstices between the QDs during the ALD process.[123]

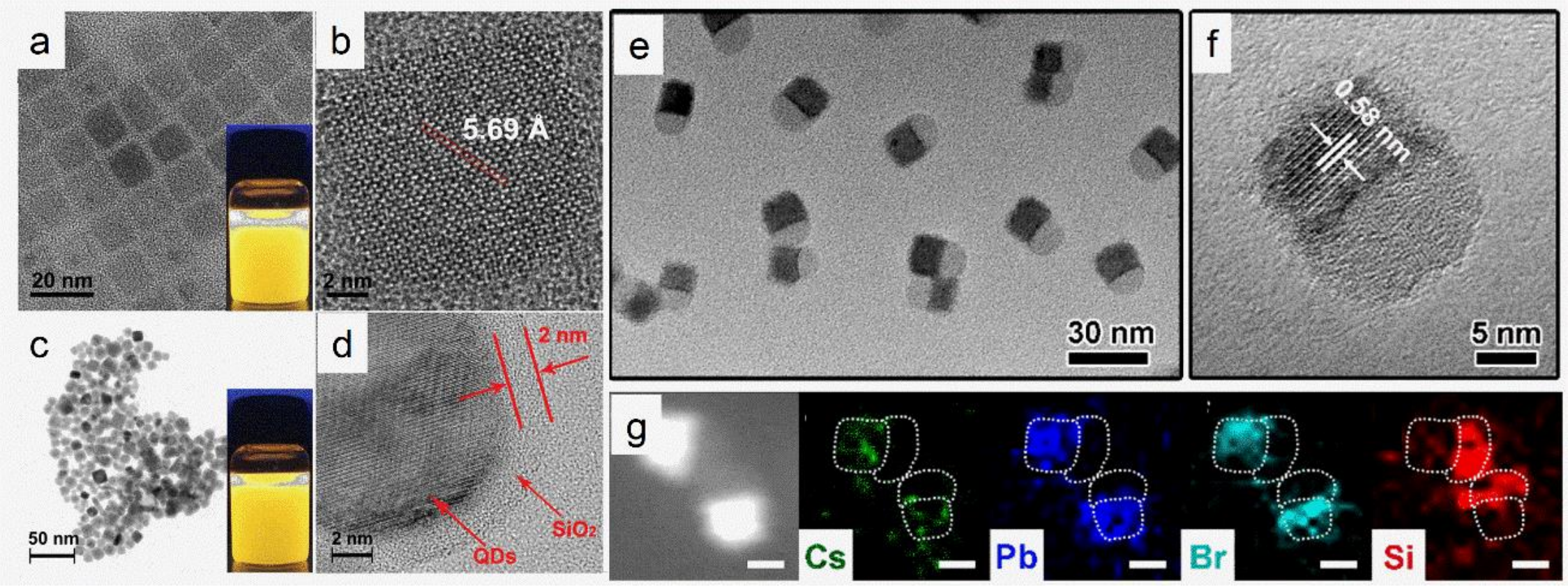

**Figure 7.** a) and b) TEM and HRTEM images of $CsPbMnX_3$ NCs; c) and d) TEM and HRTEM images of $CsPbMnX_3/SiO_2$ composites. a,b,c,d) Reproduced with permission.[74] Copyright 2019, Wiley-VCH. e) TEM, f) HRTEM and g) HAADF-STEM images of $CsPbBr_3/SiO_2$ Janus NC, and the corresponding EDS mapping of Cs, Pb, Br, and Si. Reproduced with permission.[76] Copyright 2018, American Chemical Society.

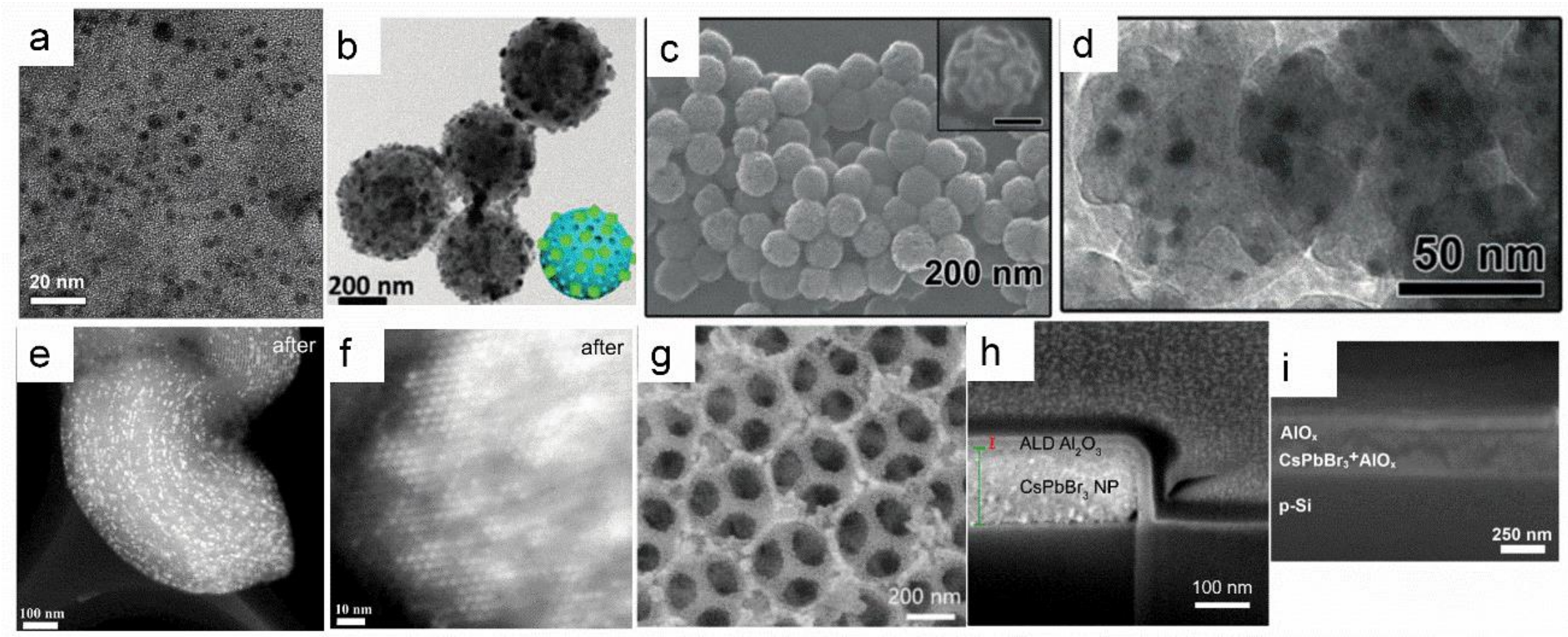

**Figure 8.** a) The TEM image of the $CsPb(Br/I)_3/SiO_2$ composite prepared through the sol-gel method. Reproduced with permission.[35] Copyright 2016, Wiley-VCH. b) The TEM image of the $CsPbBr_3$/MS nanocomposite obtained through the SILAR method. Reproduced with permission.[139] Copyright 2019, Royal Society of Chemistry. The c) SEM and d) TEM images of the $CsPbBr_3$ NCs@MSNs powder obtained via the template-assisted method. Reproduced with permission.[126] Copyright 2020, Wiley-VCH. TEM images of the MS templates with a pore size of e) ≈7.1 nm and f) ≈3.3 nm after the impregnation of MHPs. e,f) Reproduced with permission.[68] Copyright 2016, American Chemical Society. g) SEM image of the $CsPbBr_{1.5}I_{1.5}/TiO_2$ composite film obtained using the physical method. Reproduced with permission.[137] Copyright 2018, Royal Society of Chemistry. h, i) The cross-section SEM image of the $CsPbBr_3/Al_2O_3$ obtained by the ALD process. h) Reproduced with permission.[4] Copyright 2020, American Chemical Society. i) Reproduced with permission.[123] Copyright 2017, Wiley-VCH.

### 3.2. Spectroscopic Analysis

Spectroscopic analysis, like XPS, PL, UV-Vis will provide several indirect information about the MHPs/$MO_x$ composite after coating. Especially, the optical properties of the MHPs are very sensitive to the surface modification.

#### 3.2.1. *XPS Analysis*

XPS is a technique that can provide useful information for the nanomaterials like, the molecular structure, the valence state, chemical status, the element composition, and content.[176,177]

In Li's work,[77] XPS was used to detect the surface change after introducing the bifunctional passivation reagent ammonium hexafluorosilicate (AHFS, $(NH_4)_2SiF_6$). As shown in **Figure 9a-c**, the binding energies at 685.4 eV for F 1s spectra and 103.5 eV for Si 2p spectra belongs to Si−F bonds. After treating with the AHFS, an additional peak appears at *ca*. 102 eV, which is the Si-O-Si peak, indicating the formation of the $SiO_2$ shell. In addition, for the AHFS-treated MHPs, the Pb 4f spectra shifts towards higher binding energy and the shift is more obvious as the increasing amount of the AHFS (not present here). However, the F 1s shifts towards the lower bonding energy for the AHFS-treated MHPs. According to previous works,[178] the movement of the bonding energies is related to the high electronegativity of fluorine atoms, and therefore F will absorb electrons from Pb. On the other hand, the increasing movement of Pb 4f is also because that more AHFS provides more $F^-$ and therefore stronger electron-withdrawing property. These results indicate the highly close bonding between AHFS and MHPs. In addition, the surface molar ratio of Br:Pb:F changes from 1:0.54:0 to 1:0.39:0.63 was also detected by XPS after AHFS treatment, suggesting the presence of a lead-deficient surface.

As an another example,[102] the incorporation of an efficient co-reactant (CoR) ~ 2-(dibutylamino)ethanol (DBAE) in $SiO_2$ matrix was confirmed by XPS – **Figure 9d**. For the pristine $CsPbBr_3$ (CPB), only one symmetric peak located at 398.1 eV can be observed, which belongs to the primary protonated amine groups ($NH_3^+$) from oleylamine ligand.[179] For the CPB-CoR@$SiO_2$ NCs, the asymmetrical peak can be deconvoluted to two peaks. The additional peak appears at 399.1 eV is attributed to tertiary amine from DBAE, proving the successful incorporation of the DDAB to the structure.

Moreover, Zheng's group used the XPS as a co-method to prove the crystallinity state of the $TiO_2$ in the $CsPbBr_3/TiO_2$ composite. As shown in **Figures 9e** and **f**, the XPS spectra of O 1s can be deconvoluted into two different peaks. Here, the peaks located at 530.2 and 532.5 eV are assigned to the titanium oxygen bond (Ti−O) and hydrogen oxygen bond (O−H), respectively.[176,180] After calcination at 300 °C, the relative intensity of the Ti−O increases, while the peak area of O−H decreases. This result means that more water release from the $CsPbBr_3/TiO_x$ composite at higher temperature due to the dehydration process, indicating a better crystallization of the $CsPbBr_3/TiO_2$ composite.

Generally, the FTIR spectra, XRD, thermogravimetic analysis (TGA), and XPS are combined with each other to investigate the surface changes, the crystallinity state, and the composition of the nanomaterials.[181-183] Specially, several differences exist among these methods. XPS, FTIR, TGA all can be used to detect the composition of the materials. XPS and FTIR are techniques that have high sensitivity, but can only provide the information about the surface. TGA can give a full information about the composition of the samples, through observing the weight loss as the increasing temperature during the measurements; but it is very difficult to distinguish the specific elements. XPS technique is more widely used to the give the information about the valence states and binding of elements, while FTIR is used to identify chemical substances or functional groups in the materials. XRD can give a direct information of the crystallinity state of the $MO_x$, while XPS provides an indirect knowledge about the crystallinity state – the information can be obtained through deconvoluting the peaks of the oxygen and then compare the ratio between different peaks.

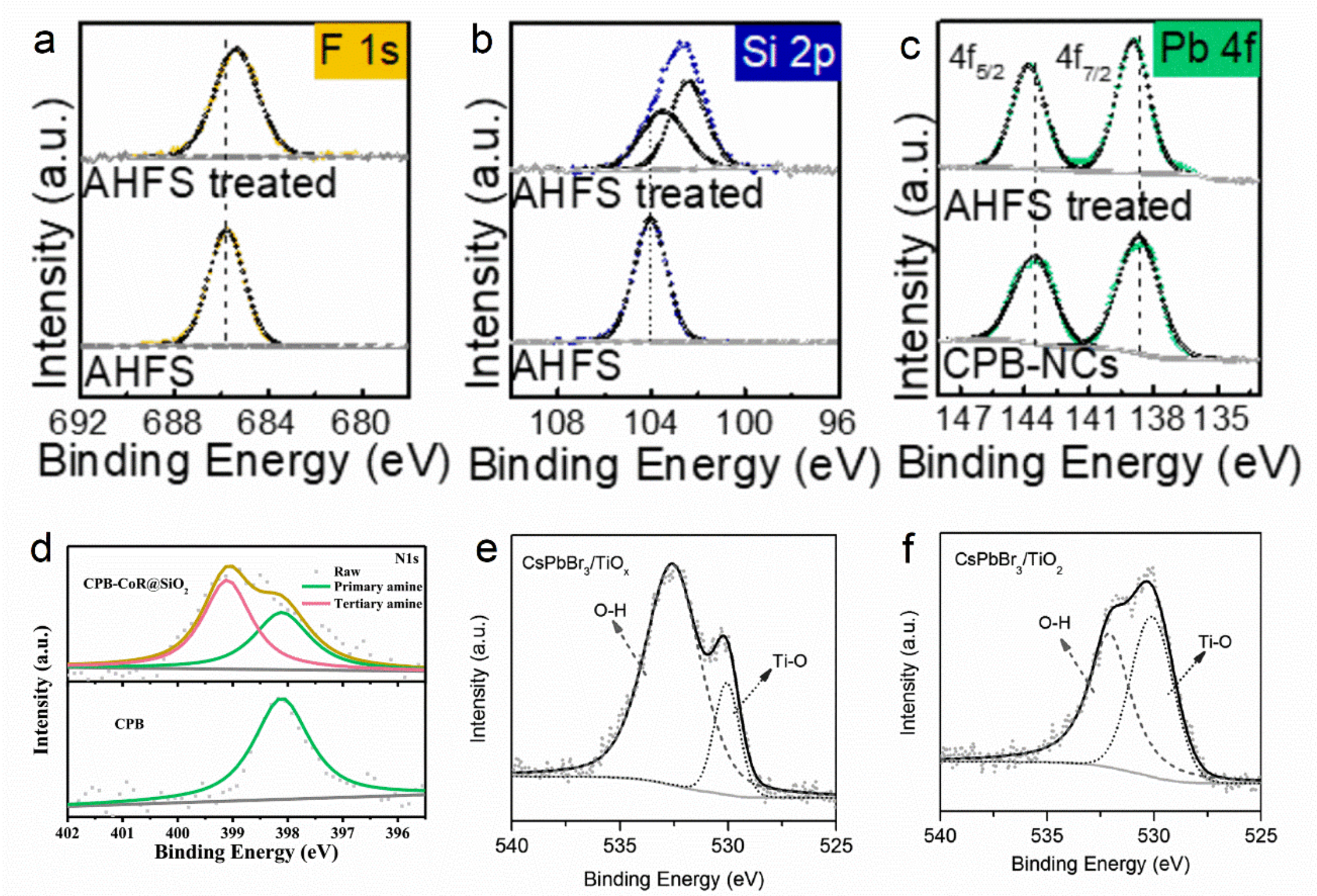

**Figure 9**. a) F 1s and b) Si 2p spectra for AHFS and AHFS-treated MHPs; c) Pb 4f spectra for the untreated and AHFS-treated MHPs. a,b,c,) Reproduced with permission.[77] Copyright 2020, American Chemical Society. d) High-resolution XPS spectra of N 1s for CPB and CPB-CoR@$SiO_2$ NCs. d) Reproduced with permission.[102] Copyright 2019, Wiley-VCH. High-resolution O 1s XPS spectra of e) $CsPbBr_3/TiO_x$ and f) $CsPbBr_3/TiO_2$ composites. Reproduced with permission.[88] Copyright 2018, Wiley-VCH.

### 3.2.2. *UV-Vis and PL Spectra*

UV-Vis spectra reflect the absorption capacity, the composition, content, and structure of the materials. For the MHP@$MO_x$ composite, UV-Vis can be used to analysis the change in the size of the MHP particles, the structure transformation, as well as the absorption capacity of the MHPs before and after coating. PL spectra and UV-Vis spectra can be recognized as the mutual complementary to each other. Fluorescence detects the transitions of the generated carriers (electrons and holes) from the excited state to the ground state, while absorption measures the transitions of the generated carriers from the ground state to the excited state. The intensity (for both the absorption and the emission) and the location of the peaks will change after the MHPs are coated with $MO_x$.

In Sun's work, UV-Vis and PL spectra were put together to compare the optical properties of the products with or without $SiO_2$ coating. As shown is **Figure 10a**, the absorption spectra of both, green- and red-emitting composites, do not show any significate change after coated by $SiO_2$. However, the maxima of the emission band is slightly red-shifted upon $SiO_2$ coating due to the induced increase in the particle size, while the FWHM typically increases due to some aggregation caused by the sol-gel process and therefore the broad size distribution.[184] Finally, the PLQYs of the MHP@$SiO_2$ powders is also typically lower than those of the respective fresh MHP solutions.[35] This is because some ligands loss as well as the aggregation of the MHPs during the purification process from solution to powders.[185]

Besides, UV−Vis and PL spectra are also used to investigate the surface modification processes.[70,76] For instance, two sharp peaks located at 230 and 314 nm can be observed for the pristine solution, which belong to the absorption spectra of the $Cs_4PbBr_6$ NCs (**Figure 10b**).[186] These two peaks decline and a new absorption peak at 507 nm appears after the modified sol-gel process, indicating the transformation from $Cs_4PbBr_6$ to $CsPbBr_3$.[78,187] Meanwhile, a sharp PL peak at 517 nm with a narrow FWHM of 18 nm also emerges, corresponding to the emission spectrum of the $CsPbBr_3$.[183,185,188,189]

Several features exist in every preparation method, apart from the above observation. The absorption, the PL emission (both the shape and location), as well as the PLQYs evolve as the reaction time and the concentration of the precursor (for the formation of the $MO_x$) during the sol-gel process. For example, a clear blue shift from 515 nm to 501 nm appear as the increasing reaction time from 2 min to 2 h (**Figure 10c**).[86] The blue shift is attributed to the decrease in the size of the MHPs and the increasing shell thickness. In **Figure 10d**, the PL emission band and the absorption onset all exhibit blue shift as the increasing precursor concentration, indication the formation of smaller size MHPs.[63] The PLQY also increases as the increasing concentration of the precursors (the maximum value of 42% is achieved for the $PNC_{APTES-16}$). However, the excess ligand will lead to the decrease of the PLQYs. At the low concentration, the ligands from the precursor will help to passivate the surface trap, the formed shell as well as the steric hindrance from the ligands helps to prevent the aggregation of the MHPs, and therefore contributes to the higher PLQYs. However, smaller particles and more uncoordinated surface atoms of the MHPs will form at a higher precursor concentration. This leads to the mismatching rate between monomer delivery through the ligand capping layer and the surface atoms. Meanwhile, the stability of the MHPs@$MO_x$ is significantly affected by the size of the MHPs. The total formation energy (Gibbs free energy, $E_{MHPs}$) of the MHPs in concert with the ligands are important parameters for the stability of the $MO_x$. Herein, $E_{MHPs}$ can be determined by the following formula,[190,191]

$$E_{MHPs} = E_{bulk} + \frac{6E_{surf}V_{\circ}}{d} \tag{4}$$

$E_{bulk}$, $E_{surf}$, Vo, and d are the formation energy of the bulk, the surface energy, the unit cell volume, and the diameter of the MHPs, respectively. The contribution from the $E_{surf}$ becomes more distinct as the decreasing MHPs size. Thus, smaller sizes will induce higher $E_{MHP}$ values. In addition, the ratio of the surface atoms to the internal atoms increases as the MHPs size decreases, while more uncoordinated surface atoms will appear at the same time. This will induce the mismatch between the surface passivation and the surface defects. These defects will be detrimental for the photoluminescence and stability of MHPs.[63] Therefore, optimizing the size of the MHPs in the MHPs@$MO_x$ composite is very important to obtain stable MHPs@$MO_x$ composite and the relative devices.

On the other hand, it is quite difficult to predict the PL properties of the composite prepared from the ALD process.[123] No peak shift is observed for the $CsPbBr_3$ QD after the ALD process, while a 50% decrease in the PL intensity is recorded (**Figure 10e**1). 85% of the PL intensity losses and 10 nm blue shift of the PL intensity for the $CsPbBr_xI_{3-x}$ QDs after the ALD process (**Figure 10e**2). However, ca.4 times higher PL intensity and no peak emission are observed for the $CsPbI_3$ QDs after the ALD process (**Figure 10e**3). Many reasons can be responsible to these changes in the PL emission, like the surface chemistry between the MHPs and the precursor for the $MO_x$, the intrinsic stability of the MHPs, the reflectance of the $MO_x$, the ligand desorption in the vacuum step induced surface trap states or the passivation of the surface traps.

For the composites that are prepared based on the $MO_x$ template, such as the template-assisted method, the physical method as well as the SILAR method, the optical properties change as the size and the morphology of the template. A gradual shift of band energy towards higher band-gap in the UV-Vis spectra means the shrinkage of the MHP size (**Figure 10f**1). Meanwhile, the emission color of the composite also changes from green to blue as the decreasing pore size of the MS template (**Figure 10f**2), along with the evolution of the PLQYs. Similarly, the blue shift is observed along with the ratio of APTES/(APTES+TEOS) for the preparation of the $CsPbBr_3@SiO_2$ through the SILAR method (**Figure 10g**).[140] Meanwhile, the PLQYs also change with the different ratios (**Figure 6d**). The diversity in the optical properties is attributed to the different morphologies and sizes of the dual-shell hollow silica spheres that will affect the growth of the $CsPbBr_3$ inside the structure.

The UV-Vis diffuse reflectance spectra (DRS) are also used to evaluate the light absorption ability of the composites.[177,192-194] For instance, **Figure 10h** shows the DRS spectra of $CsPbBr_3@TiO_2$ featuring a higher absorbance in the region from 350 to 380 nm than the pristine $CsPbBr_3$ NCs, as $TiO_2$ is an UV-activated semiconductor. Moreover, a minor red-shift is observed for the absorption edge of the $CsPbBr_3@TiO_2$ composite, the similar absorption edge means that there is no competition between the absorption of the $CsPbBr_3$ core and the $TiO_2$ shell. Regarding the light absorption ability between 550-800 nm, $CsPbBr_3@TiO_2$ is also stronger than that of the $CsPbBr_3$ NCs. This all means that the incorporation of the $TiO_2$ shell increases the light absorption ability of the $CsPbBr_3@TiO_2$, which favors the enhanced photocatalytic properties – see section 4. Indeed, the PL spectra shown in **Figure 10i**, clearly highlights a significant quenching of the PL emission peak of $CsPbBr_3@TiO_2$ compared to that of $CsPbBr_3$ NCs. This indicates the electron transfer from the $CsPbBr_3$ core to the $TiO_2$ shell that favors the better photocatalytic performance.[88,195-197]

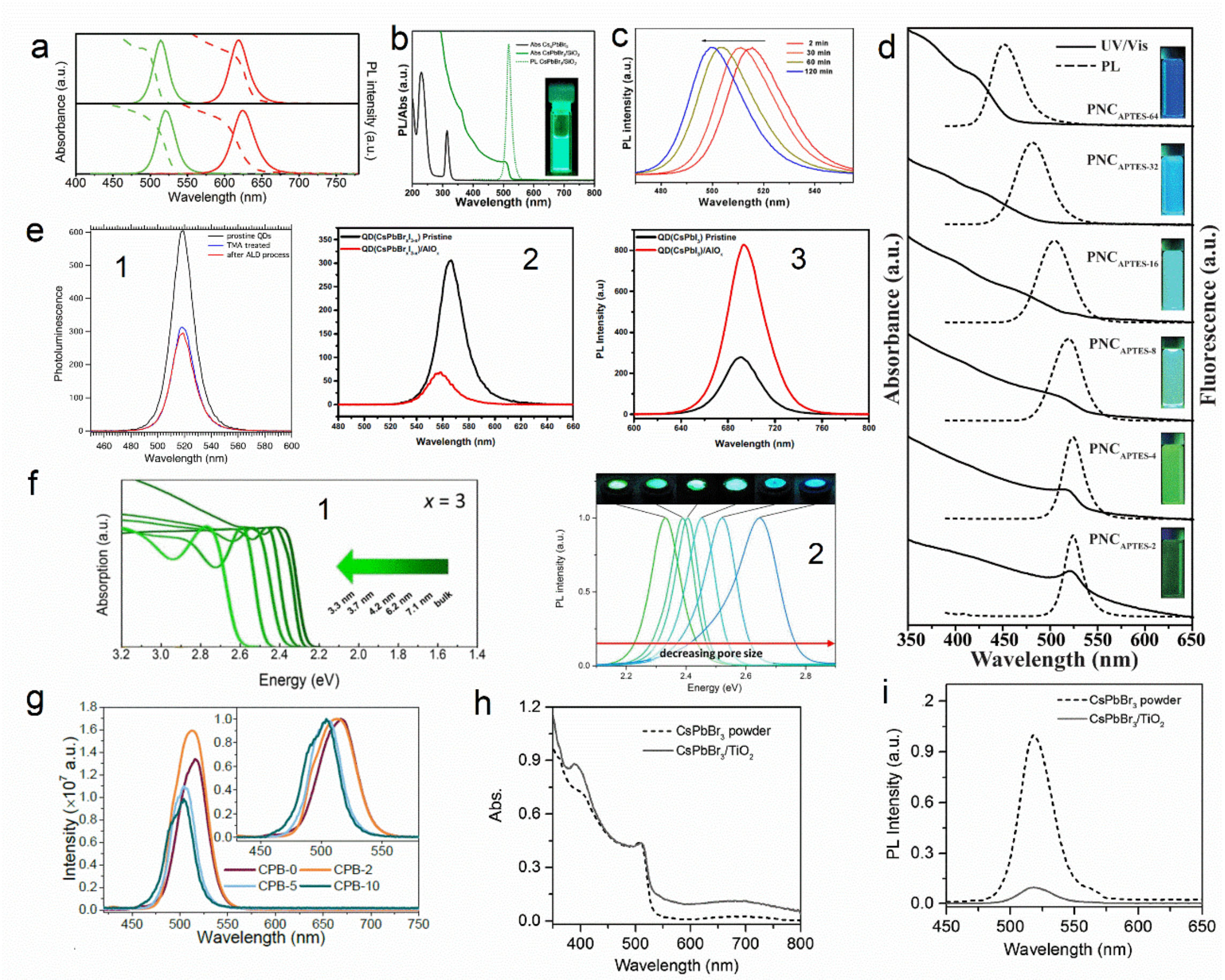

**Figure 10**. a) The UV-vis absorption (dashed lines) and PL (solid lines) spectra of the red (red line) and green (green line) upper: QDs in solution, lower: the QD/$SiO_2$ composites. Reproduced with permission.[35] Copyright 2016, Wiley-VCH. b) The UV-Vis absorption and PL spectra of $Cs_4PbBr_6$ NCs and $CsPbBr_3/SiO_2$ composite. Reproduced with permission.[76] Copyright 2018, American Chemical Society. c) The PL emission evolves as the increasing hydrolysis time for the $CsPbBr_3/SiO_2$ composite. Reproduced with permission.[86] Copyright 2018, American Chemical Society. d) The UV-Vis and PL spectra of $PNC_{APTES}$ solutions. Inset: the photographs of the solution under UV light. Reproduced with permission.[63] Copyright 2016, Wiley-VCH. e) The PL emission before and after ALD process for $CsPbBr_3$ QDs (1), $CsPbBr_xI_{3-x}$ QDs (2), $CsPbI_3$ QDs (3). Reproduced with permission.[123] Copyright 2017, Wiley-VCH. f) The UV-Vis (1) and the PL (2) spectra of the $MAPbBr_3$/MS composite as the size evolution of the MS. Reproduced with permission.[68] Copyright 2016, American Chemical Society. g) The PL spectra of the $CsPbBr_3@SiO_2$ change with the ratio of the APTES/(APTES+TEOS). Reproduced with permission.[140] Copyright 2020, Royal Society of Chemistry. h) The UV-Vis DRS and i) the PL

spectra for the $CsPbBr_3$ powder and the $CsPbBr_3/TiO_2$ powder. h,i) Reproduced with permission.[88] Copyright 2018, Wiley-VCH.

### *3.2.3. TRPL Spectra*

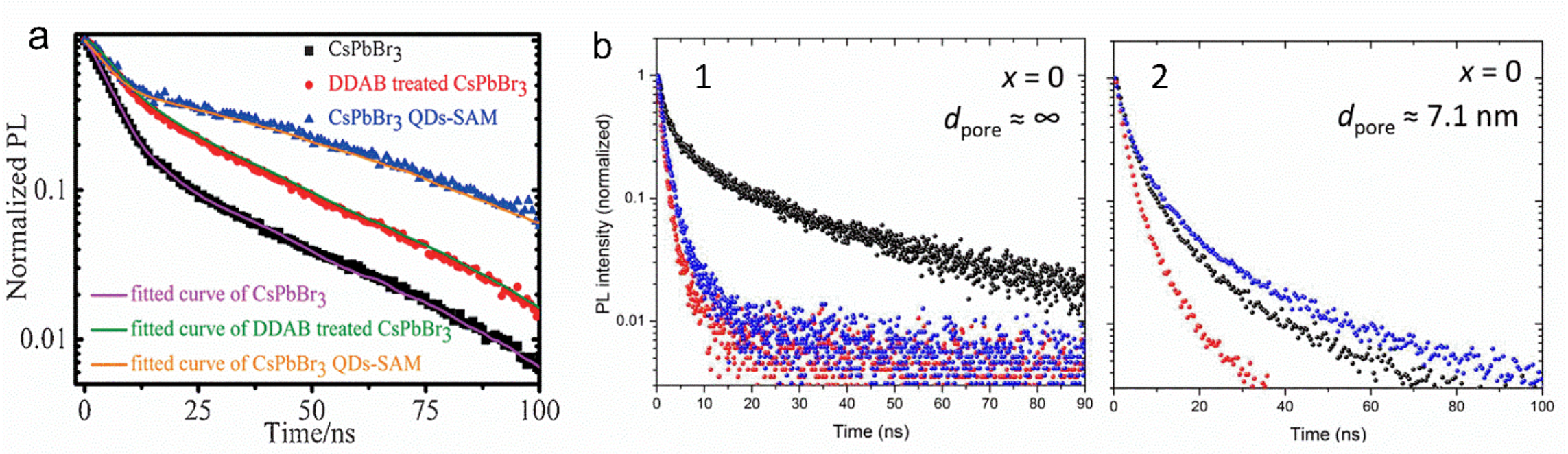

**Figure 11**. a) The PL decay curves of the $CsPbBr_3$, DDAB treated $CsPbBr_3$, and $CsPbBr_3$ QDs-SAM. Reproduced with permission.[69] Copyright 2017, Wiley-VCH. b) PL decay curves of (1) the pristine $CH_3NH_3PbI_3$ and (2) the $CH_3NH_3PbI_3$@MS (dpore ≈7.1 nm). Reproduced with permission.[68] Copyright 2016, American Chemical Society.

The TRPL spectra is an efficient tool to determine the lifetime of the MHPs based materials.[198,199] The results are usually acquired using a time-correlated single photon counting (TCSPC) technique, elucidating the charge carrier recombination behaviors. The TRPL decay curves can be fitted by the equation,[200]

$$I = \sum_n A_n \exp(-t/r_n) \qquad n = 1, 2, 3 \ldots \tag{5}$$

And the average lifetime can be determined by the following formula,

$$r_{avg} = \frac{\sum_n A_n c_n^2}{\sum_n A_n c_n} \qquad n = 1, 2, 3 \ldots \tag{6}$$

Herein, I is the normalized PL intensity, and $A_n$ is the amplitudes of the decay with related lifetime $\tau_n$.

The lifetime determined by TRPL spectra can be an effective tool to investigate the surface trap status of the MHPs and MHPs@$MO_x$ composite. As shown in **Figure 11a**, the lifetime of the $CsPbBr_3$ QDs-SAM is longer than that of other two samples.[69] The longer lifetime suggests that more efficient radiative recombination, which is in good match with the PLQY results ~ 67%, 80%, 90% for $CsPbBr_3$, DDAB treated $CsPbBr_3$, and $CsPbBr_3$ QDs-SAM, respectively. Meanwhile, the TRPL spectra also prove that DDAB can effectively passivate the surface defects of the $CsPbBr_3$ QDs as DDAB treated $CsPbBr_3$ shows longer lifetime than the $CsPbBr_3$ QDs.

On the other hand, the TRPL spectra can also be used to confirm that the $MO_x$ can protect the MHPs from the surrounding stress.[68] After be exposed to the UV light for 1 h, the pristine $MAPbBr_3$ is significantly destroyed, resulting in a decrease by factor of 13 (**Figure 11b**). And minor recovery can be observed after putting the sample in the dark for an overnight period. However, the lifetime of the $MAPbBr_3$@MS does not show significant decrease, and it can be totally recovered and even shows some improvements. The UV light causes the annealing of the pristine $MAPbBr_3$, while the "rigid" MS template provide the spatial confinement for the $MAPbBr_3$ and therefore limit the migration of the $MAPbBr_3$.

### 3.3. XRD Analysis

For the MHPs, XRD is the most widely used technique to identify the composition. The XRD peaks move to smaller angles as the increasing ratio of the I in the $CsPbBr(_{1-x}/_x)_3$ (**Figure 12a**), due to the larger radius of I than Br.[134] The structure can also be identified through comparing with the standard PDF card. In addition, XRD is also a useful tool to prove the evolution of the MHPs size as the template size for the MHPs/$MO_x$ prepared through the Template-assisted method.[68] As shown in **Figure 12b**, the characteristic peaks of the $MAPbBr_3$ become broaden as the pore size of the template decreases. This is in consistent with the decreasing $MAPbBr_3$ size. Furthermore, the crystalline state (crystal or amorphous) of the $MO_x$ shell can also be identified by XRD.[88] The XRD pattern of the $CsPbBr_3/TiO_x$ composite (drying at 25 °C) is the same as the $CsPbBr_3$ phase without new peaks belong to $TiO_2$ (**Figure 12c**). This means the formation of the amorphous $TiO_x$ phase at low temperature. New peaks belong to anatase appear in the XRD pattern of composite calcined at 300 °C, suggesting the $TiO_x$ phase transfers to $TiO_2$ at higher temperature. Importantly, the XRD can also be used as an effective way to study the stability of the MHPs/$MO_x$ compared with the pristine MHPs, through monitoring the change in the characteristic peaks of the MHPs over time.[86] Distinct characteristic peaks of the MHPs belong to the $CsPbBr_3$ can be observed for both the pristine $CsPbBr_3$ and the $CsPbBr_3@SiO_2$ at the beginning (**Figure 12d,e**). However, the peaks are almost indiscernible (even after being multiplied by 10 times) for the pristine $CsPbBr_3$ after storing for 3 days because of the degradation of the $CsPbBr_3$ without protection. However, the XRD pattern of $CsPbBr_3@SiO_2$ does not show any significant change after 3 days. And the peaks are still very clear though slight decrease is observed after 4 weeks storage (**Figure 12e**). All these results indicate that $SiO_2$ can provide effective protection for the $CsPbBr_3$ and therefore enhance the stability against storage.

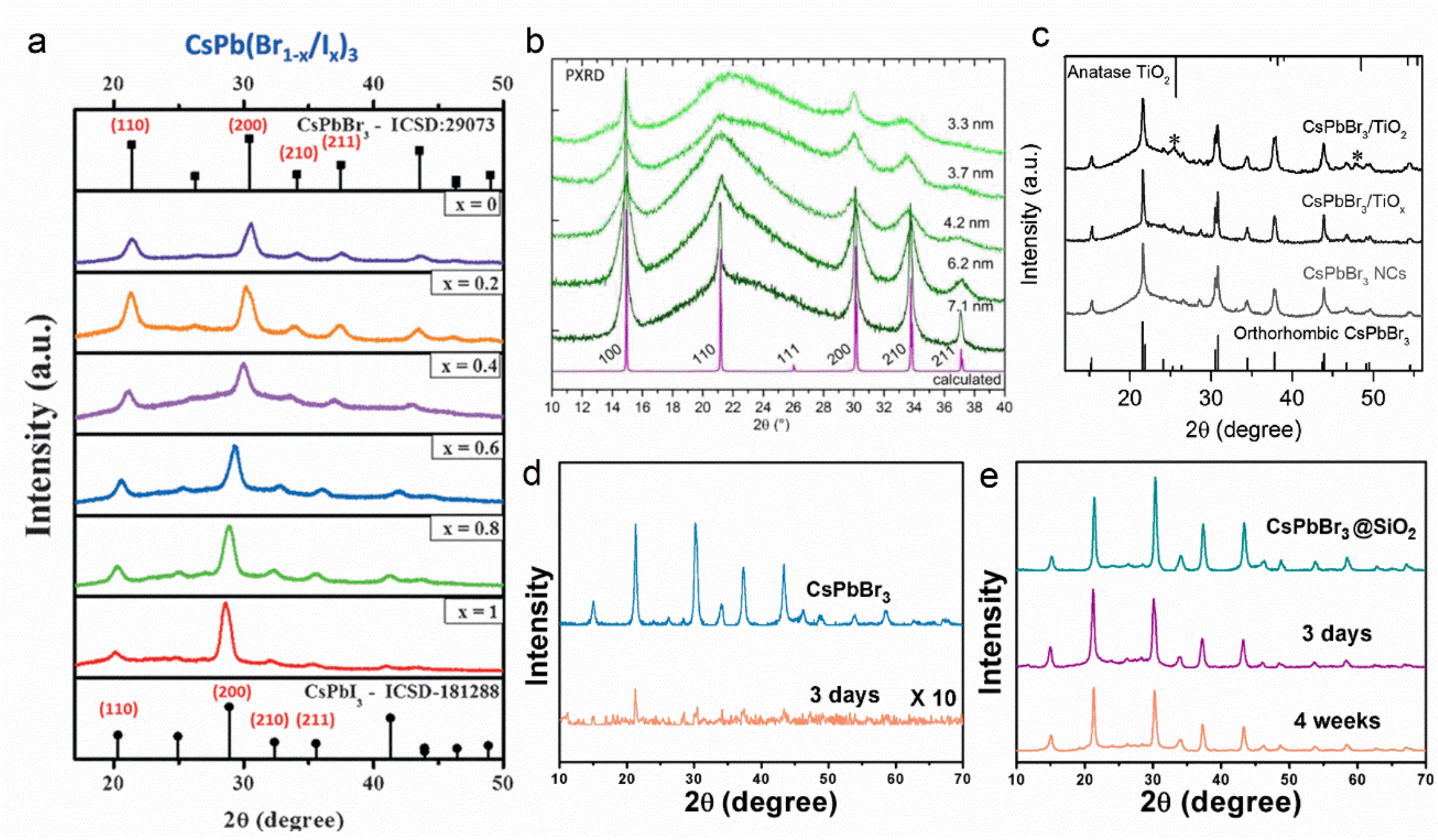

**Figure 12**. a) XRD patterns of $CsPbBr(_{1\text{-}x}/_{x})_3$. Reproduced with permission.[134] Copyright 2016, Wiley-VCH. b) XRD patterns of $MAPbBr_3$ embedded in MS template with different sizes. Reproduced with permission.[68] Copyright 2016, American Chemical Society. c) The XRD patterns of $CsPbBr_3$ NCs, $CsPbBr_3/TiO_x$ (drying at 25 °C), and $CsPbBr_3/TiO_2$ NCs (calcined at 300 °C). Reproduced with permiss ion.[88] Copyright 2018, Wiley-VCH. The XRD patterns of d) the pristine $CsPbBr_3$ and e) the $CsPbBr_3@SiO_2$ over time in air (25 °C and humidity of 75%). d,e) Reproduced with permission.[86] Copyright 2018, American Chemical Society.

### 3.4. Stability Measurement

As indicated in the introduction, highly stable MHPs still represents the milestone in this field. This includes: i) long-term storage in ambient environment (storage stability), ii) totally immersed in water (water stability), iii) exposed to the light irradiation (photo stability), and iv) treated at high temperature (thermal stability).[56,164,179,201-204]

#### *3.4.1. Storage Stability*

For the MHPs stores in the ambient conditions, the deactivation is mainly caused by the oxygen and the moisture in the air. One of the probable processes is that the oxygen-induced decomposition happens upon the irradiation. The $MAPbI_3$ degrades to $PbI_2$, $H_2O$, $I_2$ and $CH_3NH_2$ with the presence of the photo-generated carriers and $O_2$.[58,205] And the products have been verified by XRD, UV-Vis spectra, Nuclear magnetic resonance (NMR), Raman spectroscopy, and gas chromatography (GC). The degradation mechanism for the all-inorganic MHPs is still not clear, but PbO is detected by the XPS for $CsPbX_3$ after the photo-oxidation. Other explanations can be the oxygen induced aggregation of the MHPs or the surface of the MHPs are etched by the oxygen and thus the lower PLQYs.[206]

Normally, the MHPs/$MO_x$ composites show better storage stability than that of the pristine $MO_x$ because of the protecting effect provided by the coating regardless of the kind of the $MO_x$. In Sun's work,[35] they compared the storage stability between MHPs with and without the $SiO_2$ coating – **Figure 13a**. The red $CsPb(Br/I)_3$ QDs and $CsPb(Br/I)_3/SiO_2$ film obtained form through directly dropping the solution on a piece of quartz substrate and exposed to ambient environment under dark conditions. The emission of the red $CsPb(Br/I)_3$ QDs quickly quenches, most of the brightness lost after only one day, and almost no obvious emission can be observed after 2 days. In stark contrast, the $CsPb(Br/I)_3/SiO_2$ powders show an excellent stability in the air. No significate decrease in the emission can be observed after 1 day of exposure. The red $CsPb(Br/I)_3/SiO_2$ still retains the initial PL intensity even after 5 days. Indeed, PLQY of the $CsPb(Br/I)_3/SiO_2$ powders only show a minor change of ~ 5% after a storage period of 3 months.

Likewise, the storage stability is also enhanced after coating MHPs with $ZrO_2$.[70] As shown in **Figure 13b**, the PL intensity of the $CsPbBr_3/ZrO_2$ composite remain 80% of the initial value, while about 95% of the PL emission loses for the $CsPbBr_3$ NCs after 8 days storage. In our work,[89] the $CsPbBr_3@SiO_2/ZrO_2$ composites exhibit excellent storage stability, the PLQY keeps constant over 30 days under ambient conditions (**Figure 13c**). This further proves the benefit of the binary coating to protect the MHPs against the surrounding environment.

**Figure 13d** shows the comparison of the storage stability between the pristine $CsPbBr_3$ QDs and the $CsPbBr_3$ QD/$AlO_x$ composite prepared through the ALD method using the change in the maximal emission and the PL intensity as the standard.[123] The PL intensity of the pristine $CsPbBr_3$ QDs loses about 80% during the first 5 days and the PL peak shows significant red-shifts (10 nm), while only 10% of the PL intensity can be retained after 45 days. However, only small fluctuation is observed in the PL intensity and emission band shape of $CsPbBr_3$ QD/$AlO_x$ over 45 days. All these results indicate that the $AlO_x$ coatings effectively protect the $CsPbBr_3$ core, leading to the excellent storage stability.

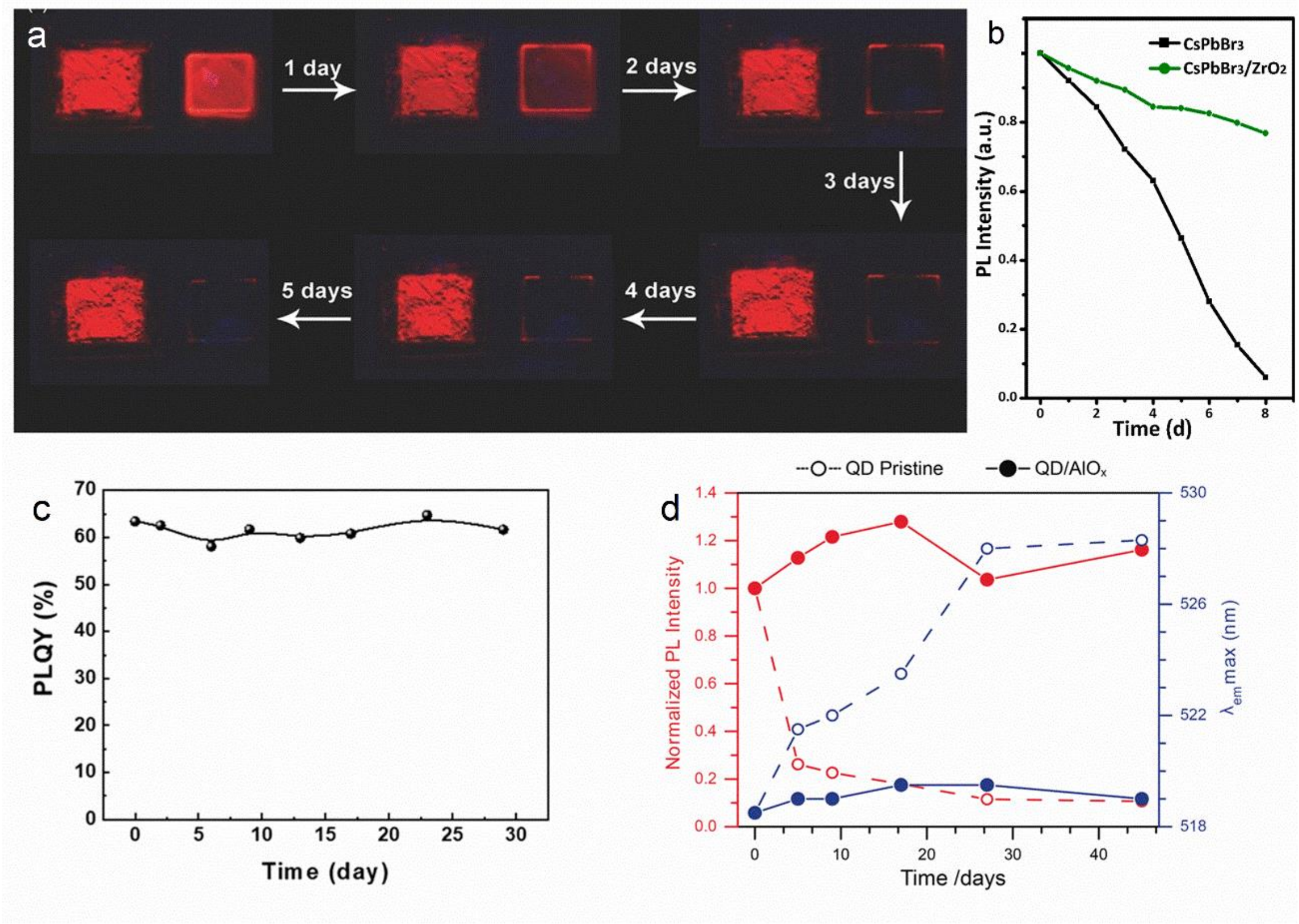

**Figure 13**. a) The storage stability of the red $CsPb(Br/I)_3/SiO_2$ powders (left) and red $CsPb(Br/I)_3$ QDs (right) under UV light. Reproduced with permission.[35] Copyright 2016, Wiley-VCH. b) The PL intensity change of $CsPbBr_3$ NCs and $CsPbBr_3/ZrO_2$ composite in a storage period of 8 days. Reproduced with permission.[70] Copyright 2019, American Chemical Society. c) Changes of the PLQYs over time for the $CsPbBr_3@SiO_2/ZrO_2$ composites under ambient storage conditions. Reproduced with permission.[89] Copyright 2020, Wiley-VCH. d) The storage stability of $CsPbBr_3$ QD/$AlO_x$ prepared by the ALD technique and the pristine $CsPbBr_3$ QDs in a period of 45 days. Reproduced with permission.[123] Copyright 2017, Wiley-VCH.

### 3.4.2. *Water Stability*

Several mechanisms have been proposed for the deactivation of the MHPs in water or under moisture.[126,207,208] For the $CsPbBr_3$ NCs@MSNs that prepared using the template-assisted method (without using the ligands and the thermal treatment under high temperature), a hypothesis is that CsBr will be soluble in water upon moisture treatment.[126] $CsPb_2Br_5$ forms because of the ionic nature of $CsPbBr_3$ NCs and therefore leads to the collapse of the $CsPbBr_3$ NCs structure. The dissolved CsBr species are still in the confined space provided by the MSNs. This process is reversible, the CsBr species will react with the $CsPb_2Br_5$ to form $CsPbBr_3$ NCs when the moisture is removed. However, the reversible phenomenon will not happen in the hot-injection prepared $CsPbBr_3$ because the hydrophobic ligands will prevent the recombination of the CsBr and $CsPb_2Br_5$ to form $CsPbBr_3$ again. Another mechanism which has been proved by both the experimental and theoretical study is that the decomposition of $MAPbI_3$. The $MAPbI_3$ combines with water vapor and the $MAPbI_3 \cdot H_2O$ forms. This phase changes can be totally reversed when the moisture is subsequently removed.[208] However, the $MAPbI_3$ will irreversibly degrade to $CH_3NH_2$, HI, and $PbI_2$ in the presence of liquid water.[207,208]

With regards to the water stability of the MHPs/$MO_x$, the MHPs/$MO_x$ composites usually show better storage stability than that of the pristine MHPs. However, the values range from several hours to several days, several weeks or even several months.[37,77,89,106,108,123,175] These values are closely related to surface status of the MHPs,[76] the thickness of the $MO_x$,[123,125] the crystalline state of the $MO_x$,[88] as well as the properties of the coating (mesoporous or dense).[37]

The surface status contributes to the enhanced water stability in addition to the protection of the $MO_x$ coating for the composite prepared through the sol-gel method. The as-prepared $CsPbBr_3/SiO_2$ NCs (I), the $CsPbBr_3$ NCs formed through water-triggered transformation (WT-$CsPbBr_3$ NCs, II), and $CsPbBr_3$ NCs obtained via hot-injection method (HI-$CsPbBr_3$ NCs, III) were dispersed in hexane solution, and equal volume of water was introduced to each vial (**Figure 14a**).[76] As a result, the $CsPbBr_3/SiO_2$ NCs solution still exhibits high emission even after 7 days. Largely decreased brightness is observed for the WT-$CsPbBr_3$ NCs solutions. However, much weaker PL intensity is observed for the HI-$CsPbBr_3$ NCs solution; the solution is almost colorless after adding water for 7 days. The surface passivation effect for the $CsPbBr_3$ NCs caused by the water-triggered process is responsible for the enhanced water stability of the WT-$CsPbBr_3$ NCs than that of the HI-$CsPbBr_3$ NCs.[209-213] The protections from both the formed $SiO_2$ particles and the oligomeric $SiO_2$ species further contributes to the improved water stability.

In addition, the thickness of the $AlO_x$ plays a key role for the stability of the MHPs@$MO_x$ composites that are fabricated through the ALD process.[123,125] The $CsPbBr_3$ QDs and $CsPbBr_3$ QDs-SLS only show lifetimes of ca. 1 and 0.5 h when they are immersed in water, respectively (**Figure 14b**).[125] The failure time increases after the deposition of the $AlO_x$ through the ALD process. And the composites show better stability as the higher ALD coating cycles. Finally, the composite obtained after 50 cycles of ALD coating shows the highest water stability-strong emission can still be observed after 20 days in water. The stability enhancement is related to the well-known diffusion barrier for water provided by the $AlO_x$ coating.[123] The same phenomenon is also proved by another work, as shown in **Figure 3d**.[123]

Meanwhile, the crystal coating shows better protection for the MHPs than that of its amorphous counterpart. In **Figure 14c**,**d**, the water stability of the $CsPbBr_3$ NC in solution, $CsPbBr_3$ NC powder, $CsPbBr_3/TiO_x$ powder, and $CsPbBr_3/TiO_2$ powder were tested through totally immersing them in Millil-Q water.[88] The $CsPbBr_3$ NCs toluene solution is very vulnerable to the water, and almost all the $CsPbBr_3$ NCs degrade in just 15 min. The $CsPbBr_3$ NC powders survive in terms of photoluminescence for more than 3 days in water. The $CsPbBr_3/TiO_x$ powders exhibit enhanced water stabilities, holding stable for more than 1 week in Millil-Q water. However, the $CsPbBr_3/TiO_2$ composite shows extremely high stability, almost no change can be detected after >12 weeks. In addition, no change in both the size and morphology, as well as the crystallites state are observed for the $CsPbBr_3/TiO_2$ composite after being immersed in water for 12 weeks as conformed by the TEM and XRD. Thus, the crystallization of the $TiO_x$ coating is critical for the enhanced stability.

On the other hand, the shape of the coating (mesoporous or dense) plays an indispensable role in the stability of the $MHPs/MO_x$ composites, this is highly reflected in the composites that are prepared through the template related methods. Usually, the porous matrixes cannot totally protect the MHPs from the surrounding water, because moisture will permeate into the porous and finally destroy the MHPs. The composites treated at lower calcination temperatures, like 400 °C and 500 °C, almost loss all their absorption spectra after water washing (**Figure 14e,f**).[37] This is because the original MS is not collapsed at the lower temperature, and the water will touch the unprotected $CsPbBr_3$ QDs and then destroy them. However, the composite shows significantly higher water resistance at higher calcination temperatures (**Figure 14g**), like 600 °C, because part of the MS has been transferred to dense $SiO_2$ shell. Furthermore, the composites exhibit outstanding water stability under calcination temperatures of more than 600 °C. No obvious decay in PLQYs can be observed after they are immersed in water for 50 days. This is attributed to the fact that a ceramic-like, dense $SiO_2$ shell has totally formed under high temperature (more than 600 °C) and therefore can effectively protect the MHPs from touching water.

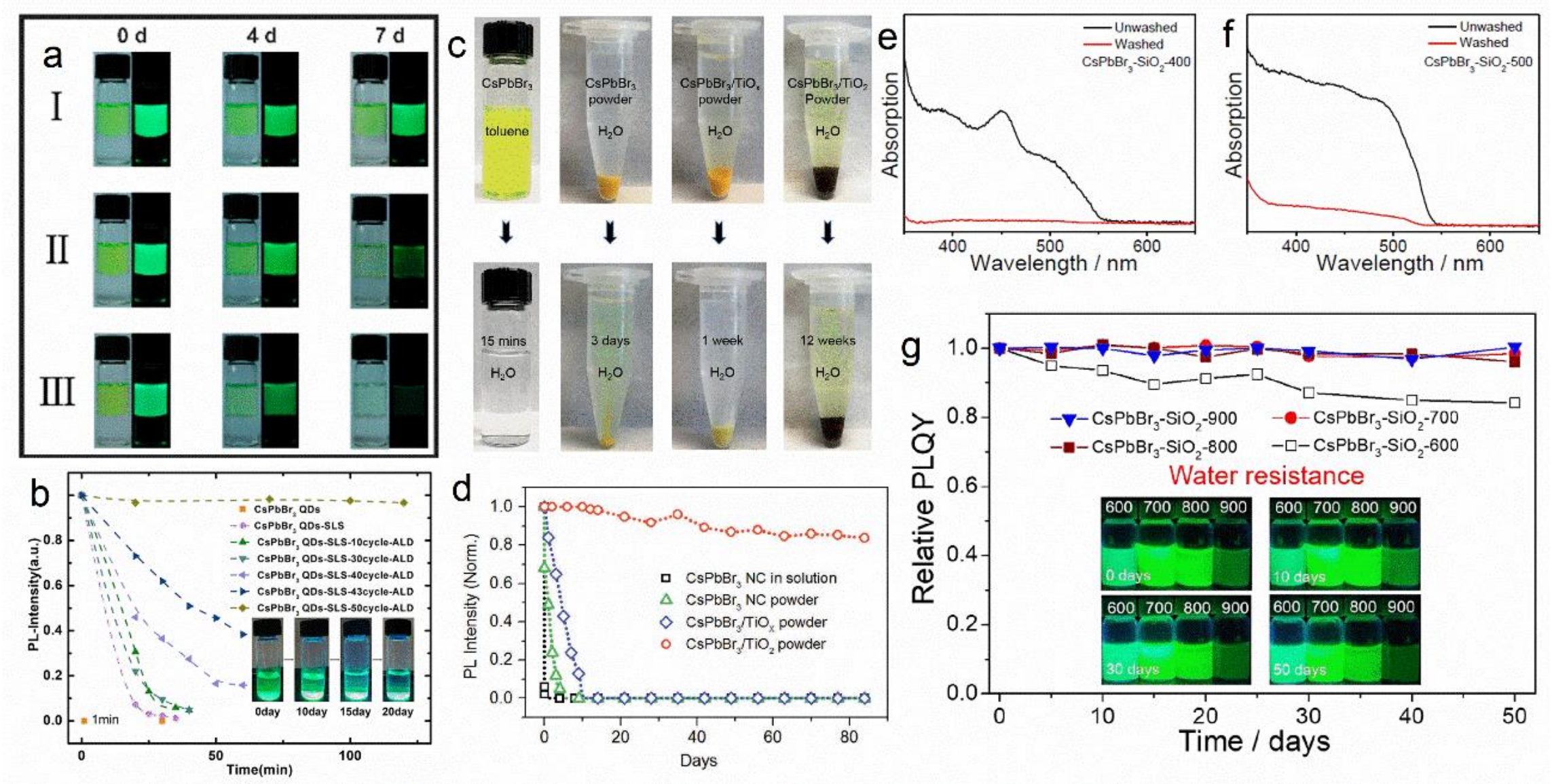

**Figure 14**. a) The water stability of $CsPbBr_3/SiO_2$ NCs (I), the WT-$CsPbBr_3$ NCs, (II), HI-$CsPbBr_3$ NCs (III). In the vials, the top layer and the bottom layer is the MHPs-hexane solution and the deionized water, respectively; and the left pictures are taken under the daylight, the right graphs are taken under the UV-light ($\lambda$ = 365 nm). Reproduced with permission.[76] Copyright 2018, American Chemical Society. b) The water stability of the $CsPbBr_3$ QDs, $CsPbBr_3$ QDs-SLS, and $CsPbBr_3$ QDs-SLS with different ALD working cycles. Inset shows the photos taken from the $CsPbBr_3$ QDs-SLS-50 cycles-ALD after immersion in water for different days under UV-light. Reproduced with permission.[125] Copyright 2018, American Chemical Society. c) The water stability and d) the PL stability of the as-prepared $CsPbBr_3$ NC in solution, $CsPbBr_3$ NC powder, $CsPbBr_3/TiO_x$ powder, and $CsPbBr_3/TiO_2$ powder. c,d) Reproduced with permission.[88] Copyright 2018, Wiley-VCH. UV-Vis absorption spectra of $CsPbBr_3$-$SiO_2$ powders calcined at e) 400 °C, f) 500 °C before and after water washing; g) the PLQYs of the composites calcined at higher temperatures immersed in water for 50 days. Inset shows the photographs taken over time. e,f,g) Reproduced with permission.[37] Copyright 2020, Nature Publishing Group.

### 3.4.3. *Photo Stability*

Light is indispensable in our daily life. As mentioned above, the MHPs will degrade under oxygen and photo illumination. However, the degradation of the MHPs happens even without the oxygen. Organic gas components ($CH_3NH_2$, HI, $CH_3I$ and $NH_3$, $I_2$) can still be detected by mass spectrometer under vacuum for the $MAPbI_3$.[214] Moreover, there is no wavelength threshold for the photodegradation of the $MAPbI_3$. And this degradation behavior under vacuum will be accelerated under higher power illumination and temperature. Another mechanism for the deactivation of the $CsPbBr_3$ QDs based on the regrowth of the QDs under illumination is proposed by Zheng's group.[215] During the illumination, the photogenerated carriers diffuse to the surface of the $CsPbBr_3$ QDs, and combined with the surface ligands of the $CsPbBr_3$ QDs. Then, some of ligands will remove from the surface and this will result in the aggregation of the unprotected $CsPbBr_3$ QDs. And the regrowth of the QDs has been proved by the TEM. The aggregation of the QDs is accompanied by the red shift of both the PL emission and the absorption spectra, as well as the color evolution.[206,215]

Embedding MHPs inside $MO_x$ is an effective way to eliminate the photo-induced aggregation of MHPs in both solution, films, powders, and under operation. Meanwhile, the $MO_x$ shell will also help the MHPs from the oxygen and water, and therefore decrease the deactivation speed under illumination. But it is very difficult to say to which kind of the $MO_x$ has the best photostability or the same $MO_x$ prepared from which method has more priority, because there is no standard to evaluate the photostability currently. The measurement conditions vary among different groups (**Table 1**). But the MHPs/$MO_x$ composites usually exhibit better stability than that of the pristine MHPs regardless of the treatment method and the $MO_x$.

As illustrated in **Figure 15a**, an instrument with 450 nm LED irradiation (175 mW $cm^{-2}$) was designed to evaluate the photostability of the MAPB-QDs solution and MAPB-QDs/$SiO_2$ solution.[83] The solutions were put in two quartz cuvettes. As shown in **Figure 15b**, improved PL intensity was observed for both solutions during the first 2 h under constant irradiation. This is known as the "photoactivation" phenomenon due to the fact that the surface trap caused by the dangling bonds are cured by the light.[203,216,217] This has been noted in many MHPs related studies regardless of the synthesis preparation, like the $CsPbBr_3$ QD/$SiO_2$/$Al_2O_3$ that prepared through the ALD process,[125] the $CsPbBr_3$/$SiO_2$ composites prepared through the SILAR method,[139,140] $CsPbBr_3$/$SiO_2$ composite prepared using the template-assisted method,[37] the composites fabricated through the sol-gel method, like $CsPbBr_3$@$SiO_2$,[86] $CsPbBr_3$@$SiO_2$/$ZrO_2$,[89] and so on. The PL intensities decrease gradually for both the solutions as the increasing irradiation time. But the PL intensity of the MAPB-QDs solution shows faster decay rate than that of the MAPB-QDs/$SiO_2$ solution. This results to only 7 % of the remnant PL for the MAPB-QDs solution, compared to a 40% PL decay for the MAPB-QDs/$SiO_2$ suspension after 49 h. Moreover, the FWHM of the MAPB-QDs exhibits stronger changes than those noted in MAPB-QDs/$SiO_2$

solution – **Figure 15b**. Finally, the MAPB-QDs solution turns yellow, showing yellow precipitates because of the photo-induced growth of the $MAPbBr_3$ particles. This indicates a severe aggregation caused by the irradiation – **Figure 15a**. However, the MAPB-QDs/$SiO_2$ solution is still green and exhibits high emission after 43 h, indicating that the $SiO_2$ coating can stop the growth of the MAPB-QDs and therefore improve the photo stability.

Similarly, the $CsPbBr_3$ QD/$AlO_x$ nanocomposites fabricated via the ALD treatment also show better photostability than the pristine $CsPbBr_3$ QD.[123] Dramatic decay in the PL intensity and significant red shift of the peak emission are observed from the pristine $CsPbBr_3$ QD after 2 h of irradiation (**Figure 15c**). The $CsPbBr_3$ QD/$AlO_x$ nanocomposites show an irreversible PL to about 80% of the initial value and slight red shift after 8 h, but the PL intensity will be stabilized after the fast decay. Further measurements from XRD and UV-Vis prove that the $CsPbBr_3$ QD does not change after photo-soaking. Therefore, the fast decay under the high irradiation is due to the formation of the carrier trapping defects caused by the photoinduced desorption of surface ligands.

Furthermore, the photostability of the $CsPbBr_3$ NC protected by the $TiO_2$ prepared from the sol-gel method is also investigated.[88] The $CsPbBr_3$ NCs in toluene solution are very vulnerable to UV light and quickly quench in 5 h under UV-light (**Figure 15d**). The $CsPbBr_3$ powers show higher stability, the PL intensity decreases to 37% of the original value after 24 h. However, the $CsPbBr_3$/$TiO_2$ powders exhibit significantly enhanced photostability under the same condition, ca.75% of the original PL intensity remains after 25 h under UV light. This proves that the $TiO_2$ shell can effectively protect the $CsPbBr_3$ NCs from the UV light and therefore the excellent stability.

Further works have also confirmed that MHP@$MO_x$ features enhanced photo-stabilities than the pristine MHP counterparts.[37,75,77,125,140] For instance, the $CsPbBr_3$ NC and $CsPbBr_3$ NC/$MO_2$ (M = Si, Ti, Sn) films were fabricated on the fluorine doped tin oxide (FTO) substrates through a centrifugally cast process.[84] These films were exposed to the 365 nm UV light to test the photostability. As shown in **Figure 15e**, the pristine $CsPbBr_3$ NCs are very sensitive to the UV light. Fast decay is observed due to serve decomposition and aggregation, as observed in TEM images. On the contrast, all these $CsPbBr_3$@$MO_2$ composites exhibit enhanced photo stability and can remain more than 60% of the initial intensity after 10 h. Here, the explanation for the superior resistance for $CsPbBr_3$ NC/$TiO_2$ and $CsPbBr_{3-x}Cl_x$ NC/$SnO_2$ compared with other composites is that they have thicker shell than the $CsPbBr_3$ NC/$SiO_2$.

The binary coating usually also shows better protection than that of the individual shell. The $CsPbBr_3$ QDs that protected by the $SiO_2/Al_2O_3$ monolith ($CsPbBr_3$ QDs-SAM) shows better photo stability under 470 nm illumination provided by a LED lamp (21 mW $cm^{-2}$) than other counterparts.[69] As shown in **Figure 15f**, the PL intensity holds at 90% of the initial value for the $CsPbBr_3$ QDs-SAM powder after being irradiated for >300 h. However, almost no PL emission is observed for the $CsPbBr_3$ QDs after 72 h, while $SiO_2$ coated $CsPbBr_3$ powder losses ca. 80% of the initial PL intensity after 200 h. Thus, the combination of a double $SiO_2$ and $Al_2O_3$ protective coatings can efficiently prevent the damage of the diffused $O_2$, limiting the photo-induced degradation in the presence of $O_2$.

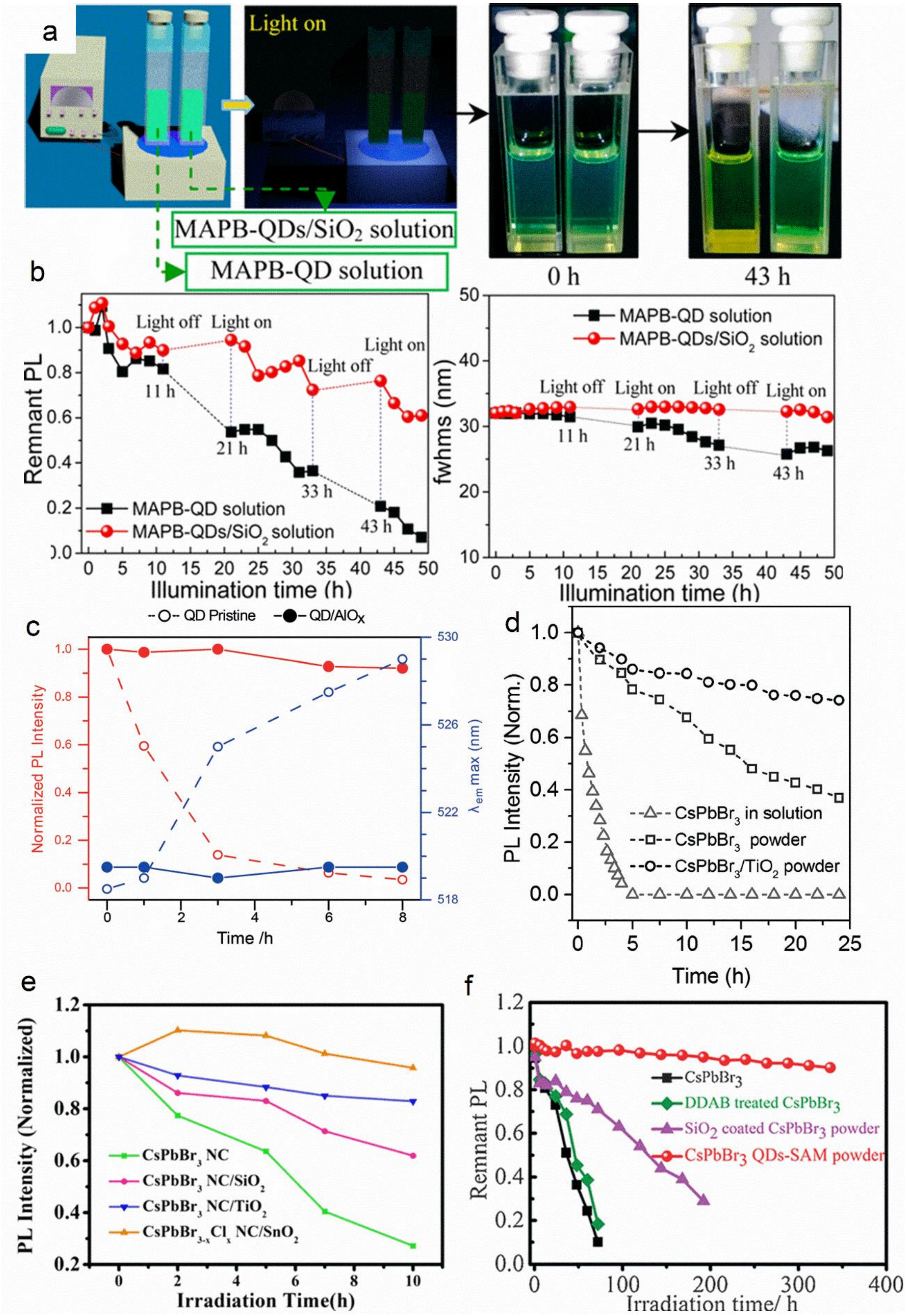

**Figure 15**. a) Schematics for the photo stability measurement of the MAPB-QDs solution and MAPB-QDs/$SiO_2$ solution under 450 nm LED irradiation (175 mW $cm^{-2}$); b) The change in the PL intensity (left) and FWHM (right) for the QD solutions vs. illumination time. Reproduced with permission.[83] Copyright 2016, American Chemical Society. c) The PL intensity and the peak location for the $CsPbBr_3$ QD/$AlO_x$ nanocomposites and the pristine $CsPbBr_3$ QD during the photostability test. The test is performed under simulated solar spectrum irradiation (100 mW $cm^{-2}$) in air; Reproduced with permission.[123] Copyright 2017, Wiley-VCH. d) The comparison of PL intensity of the samples under UV light. Reproduced with permission.[88] Copyright 2018, American Chemical Society. e) Comparison of the photo stability of the $CsPbBr_3$ NC and $CsPbBr_3$ NC/$MO_2$ (M = Si, Ti, Sn) films under 365 nm UV light irradiation. Reproduced with permission.[84] Copyright 2018, American Chemical Society. f) Photo stability of the bare $CsPbBr_3$, DDAB treated $CsPbBr_3$, $SiO_2$ coated $CsPbBr_3$, and the $CsPbBr_3$ QDs-SAM powders under 470 nm illumination provided by a LED lamp (21 mW $cm^{-2}$). Reproduced with permission.[69] Copyright 2017, Wiley-VCH.

### *3.4.4. Thermal Stability*

High temperature is unavoidable during the fabrication or operation of the MHPs based devices. There are several mechanisms about the thermal induced deactivation of the MHPs.[59,189,214,218] Cyclic heating-cooling experiments on $CsPbBr_3$ show that the PL loss can be attributed to the reversible (below 450 K) and irreversible (more than 450 K) processes (**Figure 16a**).[189] The reversible PL loos under low temperature is due to the thermally assisted trapping, like the halogen vacancy centers, while the irreversible PL loss under high temperature is caused by the loss of the organic ligands on the $CsPbBr_3$ surface and the sintering of NCs. Another work shows that $MAPbI_3$ starts to release $I_2$ under dark and inert condition when the temperature is higher than 60 °C. [214] And the thermal induced decomposition reaction under low heating temperature, vacuum, inert atmosphere is:

$$MAPbI_3 \rightarrow PbI_2 + Pb^0 + I_2 + CH_3NH_2 + CH_3I + HI + NH_3 \quad (7)$$

The common thermal degradation pathways for the widely used $MAPbX_3$ are:[59]

$$MAPbX_3 \rightarrow PbX_2 + CH_3NH_2 + HX \quad (8)$$

$$MAPbX_3 \rightarrow PbX_2 + CH_3X + NH_3 \quad (9)$$

For the $MAPbBr_3$ tested under vacuum and inert atmosphere, the degradation process will change from (8) to (9) when the temperature is higher than 300 °C. [214] Other explanation of the PL quenching under high temperature is the thermal-induced aggregation of the MHPs. As proved by embedding the MHPs into the polymer matrix and the thermal stability of the MHPs/polymer composite is significantly improved.[155,219]

However, the deactivation of MHPs caused by the oxygen, light, water will be accelerated and amplified at high temperature, and lead to the fast decay of the MHPs. Therefore, it is urgent to fabricate the core/shell structure to protect the MHPs.

The thermal stability is measured through monitoring the changes in the PL emission band either upon increasing the temperature or at constant temperature over time under ambient conditions.[29,35,78,89,202] In general, the thermal-stability of MHPs is highlighted by a linear decrease of the PL intensity upon increasing the temperature reaching 50% ($PL_{50}$) of the initial emission intensity at 50 °C ~ 70 °C (**Figure 16b**).[69,134,220] Likewise, almost all the PL intensity will loss at constant temperature of 60 ~ 70 °C for ca. 15 h ( **Figure 16c**).[80] In stark contrast, MHP@$MO_x$ composites exhibit a remarkable enhancement in thermal-stability. In average, $SiO_2$ coatings increases the temperature threshold up to 90~100 °C to reach 50 % of the initial PL intensity and stabilities of 15 h at 60 °C. [75,80,134]

The thermal quenching problem is still very severe for the MHPs even though the above significant enhancements have been achieved, as the widely adopted silica encapsulation can only help reduce the problems caused by the thermal induced aggregation. Traps caused by the thermal induced halogen vacancies are also responsible for the thermal quenching.[189] In the previous work, fluoride ions have been proved to have the function of passivating the vacancies for MHPs films.[221] Therefore, a very recent work describes the use of AHFS water solution as both the precursor for the hydrolysis of $SiO_2$ and the source for fluoride ions to fabricate the $CsPbBr_3$@$SiO_2$.[77] As shown in **Figure 16d**, five sealed vials with the pristine $CsPbBr_3$ NCs, water treated, and 0.1, 0.5, 1 M AHFS treated samples were put on a hot plate in an oil bath at RT, and the maximal absorption was kept at the same value. As the temperature increases, the PL intensities of the water treated and the pristine CPB-NCs decrease quickly in the similar trend, and most of the emission loses at 363 K. However, all these AHFS treated samples show higher resistibility against thermal quenching (RATQ) under the same condition, and more than 80% of the initial emission can be retained at 363 K. When these samples were further treated at a higher temperature of 423 K, the solution of the water treated and the pristine CPB-NCs become yellow and only minor emission can be observed under UV light (right in **Figure 16d**). On a striking contrast, the AHFS-treated solutions are still green and exhibit bright green emission under UV-light. At the same time, the FWHMs of the samples without AHFS treated change largely, especially the pristine CPB-NCs that shows a big change of ca. 25 nm because of the serve aggregation. However, all these modified samples show a narrower range of 3 nm, indicating better thermal resistance. The TEM images also prove the serious aggregation and growth for the unmodified CPB-NCs after 353 K (80 °C) treatment.

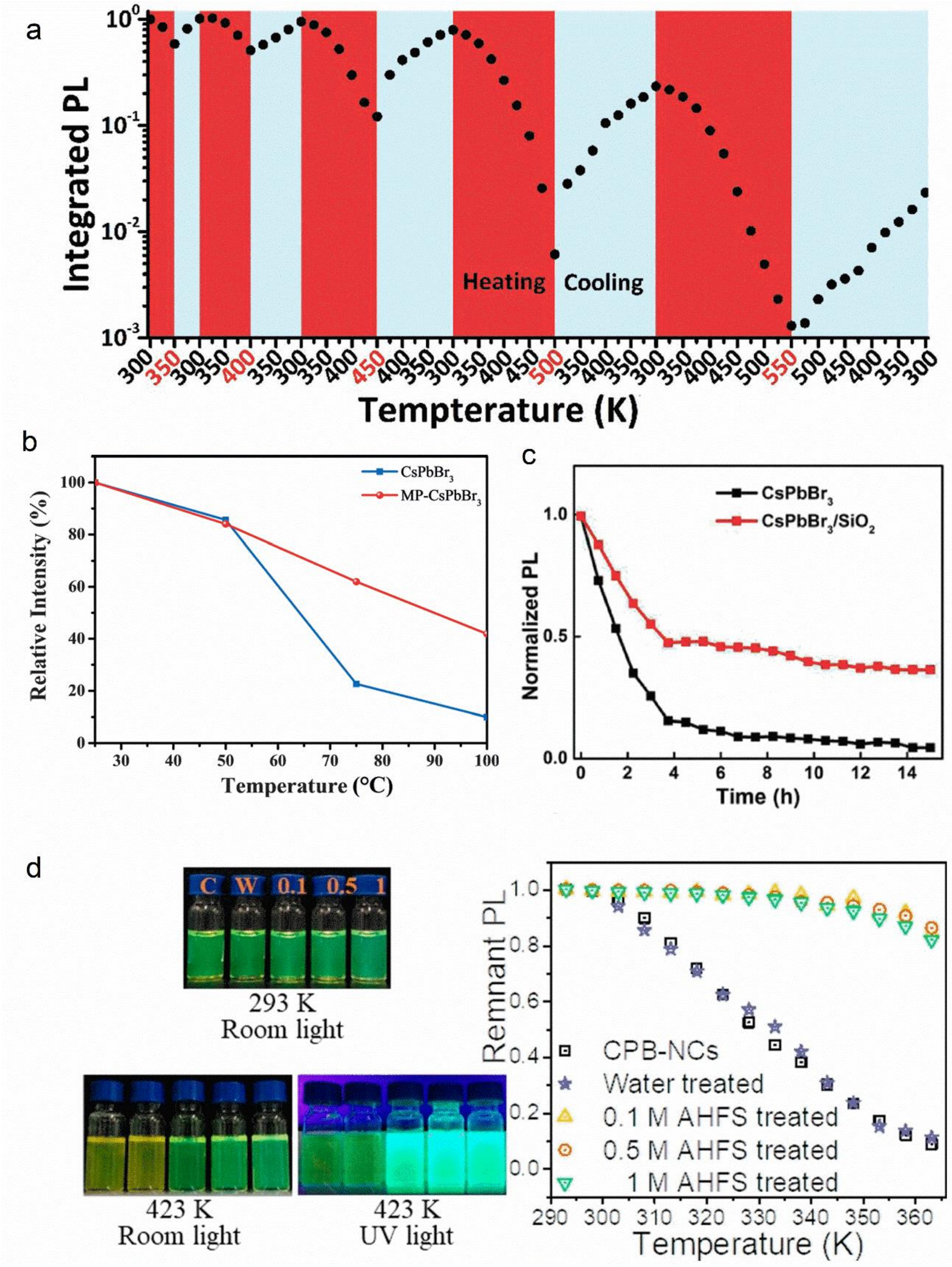

**Figure 16**. a) Thermal cycling measurements of $CsPbBr_3$. Reproduced with permission.[189] Copyright 2017, Wiley-VCH. b) The thermal stability of the $CsPbBr_3$ and MP-$CsPbBr_3$. Reproduced with permission.[134] Copyright 2016, Wiley-VCH. c) The thermal stability of the $CsPbBr_3$ and $CsPbBr_3/SiO_2$ treated at constant 60 ºC, ambient environment with 75% of RH.

Reproduced with permission.[80] Copyright 2018, Wiley-VCH. d) Photographs (left) and the change of the PL intensity (right) of the pristine CPB-NCs, water treated, and 0.1, 0.5, 1 M AHFS treated samples at t 293 and 423 K from 293 to 363 K. d) Reproduced with permission.[77] Copyright 2020, American Chemical Society.

### 3.5. Other Technique Methods

Besides, many other technique methods, such as the FTIR are also widely used to investigate the $MO_x$ formation process and the surface species of the composite;[35,88] and TGA can be used to determine the composition as well as the thermal stability of the composites.[88,222] In addition, NMR is also essential to explore the interaction of the precursor of the $MO_x$, like TMA with the MHPs surface during the initial phases of nucleation.[123,125]

### 3.6. Take Home Messages about Main Features of MHP@$MO_x$ with Respect to the Type of $MO_x$ Coating

The characterization methods should be combined with each other to fully understand the properties of MHP@$MO_x$. Specifically, the SEM and TEM can provide the morphology of the composites, the size of the MHPs as well as the thickness of the shell, the distribution of the MHPs in the $MO_x$ (homogenous or not), the properties of the MHPs/$MO_x$ (aggregation or monodisperse), and the structure of the core-shell, like Janus structure or totally coverage. Moreover, the cross-section SEM is a good tool to detect the interface between $MO_x$ and MHPs for the composite prepared by the ALD method. XPS can be used to analysis the bonding interactions between MHPs and $MO_x$ through comparing the binding energy of the elements before and after the coating. It can also provide surface information, like Pb-rich or halide-rich surface, through studying elemental ratio. And it also be used to reveal the crystallinity state of the $MO_x$. XRD is widely used to identify the composition of the MHPs and the $MO_x$, the crystalline state (crystal or amorphous) of the $MO_x$ shell, as well as the size evolution of the MHPs in the template as the size of the template. Furthermore, the stability of the MHPs/$MO_x$ can also be studied through monitoring the change in the characteristic peaks.

**Table 3**. The comparison of the characterization method for the MHPs@$MO_x$.

| **Characterization method** | | **Obtained information for MHPs and MHPs@$MO_x$** | **Comments** |
|---|---|---|---|
| **Microscopic analysis** | SEM | Distribution of the particles, composition and the elemental distribution (combined with EDS); interface between MHPs and $MO_x$ using cross-section SEM | SEM and TEM are always combined with each other to get clear |
| | TEM | Distribution of the particles; the size and structure of both MHPs and $MO_x$; confirming the boundary between MHPs and $MO_x$; composition and the elemental distribution (combined with EDS); evolution of the sizes | information of the MHPs@$MO_x$ |
| | XPS | Valence state; chemical status; elemental composition; crystallinity state of $MO_x$ | FTIR spectra, XRD, TGA, and XPS should be combined with each other to investigate the surface changes, the crystallinity state, and the composition of the nanomaterials |
| **Spectroscopic analysis** | UV-Vis spectra | Size evolution of the MHP particles; structural transformation; absorption capacity; qualitative determination; band-gap | PL spectra and UV-Vis spectra can be recognized as the mutual complementary to each other |
| | PL spectra | Size evolution of the MHP particles; structural transformation; qualitative determination; PLQYs; radiative recombination | |
| | TRPL spectra | Lifetime; surface trap status; radiative recombination and non-radiative recombination | |
| **XRD analysis** | | Identification of the composition; evolution of the MHPs size; crystalline state (crystal or amorphous); stability of MHPs and MHPs@$MO_x$ | |
| | Storage stability | The decay process with the storage period | The thickness, the surface status (passivated or not, crystal or amorphous), shape (mesoporous or dense) of $MO_x$ have very important effects on the stability of the MHPs@$MO_x$; the distribution of the MHPs inside $MO_x$ also maters for the stability of the MHPs@$MO_x$ |
| | Water Stability | The deactivation of the MHPs in water or under moisture | |
| **Stability measurements** | Photo-stability | The stability against light | |
| | Thermal Stability | The stability against heating | |
| **Other technique methods** | FTIR, TGA, NMR... | The surface species, chemical status, composition, interaction between MHPs and $MO_x$… | The techniques should be combined with each other to have a better understanding of the MHPs@$MO_x$ |

PL spectra and UV-Vis spectra can be recognized as the mutual complementary to each other. The intensity (for both the absorption and the emission) and the location of the peaks change after the MHPs are coated with $MO_x$. They can be used to identify the phase change of the MHPs, for example, the transformation from $Cs_4PbBr_6$ to $CsPbBr_3$. However, the absorption, the PL emission (both the shape and location), as well as the PLQYs evolve as the reaction time, the concentration of the precursor (for the formation of the $MO_x$), the size and the morphology of the template used in the fabrication of the MHPs/$MO_x$. The parameters that affect the properties of the MHPs/$MO_x$ are very complicated, it is very hard to say we can get the best products under which condition. Therefore, all these above parameters should be kept in mind during the optimization process. But the MHPs that prepared using the ligands normally show better PLQYs than that do not adopt. For the photocatalysis, the UV-Vis DRS and PL spectra can give valuable information about the light absorption ability and the photogenerated electron transfer of the composite.

The deactivation of the MHPs is related to the oxygen-induced decomposition happens upon the irradiation, the water/photo/thermal induced degradation, the thermal/photo/oxygen induced aggregation, the surface etched by the oxygen, the thermally assisted trapping (like the halogen vacancy centers), the loss of the surface ligands under thermal treatment. And these deactivation behaviors will be accelerated when they face each other. Usually, the $MO_x$ can give protection to the MHPs against storage, moisture, heat, as well as light regardless of the kind of the $MO_x$ and the method to fabricate the composite. However, a passivated MHPs surface, thicker coating, highly crystallized $MO_x$ favor the better water stability. Moreover, the binary coating can also provide better protection for the MHPs than that of the individual ones. Importantly, the $MO_x$ with a dense surface can simultaneously provide efficient barrier against the surrounding stress. And the better understand the deactivation process is closely related to the employ the basic analysis technique.

## 4. The Application of the MHPs@$MO_x$

As the enhanced stability and easier synthesis process are achieved for the MHPs@$MO_x$ composites, a lot of promising applications are promoted in many fields with excellent performance. In this section, we will present the application of MHP@ $MO_x$ in LEDs, lasers, photoelectrochemistry, and cells imaging, and triple-modal anti-counterfeiting codes, drug delivery, and ECL, and other applications, like daytime radiative cooling, detecting, and so on. These applications are classified by the type of the $MO_x$ coatings.

### 4.1. MHPs@$SiO_2$

$SiO_2$ is the most widely used coating materials for the encapsulation of MHPs, because of its easy accessibly, good environmental stability, as well as high transparency.

#### 4.1.1. *Filters for Phosphor-Converted White Light-Emitting Diodes (pc-WLEDs) and Backlight Display*

Currently, the pc-LEDs are the mostly and widely used WLEDs technology in modern lighting technology and backlight displays.[223, 224] The working mechanism consists in using a luminescent coating that is directly excited by the UV or blue LED chips partially converting the high-energy emission into yellow, orange, and red lights. The combination of the blue/UV emission of the chip and the low-energy emission of the coating leads to the desired white light.

**Table 4.** Summary of the device performance for MHPs/$MO_x$ based pc-WLEDs.

| No. | Composite | Device structure | Luminous efficiency ($lm\ W^{-1}$) | CRI | CCT (K) | CIE (x,y) | Color gamut | Stability | Ref. |
|---|---|---|---|---|---|---|---|---|---|
| 1 | $CsPbBr_3@SiO_2$; $CsPb(Br/I)_3@SiO_2$ | Blue LED chip/$CsPbBr_3@SiO_2$/$CsPb(Br/I)_3@SiO_2$ | 61.2 | | | (0.33, 0.33) | 120% of NTSC | Half lifetime is 227 h with an initial luminance of 100 cd $m^{-2}$ | [35] |
| 2 | $CsPbBr_3@SiO_2$ | Blue LED chip/$CsPbBr_3@SiO_2$/CdSe | 56 at 5 mA | 63 | | (0.3, 0.32) | 138% of NTSC | The color coordinate is still in the white emission area after 1 h | [76] |
| 3 | $CsPbBr_3@ZrO_2$ | Blue LED chip/$CsPbBr_3@ZrO_2$/CdSe | 55 at 1 mA | | | (0.28, 0.33) | | PL intensity of green emission remains above 90% after 2 h | [70] |
| 4 | $CsPbBr_3$@silicone | Blue LED chip/ CsPbBr3@silicone/ $CsPb(Br_{0.3}I_{0.7})_3$@silicone | | | | (0.32, 0.30) | | | [93] |
| 5 | $CsPbBr_3@SiO_2$ | Blue LED chip/$CsPbBr_3@SiO_2$/KSF | 63.5 | 83.3 | 7425 | (0.32, 0.30) | | Emission peak and spectra shape of the WLED remain unchanged after 13 h continuous operation at 6 mA | [101] |
| 6 | $MAPbBr_3@SiO_2$ | GaN LED chip/$MAPbBr_3@SiO_2$/KSF | | | | (0.32, 0.33) | | Green emission drops 10% after 25 h under 20 mA | [75] |
| 7 | $CsPbBr_3/SiO_2$ $CsPb(Br/I)_3/SiO_2$ | Blue LED chip/$CsPbBr_3/SiO_2$/$CsPb(Br/I)_3/SiO_2$ | 35.32 | | | (0.30, 0.31) | | The device is stable after 40 h | [79] |
| 8 | $CsPbMnCl_3@SiO_2$ | UV GaN LED/$CsPbBr_3$/$CsPbMnCl_3@SiO_2$ | 68.4 at 10 mA | 91 | 3857 | | | | [74] |
| 9 | Mn-doped $CsPbCl_3$-$SiO_2/Al_2O_3$ | Blue LED chip/Ce-PiG/Mn-doped $CsPbCl_3$-$SiO_2/Al_2O_3$ | 80.91 | 83.8 | 4082 | (0.38, 0.37) | | | [104] |
| 10 | $CsPbBr_3$-$SiO_2$ | Blue LED/$CsPbBr_3$-$SiO_2$/$CsPbBr_{1.2}I_{1.8}$ | 35.4 | | 5623 | (0.33, 0.36) | 127% of NTSC; | | [107] |
| | | | | | | | 95% of Rec. 2020 | | |
| 11 | $Cs(Pb_{0.66}/Mn_{0.34})Cl_3$@MSNs | UV LED chip/$CsPb(Br/Cl)_3$@MSNs/$Cs(Pb_{0.66}/Mn_{0.34})Cl_3$@MSNs | 62.5 | 82 | 5677 | (0.34, 0.36) | | 65% of the initial intensity after 30 min at 10 mA | [131] |
| 12 | $CsPbBr_3$/MS | Blue LED chip/$CsPbBr_3$/MS/$CsPb(Br_{0.4}I_{0.6})_3$ | 30 | | | (0.24, 0.28) | 113% of NTSC; 85% of Rec. 2020 | | [134] |
| 13 | $CsPbBr_3$/MS | Blue LED chip/$CsPbBr_3$/MS/$Sr_2Si_5N_8$:$Eu^{2+}$ | 47.6 | 72.3 | 5318 | (0.34, 0.34) | | chromaticity coordinates move from (0.34, 0.34) to (0.32, 0.32) as current increase from 20 mA to 120 mA | [135] |
| 14 | $CsPbBr_3/SiO_2$ | Blue LED chip/$CsPbBr_3/SiO_2/CaAlSiN_3$:$Eu^{2+}$ | 94 | 82 | 4448 | (0.36, 0.35) | 136% of NTSC | | [140] |
| 15 | $CsPb(Br/Cl)_3/SiO_2$; $CsPbBr_3/SiO_2$; $CsPb(Br/I)_3/SiO_2$ | UV LED chip/ mixture of $CsPb(Br/Cl)_3/SiO_2$, $CsPbBr_3/SiO_2$, and $CsPb(Br/I)_3/SiO_2$ | | | | (0.33, 0.29) | | No conspicuous change of its PL spectrum after working 8 h at 20 mA | [130] |
| 16 | $CsPbBr_3$@$SiO_2$ | Blue GaN chip/$CsPbBr_3$@$SiO_2$/$CsPbBr_{0.6}I_{2.4}$@$SiO_2$ | 65 | ~90 | 5993–6588 K | (0.32, 0.33) | | | [113] |
| 17 | $Cs_2AgInCl_6/SiO_2$ | UV LED chip/$Cs_2AgInCl_6/SiO_2$ | | | | | | | [119] |
| 18 | DDAB-$CsPbBr_3$@$SiO_2$ | Blue LED chip/DDAB-$CsPbBr_3$@$SiO_2$/DDAB-$CsPbBr_1I_2$@$SiO_2$ | | 85.3 | 5274 | (0.35, 0.35) | | The EL spectra are stable after continuous work for 20 h at 30 mA | [103] |
| 19 | PMMA-capped $CsPbBr_3/TiO_2$ | Blue LED chip/PMMA-capped $CsPbBr_3/TiO_2$/N620 phosphor | 40.5 | | | (0.33, 0.32) | | Manifests a superior stability under ambient condition with the CIE color coordinate of (0.33, 0.31) after a 210-day storage | [105] |
| 20 | $MAPbBr_3$@$SiO_2$/ PVDF | Blue LED chip/$MAPbBr_3$@$SiO_2$/ PVDF/KSF adhesive | 147.5 | | 7968 | (0.29, 0.32) | 120% of NTSC | | [129] |
| 21 | $CsPbBr_3$ QDs/$FSiO_2$ | 400 nm UV chip/ $BaMgAl_{10}O_{17}$:Eu/ $CsPbBr_3$ QDs/$FSiO_2$/(Sr,Ca)$AlSiN_3$:Eu | 10.7 | 52.9 | 5235 | (0.33, 0.33) | | | [95] |
| 22 | $CsPbBr_3$–$CsPb_2Br_5$@$SiO_2$ | Blue LED chip/$CsPbBr_3$–$CsPb_2Br_5$@$SiO_2$/$CaAlSiN_3$:$Eu^{2+}$ | 57.65 | 91 | 5334 | (0.34, 0.34) | | 87.4% of the initial luminous efficiency after 120 min of excitation | [98] |
| 23 | $CsPbBr_3$/MS@$SiO_2$ | Blue LED chip/$CsPbBr_3$/MS@$SiO_2$/KSF | 85 | | | (0.33, 0.32) | 128% of NTSC; 96% of Rec. 2020 | | [133] |
| 24 | $CsPbBr_3$/mesoporous-@$SiO_2$ | Blue LED chip/$CsPbBr_3$/mesoporous-@$SiO_2$/$K_2SiF_6$:Mn | | | 7692 | (0.30, 0.31) | 87% of Rec.2020 | | [143] |
| 25 | DDAB-$CsPbBr_3$/$SiO_2$ QDs | Blue LED chip/DDAB-$CsPbBr_3$/$SiO_2$ QDs/AgInZnS | 63.4 | 88 | 3209 | (0.41, 0.38) | | CRI slightly changes as the applied current increases from 2.5 to 3 V | [122] |

MHPs@$SiO_2$ composites, which show excellent environmental stability, high PLQYs, good spectral tunability, narrow FWHM, are ideal candidates for the application in the pc-WLEDs.[74,75] **Table 4** is the summary of the device performance for MHPs/$MO_x$ based pc-WLEDs. The representative ways can be used to realize the MHPs@$SiO_2$ based pc-WLEDs are: i) combining the blue LED chip, MHPs@$SiO_2$ with another MHP with different colors (first type),[35,79,93,130,134] ii) combining the blue LED chip, the MHPs@$SiO_2$ with other kind of luminescent materials, such as the traditional inorganic CdSe QDs or commercial red phosphors, like KSF, $Sr_2Si_5N_8:Eu^{2+}$, $CaAlSiN_3:Eu^{2+}$ (second type);[70,75,76,101,135,140] iii) achieved the pc-WLEDs by using a UV LED chip in combination with yellow and green or blue phosphor at controllable wavelength (third type).[74,131]

For the MHPs/$MO_x$ based pc-WLEDs (first type), some polymers, like PMMA, or silicone resin are usually used to encapsulate the MHPs/$MO_x$ or MHPs powders to fabricate the filters for the pc-WLEDs.[35,79,134] Herein, usually one filter with the mixture is achieved. In Sun's work, a remote configuration of pc-WLEDs was fabricated through placing a piece of quartz glass with the mixture of the green emission $CsPbBr_3/SiO_2$ and red emission $CsPb(Br/I)_3/SiO_2$ embedded in the PMMA matrix above the blue LED chip.[35] Three discrete emission peaks located at 458, 522, and 624 nm correspond to the emission from the blue LED chip, green emission $CsPbBr_3/SiO_2$, and red emission $CsPb(Br/I)_3/SiO_2$, respectively (**Figure 17a**). Benefited from the excellent protection, no anion exchange reactions happen between the $CsPbBr_3/SiO_2$ and red emission $CsPb(Br/I)_3/SiO_2$ as no intermediate emission peak is detected. The pc-WLEDs show a CIE chromaticity coordinates of (0.33, 0.33) with a high luminous efficiency of 61.2 lm $W^{-1}$. The devices also exhibit a wide color gamut, covering 120% of the NTSC color standard. And the devices also show outstanding operational stability of 227 h ($T_{1/2}$) with an initial luminance of 100 cd $m^{-2}$ (**Figure 17b**). Other work involves in adopting the green emission $CsPbBr_3$/MS with the red $CsPb(Br_{0.4}I_{0.6})_3$ in silicone resin to realize the pc-WLEDs with a CIE chromaticity coordinates of (0.24, 0.28).[134] The obtained pc-WLEDs covers up 113 % of the NTSC color standard and ca. 85% of the Rec 2020. This number is higher than that of the traditional phosphor WLEDs (86 % of the NTSC standard) and Cd QDs based WLEDs (104 % of the NTSC standard), attributing to the narrow emission of the green and red MHPs.

For the second type MHPs/$MO_x$ based pc-WLEDs, the MHPs/$MO_x$ and the other luminescent materials are usually mixed with each other in the resins[101,135,140] or the MHPs/$MO_x$ and other luminescent materials are separately mixed with polymer or resins and then put on the blue LED chip layer by layer to realize the white light.[70,75,76] Zhang's group fabricated a pc-WLED device via combining a blue-emissive GaN chip, the green emissive $CsPbBr_3$@$SiO_2$-PMMA layer, and a red-emissive CdSe-PMMA layer – **Figure 17c**.[76] The green layer is put on top of the red layer. This pc-WLED device features a luminous efficiency of 56 lm $W^{-1}$ at 5 mA with a CIE chromaticity coordinates of (0.30, 0.32). And the device also shows a wide color gamut for the backlight display, covering up to about 138% of NTSC color standard and a color rendering index (CRI) value of 63 (**Figure 17d**). On the other hand, Li's group uses the green emissive $CsPbBr_3/SiO_2$ combined the commercial red emission KSF and the blue LED chip to achieve the pc-WLED (**Figure 17e**).[101] The mixture of the $CsPbBr_3/SiO_2$ and KSF are encapsulated inside the organic polymer silica gel and then put above the blue LED chip in a non-contact configuration. The chromaticity coordinate of the obtained pc-WLED device is optimized to be (0.32, 0.30), close to the standard white emission coordinate (0.33, 0.33). And other parameters, like CRI, correlated color temperature (CCT), and luminous efficiency are measured to be 83.3, 7425 K, and 63.5 lm $W^{-1}$, respectively. Finally, both the emission peak and spectra shape from the three emission parts of the pc-WLED remain unchanged after 13 h operated at 6 mA under ambient air (25 °C, 35 –50% humidity) (**Figure 17f**). And the pc-WLED with a structure of Blue LED chip/$CsPbBr_3/SiO_2/CaAlSiN_3:Eu^{2+}$ was also fabricated through embedding the mixture of the $CaAlSiN_3:Eu^{2+}$ and $CsPbBr_3/SiO_2$ in the silicone resin.[140] The device shows high luminous efficacy of 94 lm $W^{-1}$, CRI of 82, CCT of 4418 K, CIE color coordinates of (0.36, 0.35), as well as a wide color gamut of covering 136% of the NTSC color standard.

For the third type MHPs/$MO_x$ based pc-WLEDs, the MHPs/$MO_x$ and the other kind of MHPs/$MO_x$ or MHPs or other materials are embedded together or separately inside the resin and then put them above the UV chip to get the white light.[74,130,131] For example, Liu's group used blue emissive $CsPb(Br/Cl)_3$@MSNs and yellow emissive $Cs(Pb_{0.66}/Mn_{0.34})Cl_3$@MSNs combined the UV LED chip to achieve a pc-WLED device.[131] During the fabrication process, the $CsPb(Br/Cl)_3$@MSNs and $Cs(Pb_{0.66}/Mn_{0.34})Cl_3$@MSNs are separately dispersed in silicon resin and thermally cured. And then, the yellow part is put on the upper surface of blue-based LED devices, and thermally cured again to get the final device (**Figure 17g**). The device can generate white emission with a color coordinate of (0.34, 0.36), luminous efficacy of 62.5 lm $W^{-1}$, CRI of 82, CCT of 5677 K. In addition, ca. 90% and 65% of the original intensity remains after being exposed to the ambient condition for 10 days or working for 30 min, respectively (**Figure 17h**). Other example about using the UV LED chip and the yellow emissive MHPs/$MO_x$ is that the pc-WLED device fabricated with the UV LED chip, the $CsPbBr_3$ QDs, and the $CsPbMnCl_3$@$SiO_2$ composite.[74] Moreover, the device also exhibits a luminous efficacy of 68.4 lm $W^{-1}$, CRI of 91, CCT of 3857 K.

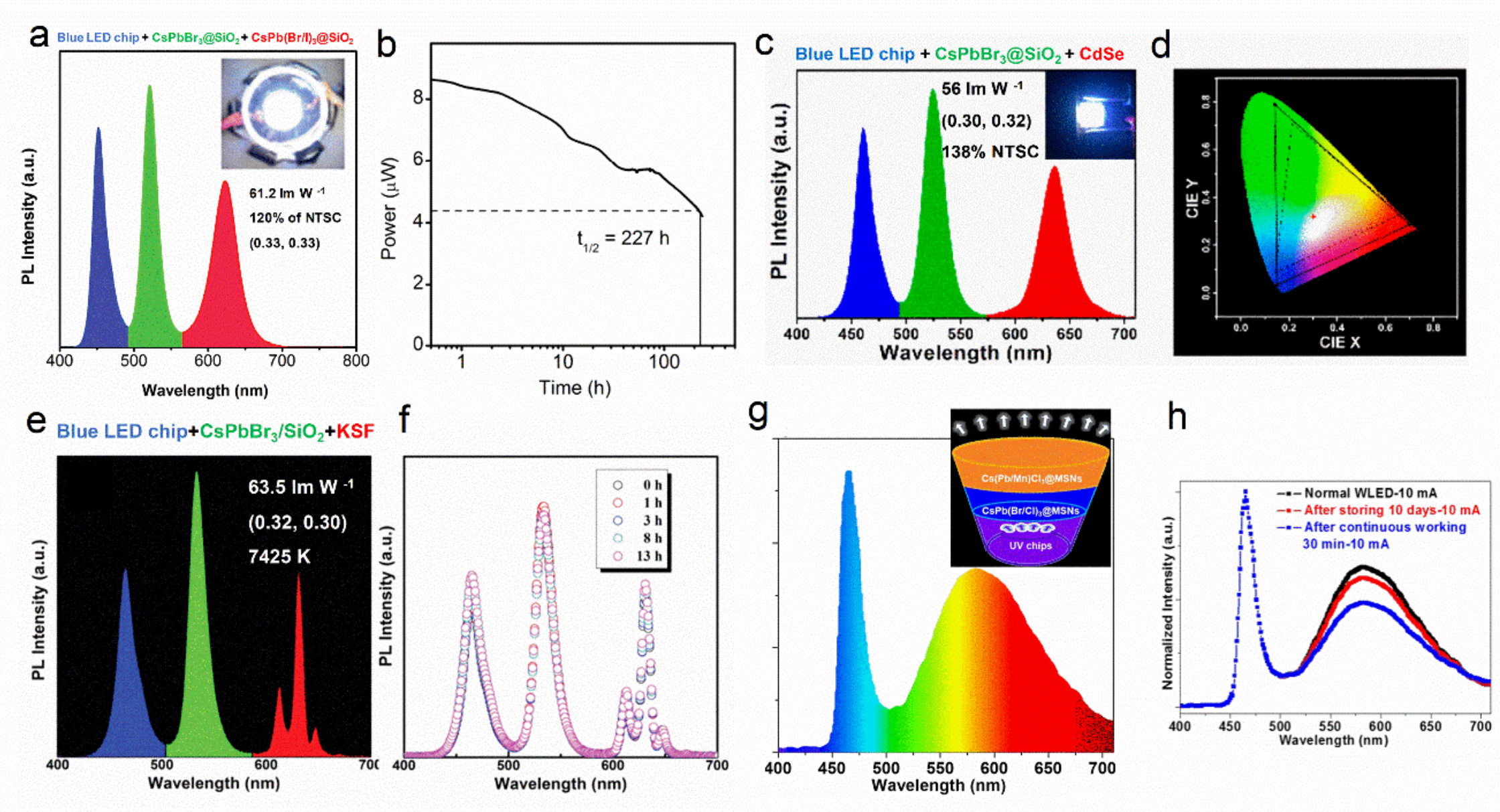

**Figure 17**. a) The emission spectra of the pc-WLED based on the blue LED chip, the $CsPbBr_3/SiO_2$, and red emission $CsPb(Br/I)_3/SiO_2$ (inset: the photograph of the device operated at 20 mA); b) Half-life of the pc-WLED with the initial luminance of 100 cd $m^{-2}$. a,b) Reproduced with permission.[35] Copyright 2016, Wiley-VCH. c) The emission spectra of pc-WLED based on the blue LED chip, the green emission $CsPbBr_3$@$SiO_2$, and red emission CdSe (inset: the photograph of the device operated at 5 mA). d) The CIE color gamut of the above pc-WLED (solid line) and NTSC standard (dashed line);[76]. c,d) Reproduced with permission. Copyright 2018, American Chemical Society. e) The emission spectra of the pc-WLED based on the blue LED chip, the $CsPbBr_3/SiO_2$, and red emission KSF operated at 6 mA. f) The spectra of the device operating at 6 mA over time under ambient air (25 °C, 35 –50% humidity). e,f) Reproduced with permission.[101] Copyright 2018, Royal Society of Chemistry. g) The emission spectra of the pc-WLED based on the UV LED chip, the $CsPb(Br/Cl)_3$@MSNs and $Cs(Pb_{0.66}/Mn_{0.34})Cl_3$@MSNs operated at 20 mA, inset is the configuration of device. h) The spectra of the PC-WLED device: the original device (black curve), the device stored under ambient condition for 10 days (red curve) and after operating

for 30 min (blue curve). g,h) Reproduced with permission.[131] Copyright 2019, American Chemical Society.

### *4.1.2. Cells Imaging*

The properties like high PLQY, narrow FWHM, low fluorescence blinking make MHPs as one of the most popular materials in the cells imaging.[66,225,226] However, the MHPs are vulnerable in water, and easy to hydrolysis in aqueous media, and the toxicity issues from the MHPs are also harmful to the cells. On the other hand, the biocompatibility is also very important for the luminescent probes to inter the cells efficiently. In this case, the $SiO_2$ coating which has good environment stability, excellent biocompatibility, nontoxicity is a good choice.

Song's group incorporated $CsPbX_3$ QDs into $SiO_2$ to fabricate the monodisperse $CsPbX_3/SiO_2$ nanocomposite.[79] The composite demonstrates enhanced water-solubility and water stability, and environmental stability ~ retention of high PLQYs for more than 1 month in air. MCF-7 cells were chosen to realize the cells imaging. High blue, green, and red emissions were achieved based on $CsPb(Cl_{0.5}/Br_{0.5})_3/SiO_2$, $CsPbBr_3/SiO_2$, and $CsPb(Br_{0.3}/I_{0.7})_3/SiO_2$ nanocomposites in the middle column of **Figure 18a**. As shown in the right column of **Figure 18a**, the confocal fluorescence images suggest that all these $CsPbX_3/SiO_2$ composites have entered the living cells. On the other hand, these $CsPbX_3/SiO_2$ particles all disperse in the middle of the cells, suggesting good biocompatibility. This is different from the other works that the MHPs all disperse in the edge of the tested cells.[66,226] In addition, a measurement to evaluate the cytotoxicity of $CsPbBr_3/SiO_2$ composite was carried out using the standard 3-(4, 5-dimethylthiazol-2-yl)-2, 5-diphenyltetrazolium bromide (MTT) assay. As presented in **Figure 18b**, the cellular viability is still as high as 98.7% after MCF-7 cells were incubated in the green $CsPbBr_3/SiO_2$ composites medium with a high concentration of 70 μg $mL^{-1}$ for 24 h, indicating that the $CsPbX_3/SiO_2$ composites are highly stable and no traced lead leakage happened during the cell incubation. Therefore, the $CsPbX_3/SiO_2$ composites are promising cellular labeling probes for imaging in live cells with high stability, excellent biocompatibility, and non-toxicity.

In Zou's work, the $CsPbBr_3@SiO_2$ composites were used as the fluorescent probes for CT26 tumor cell imaging.[108] 4',6-diamidino-2-phenylindole (DAPI) was also attached as a fluorescent DNA stain. In the confocal images in **Figure 18c,** DAPI (series 2) and $CsPbBr_3@SiO_2$ (series 3) exhibit blue and green emission, respectively. Notably, the emissions from DAPI and $CsPbBr_3@SiO_2$ mostly disperse at the nucleus regions and cytoplasmic regions, respectively. This indicates that the $CsPbBr_3@SiO_2$ nanoparticles have successfully passed through the cell membranes. Furthermore, the emission of $CsPbBr_3@SiO_2$ is still highly visible in the merged images with DAPI (**Figure 18c**, series 4), suggesting the feasibility of detecting multiple simultaneous labels. In addition, highly emission from $CsPbBr_3@SiO_2$ can still be observed even after 24 h in the live cells (**Figure 18c**). Moreover, the cytotoxicity tests of different concentration of $CsPbBr_3@SiO_2$@human serum albumin (HSA) were conducted using a standard CCK-8 assay with CT26 tumor cells. The viabilities of treated cells are all very high, ranging from 96.5% to 109.4% compared to the untreated cells after 3 h of cytotoxicity tests (**Figure 18d**). These results suggest that the ignorable decomposition of the $CsPbBr_3$ NCs, furthers proves the excellent protection of the $SiO_2$ shell.

Benefiting from the excellent biocompatibility and efficient protection of the $SiO_2$ shells, the cytotoxicities of different concentration of $CsPbBr_3@SiO_2$ particles were also tested in the HeLa cells using a standard CCK-8 assay. No obvious difference in the cell viability can be detected after 24 and 48 h compared with the untreated cells (**Figure 18e**), indicating that the obtained $CsPbBr_3@SiO_2$ particles have low toxicity for the HeLa cells. The confocal images in **Figure 18f** proves that the $CsPbBr_3@SiO_2$ nanocrystals were internalized by the HeLa cells and mostly disperse in cytoplasm part (**Figure 18f**).

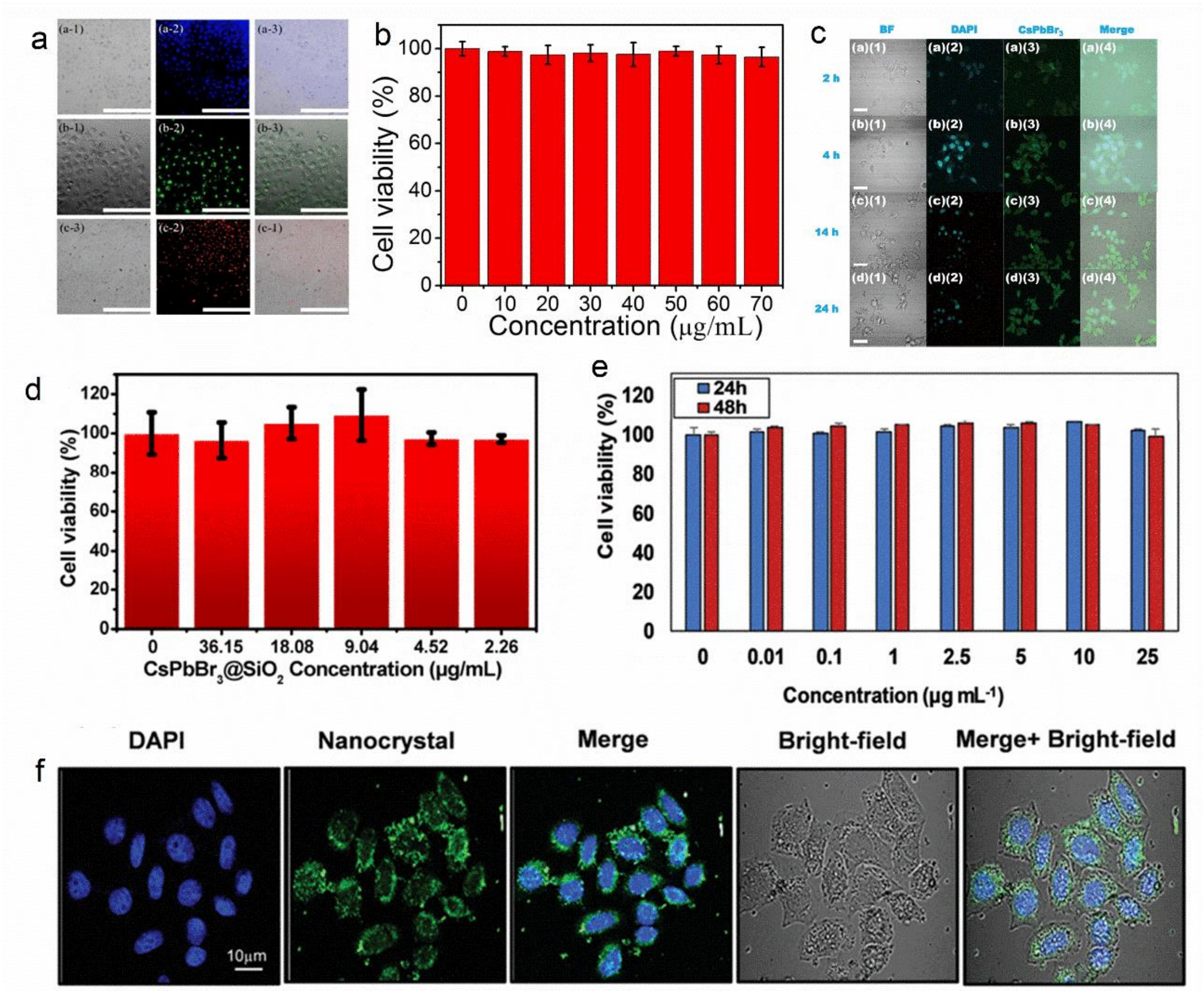

**Figure 18**. a) Fluorescence microscope images of MCF-7 cells incubated with $CsPb(Cl_{0.5}/Br_{0.5})_3/SiO_2$ (series a), $CsPbBr_3/SiO_2$ (series b), and $CsPb(Br_{0.3}/I_{0.7})_3/SiO_2$ (series c). The left side are the bright - field images, in the middle are the luminescence images, the right side are the overlaid pictures of the left images and the middle images, all scale bars are 30 μm; b) The viability of MCF-7 cells under various $CsPbBr_3/SiO_2$ concentrations. a,b) Reproduced with permission.[79]. Copyright 2018, IOP Publishing. c) Confocal images of CT26 tumor cells incubated with $CsPbBr_3@SiO_2$. The concentration of $CsPbBr_3$ NCs is fixed at 18.08 μg/mL. d) The cell cytotoxicity tests of a standard CCK-8 assay with CT26 tumor cells incubated with different concentration of $CsPbBr_3@SiO_2@HSA$ for 3 h. c,d) Reproduced with permission.[108] Copyright 2020, Springer Nature. e) The cell cytotoxicity tests of a standard CCK-8 assay with Hela cells incubated with different concentration of $CsPbBr_3@SiO_2$ for different time. f) Confocal images of Hela cells incubated with $CsPbBr_3@SiO_2$. Reproduced with permission.[110] Copyright 2020, Royal Society of Chemistry.

*4.1.3. Lasers*

Lasers are devices which can emit coherent and spatial light with narrow FWHM through the optical amplification process. And lasers widely exist in both industry production and daily life.[227-229] MHPs, which always show tunable emission, low-threshold, strong multi photon absorption, and ultra-stable stimulated emission, have been used as the gain media to form laser devices. For example, the room-temperature optical amplification based on the $CsPbBr_3$ has been realized over the range from 440 to 700 nm with a low pump threshold of about 5 μJ $cm^{-2}$ and high values of modal net gain of about 500 $cm^{-1}$.[227] However, the further development of this perspective is always impeded by the severe stability of the MHPs, from both the chemical and optical degradations. Meanwhile, previous works have proved that coating with transparent metal oxides, such as $SiO_2$, is an effective strategy to improve the stability of MHPs as well as keep the excellent optical properties of the MHPs.[175]

$CsPbBr_3$ was embedded into a $SiO_2$ sphere to act as a gain media for frequency up-conversion.[80] Enhanced ASE from $CsPbBr_3/SiO_2$ under two-photon (800 nm femtosecond laser pulses, 1 kHz, 100 fs) excitation at ambient environment was observed (**Figure 19a**). The $CsPbBr_3/SiO_2$ (230.8 μJ $cm^{-2}$) shows more than two times lower threshold intensity than that of the pristine $CsPbBr_3$ NCs (480.8 μJ $cm^{-2}$). For both $CsPbBr_3$ and $CsPbBr_3/SiO_2$, the spontaneous emission centered at 534 nm with a FWHM of 24 nm. After reaching the threshold pump intensity, both devices show dramatically increased ASE and a sharp redshifted peak at 537 nm with a FWHM of 5 nm. And a liner relation shows between the output intensity and the pump intensity under 800 nm excitation after reaching the threshold intensity (**Figure 19b**). Further comparison proves that the slope of the $CsPbBr_3/SiO_2$ based line is much higher than that of the $CsPbBr_3$ based line. Therefore, these results mean that the ASE threshold decreases significantly and the efficiency improves dramatically after the $CsPbBr_3$ is coated with the $SiO_2$ shell. As expected, the $CsPbBr_3/SiO_2$ based laser device shows excellent stability (**Figure 19c**). The device can keep about 95% of its initial emission after 12 h (up to $4.32 \times 10^7$ excitation cycles) under the pump intensity of 600 μJ $cm^{-2}$ excitation using 800 nm femtosecond laser. However, the pristine $CsPbBr_3$ based devices losses 15% of its emission under the same condition. What is more, the $CsPbBr_3/SiO_2$ composite keeps stable ASE after being exposed to air for 2 months (the inset in **Figure 19c**). All the above results prove that the $SiO_2$ coating can significantly improve the properties while does not impair the emission performance of the $CsPbBr_3$ NCs. A $CsPbBr_3/SiO_2$ composite structure was also fabricated by the same group using the analogously strategy and incorporated into a cylindrical microtubule to generate pure whispering-gallery mode (WGM) lasing.[112] The interface reflection induced by the cylindrical microtubule will confirm large amount of photons to work as WGM (**Figure 19d**). Only one broad emission located at ~ 530 nm with a FWHM of 18 nm is observed when the pump intensity is low (less than 434 μJ/cm$^2$), this is attributed to the spontaneous emission. Several peaks with narrow FWHM of ~ 1 nm appear under high pump intensity (**Figure 19e**). **Figure 19f** shows the device has a clear low pump threshold (the evolution from spontaneous emission to lasing) of ~430 μJ/cm$^2$, $P_{th}$. After the $CsPbBr_3$ QDs were embedded into the $SiO_2$, the laser medium changes from air (n~1) to $SiO_2$ (n~1.5) for the $CsPbBr_3$ QDs. The higher n is better for reaching population inversion and therefore low pump threshold.

Qu's group embedded $CsPbBr_3$ NCs in the dual MS with gold nanocore (DMSP) to act as two-photon-pumped plasmonic nano lasers.[175] DMSP hexane solution was loaded in cuvettes and pumped by an 800 nm femtosecond pulses,1 kHz, 100 fs. As shown in **Figure 19g**, a narrow emission peak located at λ = 535 nm appears when the pumping energy reaches the threshold pump intensity (1.08 mJ cm$^{-2}$). But spontaneous emissions are simultaneously enhanced with the increasing pump power, because part of the $CsPbBr_3$ NCs disperse in the outmost of the MS layer and are not involved in the stimulated emission mode. The insert in **Figure 19g** shows a typical characteristic of lasers - input-output curve with a threshold. On the other hand, the PL emission shift from 524 nm to 511 nm when the concentration of the $CsPbBr_3$ precursor (for the synthesis of the DMSP) decrease from 0.25 mol L$^{-1}$ to 0.025 mol L$^{-1}$ (**Figure 19h**). Meanwhile, the corresponding lasing spectra show a blue shift from 535 nm to 529 nm, because the plasmon resonance of the gold nanocore contributes to the selective PL enhancement (**Figure 19i**). This indicates that the highly tunable two-photon pumped nano laser depends on both the surface plasmon of the gold nanocore and the $CsPbBr_3$ NCs, providing more opportunities for designing new kinds of two-photon pumped nano lasers.

Recently, a water-resistant $CsPbBr_3$ PQDs@$SiO_2$ nanodots (wr-PNDs) was developed and used in a two-photon WGM laser (fabricated by introducing the nanodots into a capillary-like whispery gallery microcavity) and a two-photon random laser device (using the powder directly).[106] **Figure 19j** is the setup used to collect the ASE and lasing response under 800 nm fs laser excitation (Libra Coherent, 1 kHz, 50 fs). Firstly, the water stability of the devices based on the wr-PNDs and the pristine $CsPbBr_3$ PQDs were investigated. When both of the devices are immersed in water and measured under 375 nm laser excitation, 80% of the initial PLQY can be retained for the wr-PNDs based device after 13 h, while less than 10% for the pristine $CsPbBr_3$ PQDs based device after 3 h (**Figure 19k**). For the WGM laser performance, only broad emission peak can be detected when low pumping intensities are applied (less than 0.91 mJ $cm^{-2}$), whereas sharp peaks (with a line width of 0.3–0.5 nm) decorated on a narrowed ASE peak emerged under higher pumping intensity (**Figure 19l**). Herein, the wr-PNDs based device exhibits a low lasing threshold of ~1.12 mJ $cm^{-2}$. Then, a two-photon random laser device can be operated by collecting the wr-PNDs powder that had been dispersed in water for 15 days. Here, the powder can directly work as the random laser.[230] Under low-power intensity (less than 0.85 mJ $cm^{-2}$), broad emission at ~ 525 nm is detected (**Figure 19m**). Then, different new peaks located at different wavelength gradually appear under increasing pump power (**Figure 19m**). This work extends the application of the PQDs from solid-state laser to random laser in an aqueous medium.

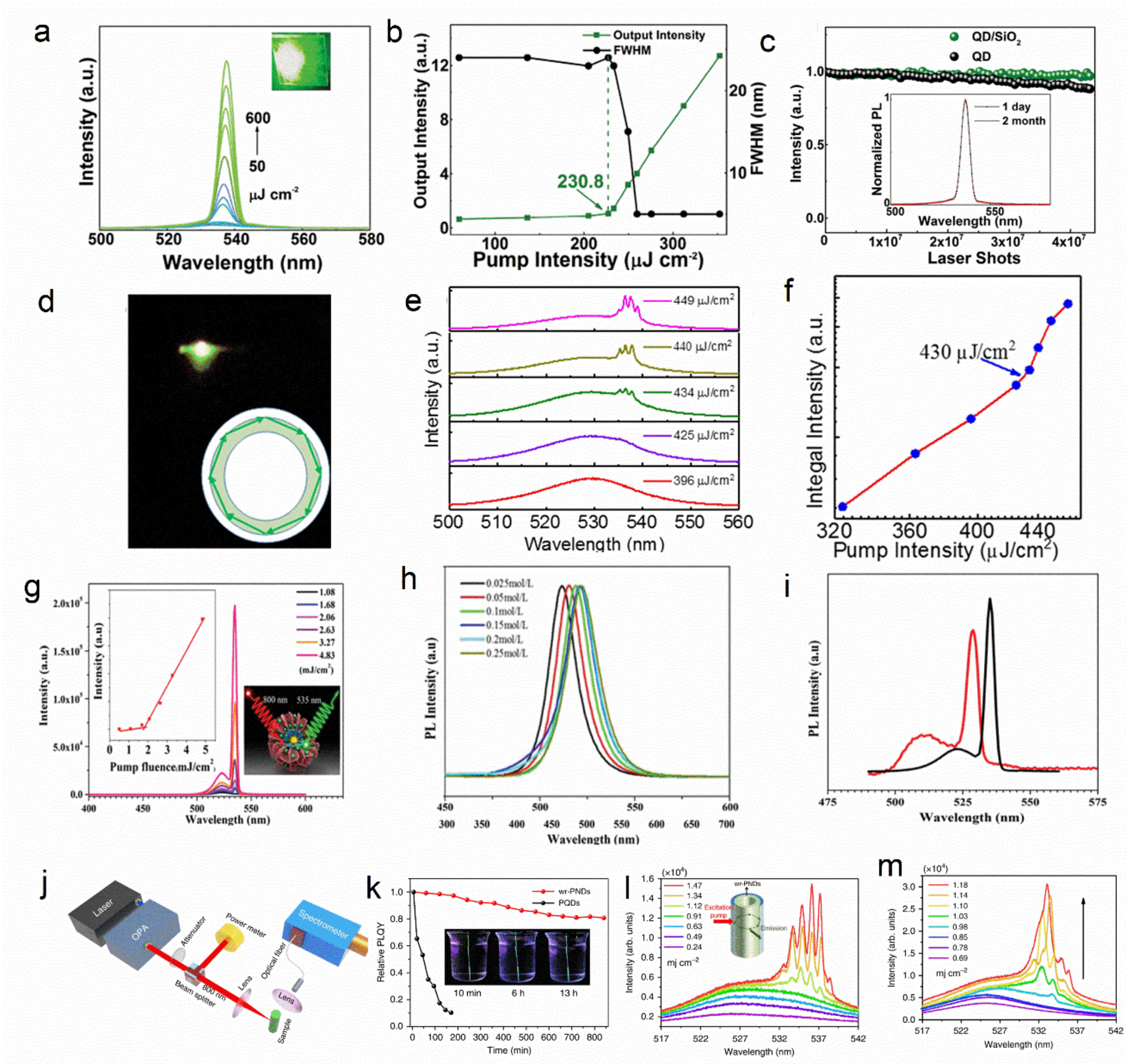

**Figure 19**. a) The relation between ASE pump-intensity and the light emission for $CsPbBr_3/SiO_2$ composite under two-photon (800 nm femtosecond laser pulses, 1 kHz, 100 fs) excitation, the inset is a photograph obtained above the ASE threshold; b) The output intensity (green) and FWHM (black) vs. pump energy density for $CsPbBr_3/SiO_2$ based lasers; c) Plot of ASE intensity measured under continuous pulsed 800 nm femtosecond laser, 600 μJ $cm^{-2}$, 12 h, 4.32 × 107 excitation cycles, the inset is the comparison of the ASE spectra between working for 1 d and 2 months. a,b,c) Reproduced with permission.[80] Copyright 2018, Wiley-VCH. d) Lasing image of the $CsPbBr_3/SiO_2$ QDs in a cylindrical microtubule based device; e) The spectra obtained from increasing pump intensity; f) The integral emission intensity vs. the increasing pump intensity curve. d,e,f) Reproduced with permission.[112] Copyright 2019, The Optical Society. g) The light emission spectra and the ASE pump intensity of DMSP hexane solution under an 800 nm

femtosecond pulses,1 kHz, 100 fs, the inset is the integrated lasing intensity vs. pump intensity; h) The spontaneous emission and i) the corresponding lasing spectra move with the adjustment of the concentration of the $CsPbBr_3$ precursor in DMSP solution. g,h,i) Reproduced with permission.[175] Copyright 2018, Royal Society of Chemistry. j) The setup for two-photon lasing measurements; k) The change in the PLQYs of the laser devices; l) The WGM lasing performance of the wr-PNDs-based devices under increasing pumping intensity after immersed in water for 13 h; m) The two-photon random lasing performance of the wr-PNDs-based devices under increasing pumping intensity after immersed in water for 15 days. j,k,l,m) Reproduced with permission.[106] Copyright 2020, Nature Publishing Group.

#### *4.1.4. Anti-Counterfeiting Codes*

Anti-Counterfeiting is very important for both the communities and the individuals, as more and more criminal activities related with the counterfeiting threat to the economy and the security.[231,232] Some aspects which are closely related to our daily life, like tickets, certificates, banknotes, have a high risk of being replicated. Materials with fluorescence are very promising to defend this problem, because of their multimodal emission properties.[233] $CsPbBr_3$ exhibits high PLQY under UV-light and has two-photon induced up-conversion PL (UC-PL) property;[234,235] meanwhile, it is also high sensitive to temperature, the emission decreases as the temperature increases and no emission can be observed when heated to a certain temperature.[204,220,236] Therefore, $CsPbBr_3$ shows triple-modal functions for the application of the anti-counterfeiting. However, the pristine $CsPbBr_3$ usually exhibits a relatively low stability towards light and heating along with the PL quenching caused by the humidity, this will limit the real application.

Zeng's group successfully fabricated the $CsPbBr_3@Cs_4PbBr_6/SiO_2$ composites with triple-modal (UV, IR, and thermal) fluorescent anti-counterfeiting properties.[81] **Figure 20a** is a small logo that is used to simulate how the triple-modal anti-counterfeiting codes work. The logo consists of a Chinese character "中" and a English character "0", written using the $CsPbBr_3@Cs_4PbBr_6/SiO_2$ based inks ($CsPbBr_3@Cs_4PbBr_6/SiO_2$ powders dispersed homogeneously in polydimethylsiloxane, PDMS), the number 100 is a traditional anti-counterfeiting patterns written using the $Cs_4PbBr_6$-based inks ($Cs_4PbBr_6$ powders dispersed homogeneously in PDMS). And all the outlines from these codes were marked with dotted lines. $CsPbBr_3@Cs_4PbBr_6/SiO_2$-based characters "中" and "0" are all invisible under daylight. Clear green characters with high brightness are observed under 365 nm UV lamp at RT (PL mode), and yellow emission can be seen from the number "100". On the other hand, green characters "中" and "0" can still be observed (UC-PL mode), while the number "100" disappear under a 800 nm femtosecond pulse laser. What is more, these characters can be recovered to their initial states after moving the UV and IR light (I and II). Moreover, $CsPbBr_3@Cs_4PbBr_6/SiO_2$ based inks show high thermal sensitivity, the emission quenches to none when exposed to 150 °C and strong emission can be recovered again after cooling down to RT under both UV light (III) and IR light (III′). However, no obvious changes can be observed for the commercial "100" label under the same condition. Therefore, the $CsPbBr_3@Cs_4PbBr_6/SiO_2$ shows excellent triple-modal anti-counterfeiting property and this property cannot be easily replicated. Moreover, this property still works well after being exposed to the ambient environment for 2 months, means great potential for the practical application in advanced anti-counterfeiting.

In a recent work published by Zhang's group, reversible transformation between luminescent $CsPbBr_3$ and non-luminescent $CsPb_2Br_5$ is achieved through exposing/removing moisture.[126] In the work, $CsPbBr_3$ was embedded into the MSNs through the template method (**Figure 4f**). As illustrated in **Figure 20b**, the green emission cannot be recovered after pristine $CsPbBr_3$ films (from spin-coating) is treated with moisture and then dry in air for only 1 time. However, the green emission can be recovered through switching on/off the $CsPbBr_3$/MSNs films (from spin-coating) by treated with moisture and dry in air (**Figure 20c**). Similar with the pristine $CsPbBr_3$ film, the yellowish $CsPbBr_3$/MSNs film become grayish along with the losing emission under UV light when treated by moisture. And then the non-emission grayish films recover to yellowish film and the green emission also appears when drying in the air. Moreover, the optical properties of the $CsPbBr_3$/MSNs film, such as the PL intensity, peak location, FWHM, do not change even after 10 cycles of the on/off process (**Figure 20d**). It should be noted that the ligand-free surface of $CsPbBr_3$ NCs plays an important role for the reversible switching between fluorescence and non-fluorescence. As a comparison, the green emission of the $CsPbBr_3$ NCs that was prepared through hot-injection quickly disappear and could not recover (**Figure 20e**). For the $CsPbBr_3$/MSNs, the structure of the fluorescent $CsPbBr_3$ breaks and the non-fluorescent $CsPb_2Br_5$ forms upon touching water because of high solubility of CsBr in water. The $CsPb_2Br_5$ NCs will react with CsBr to form $CsPbBr_3$ when the moisture was removed from the system. This is because the dissolved CsBr is confined within the space provided by MSNs matrix. However, the hydrophobic ligands, like OA and oleylamine prevent the dissolved CsBr touching the non-fluorescent $CsPb_2Br_5$ for the hot-injecting prepared $CsPbBr_3$ NCs. Furthermore, the $CsPbBr_3$ NCs@MSNs powder was directly put inside the empty laser printer toner. Then, different printed patterns with high resolution and the characters 'FUNSOM' with different thickness can be obtained (**Figure 20f**). And the information encryption and decryption processes can be realized through moisture treatment and evaporation, respectively. The printed pattern can be hidden by the moisture treatment and recovered by evaporation (information) (**Figure 20g**). In addition, bright emission can be retained after 10 times erasing-recovering cycles (**Figure 20h**).

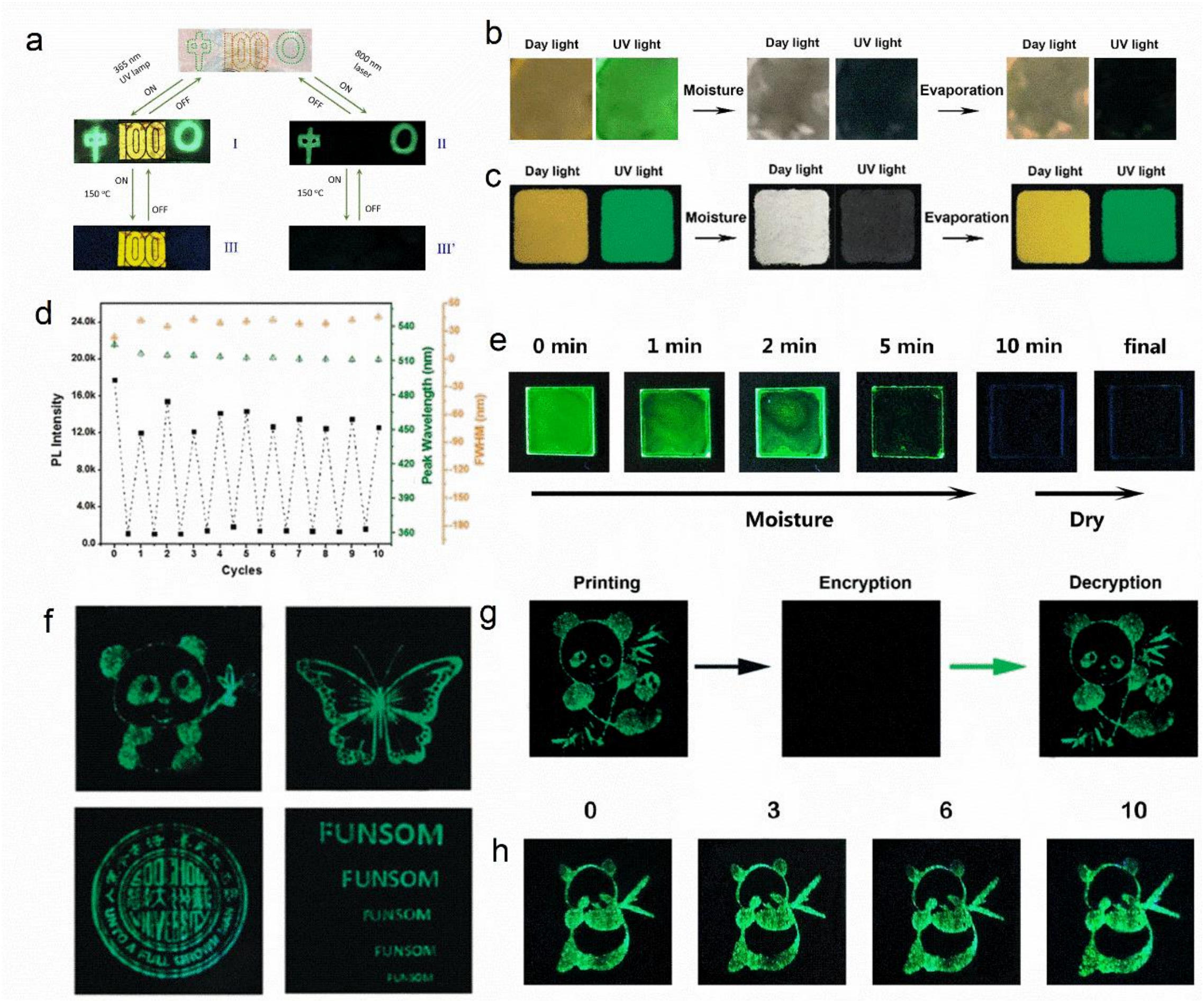

**Figure 20**. Graph shows how $CsPbBr_3@Cs_4PbBr_6/SiO_2$ works as the triple-mode anti-counterfeiting codes: excited with (turned ON) or without (turned OFF) (I) a 365 nm UV lamp and (II) a 800 nm femtosecond pulse laser; (III) and (III′) treated at 150 °C (ON) and RT (OFF) and excitated with a 365 nm UV lamp and a 800 nm laser, respectively. Reproduced with permission.[81] Copyright 2017, American Chemical Society. b) The pristine $CsPbBr_3$ film and c) the $CsPbBr_3$ NCs@MSNs film under moisture-evaporation treatment; d) The change in the PL intensity, the peak location, and the FWHM of $CsPbBr_3$ NCs@MSNs during 10 cycles moisture-

evaporation test; e) The change of the $CsPbBr_3$ (prepared through hot-injection) film during the moisture-dry process under UV-light; f) Different patterns obtained by laser-jet printing using $CsPbBr_3$ NCs@MSNs under UV-light; g) The encryption and decryption processes of the printed panda pattern in one cycle; h) The printed panda under different testing cycles. b,c,d,e,f,g,h) Reproduced with permission.[126] Copyright 2020, Wiley-VCH.

*4.1.5. Electrochemiluminescent (ECL)*

Si nanocrystals, which can store both positive and negative charges under electrochemical conditions, were firstly investigated in the ECL technology.[237] However, the ECL efficiency is still limited. Therefore, MHPs with excellent physic-optical properties, especially the high PLQYs, can be recognized as a promising candidate to achieve highly efficient ECL. Significant progresses about using MHPs in ECL have been realized in air-free dichloromethane[238] and aqueous medium.[239] But the long term using of ECL is still big behind. On the other hand, the efficiency of MHPs based ECL still needs to be improved.

In Zhu's work, a novel CPB-CoR@$SiO_2$ nanostructure was fabricated. Herein, the CoR and CPB are encapsulated inside the $SiO_2$ shell. As a reference, the CPB@$SiO_2$- - -CoR were also prepared through physical mixing the obtained CPB@$SiO_2$ with the coreactants. Typically, the working mechanism (**Figure 21a**) for CPB-2-DBAE@$SiO_2$ NCs based ECL follows these equations:

$$\text{CBP−DBAE@SiO}_2 - 2\,e^- \rightarrow \text{CBP}^+\text{−DBAE}^{+\bullet}\text{@SiO}_2 \quad (10)$$

Both CPB and DBAE are electrooxidized to $CPB^+$ and $DBAE^{+\bullet}$;

$$\text{CBP}^+\text{−DBAE}^{+\bullet}\text{@SiO}_2 \rightarrow \text{CBP}^+\text{−DBAE}^{\bullet}\text{@SiO}_2 + H^+ \quad (11)$$

$DBAE^{+\bullet}$ losses protonation to produce a highly reductive radical $DBAE^{\bullet}$;

$$\text{CBP}^+\text{−DBAE}^{\bullet}\text{@SiO}_2 \rightarrow \text{CBP*−DBAE@SiO}_2 \quad (12)$$

$$\text{CBP*−DBAE@SiO}_2 \rightarrow \text{CBP−DBAE@SiO}_2 + h\nu\ (\text{ECL}) \quad (13)$$

$DBAE^{\bullet}$ and $CPB^+$ radiatively recombine with each other and then the generation of the ECL.

As excepted, almost all these CPB-CoR@$SiO_2$ show significantly enhanced ECL performance than that of the CPB@$SiO_2$- - -CoR (**Figure 21b**). And CPB-2,2′-(butylimino) diethanol (BIDE)@$SiO_2$ shows the highest ECL intensity among all these CPB-CoR@$SiO_2$ based ECL. These results show that the indispensable role of introducing coreactants for emitting higher ECL. And the decrease in the CPB-tripropylamine (TPA)@$SiO_2$ based ECL is attributed to the lack of functional groups to incorporate TPA into the $SiO_2$ matrix. And the two -OH groups in the BIDE favor the easy incorporation of the BIDE to the $SiO_2$ matrix and therefore the excellent ECL intensity. The work further shows that the CPB-CoR@$SiO_2$ exhibits outstanding ECL stability under 10 cycles test, compared to the decreasing ECL intensity for the CPB---DBAE under the same condition (**Figure 21c**). The high stability owes to the efficient $SiO_2$ coating.

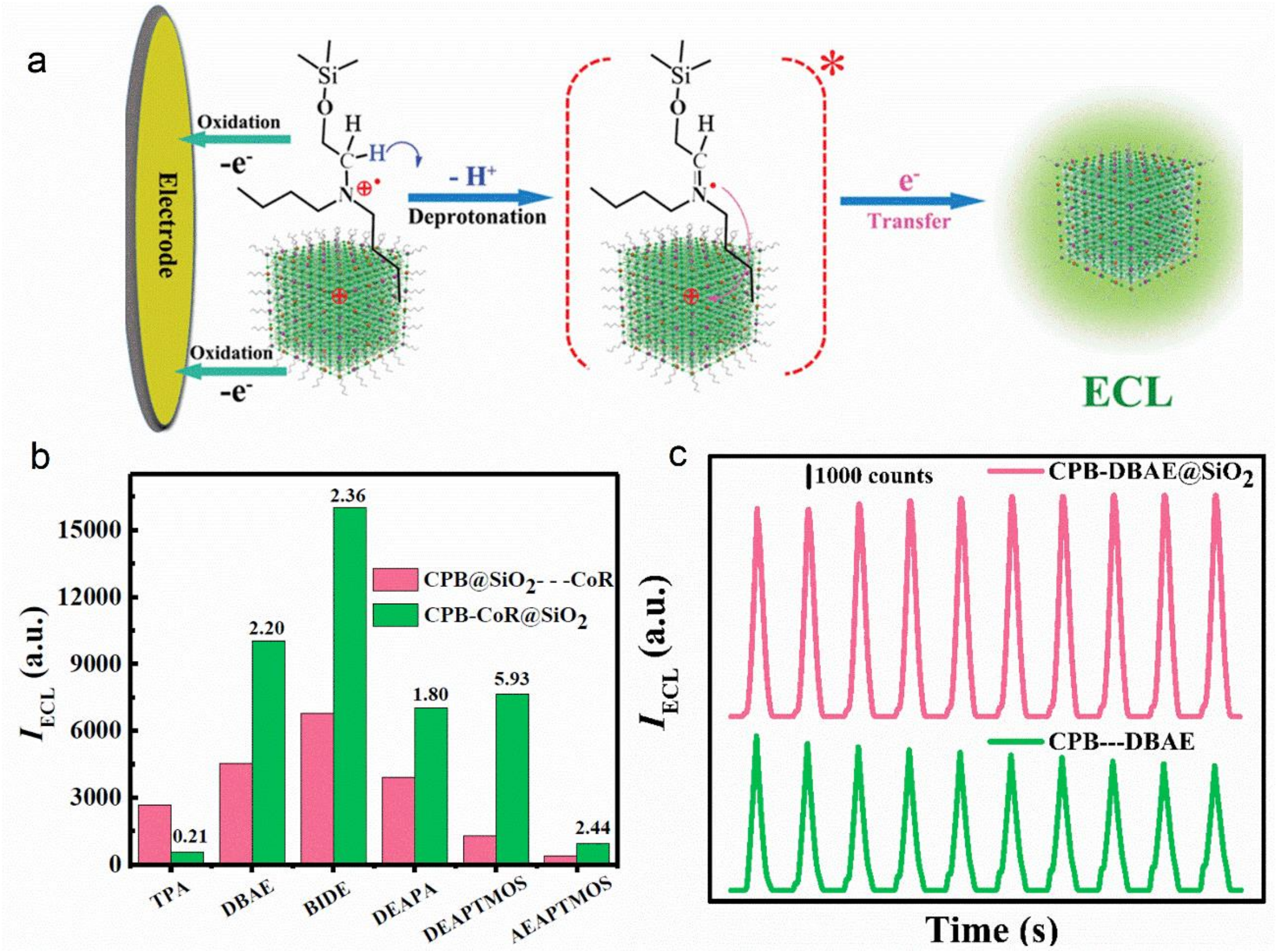

**Figure 21**. a) The working mechanism for CPB-DBAE@$SiO_2$ NCs based ECL; b) comparison of ECL intensity between CPB-CoR@$SiO_2$ and CPB@$SiO_2$---CoR systems; c) comparison of ECL stability of CPB-DBAE@$SiO_2$ and CPB---DBAE. a,b,c) Reproduced with permission.[102] Copyright 2019, Wiley-VCH.

### *4.1.6. Drug Delivery*

Drug Delivery is a process that the pharmaceutical compounds are delivered by a carrier to safely arrive the desired body site to achieve the therapeutic effect. Nanoparticles are usually recognized as good candidate for drug delivery, because the enhanced efficiency and the ignorable side effects.[240] $CsPbBr_3$@$SiO_2$ core–shell were used as the carrier to deliver the Doxorubicin (Dox) to the HeLa cells.[110] The tests were performed in a phosphatebuffered saline (PBS, pH is 7.4) solution. The process of the drug delivery is shown in **Figure 22a**. The drugs are absorbed on the surface of $CsPbBr_3@SiO_2$ through the electrostatic interaction or the pores of the $SiO_2$ shell. Then, the drugs will be delivered by the $CsPbBr_3@SiO_2$ into the cells through endocytosis and release in the cells. The drugs can be totally released in 14 days as shown in the cumulative release profile of the drug in **Figure 22b.** This slower release will increase the bioavailability of the drug and therefore improve the therapeutic effect. The endocytosis of Dox-loaded $CsPbBr_3@SiO_2$ was proved by the confocal imaging (**Figure 22c**). The bright red dots in the cytoplasm region and the homogeneous red emission in the nucleus region belong to the aggregated $CsPbBr_3@SiO_2$ and the free drugs, respectively.

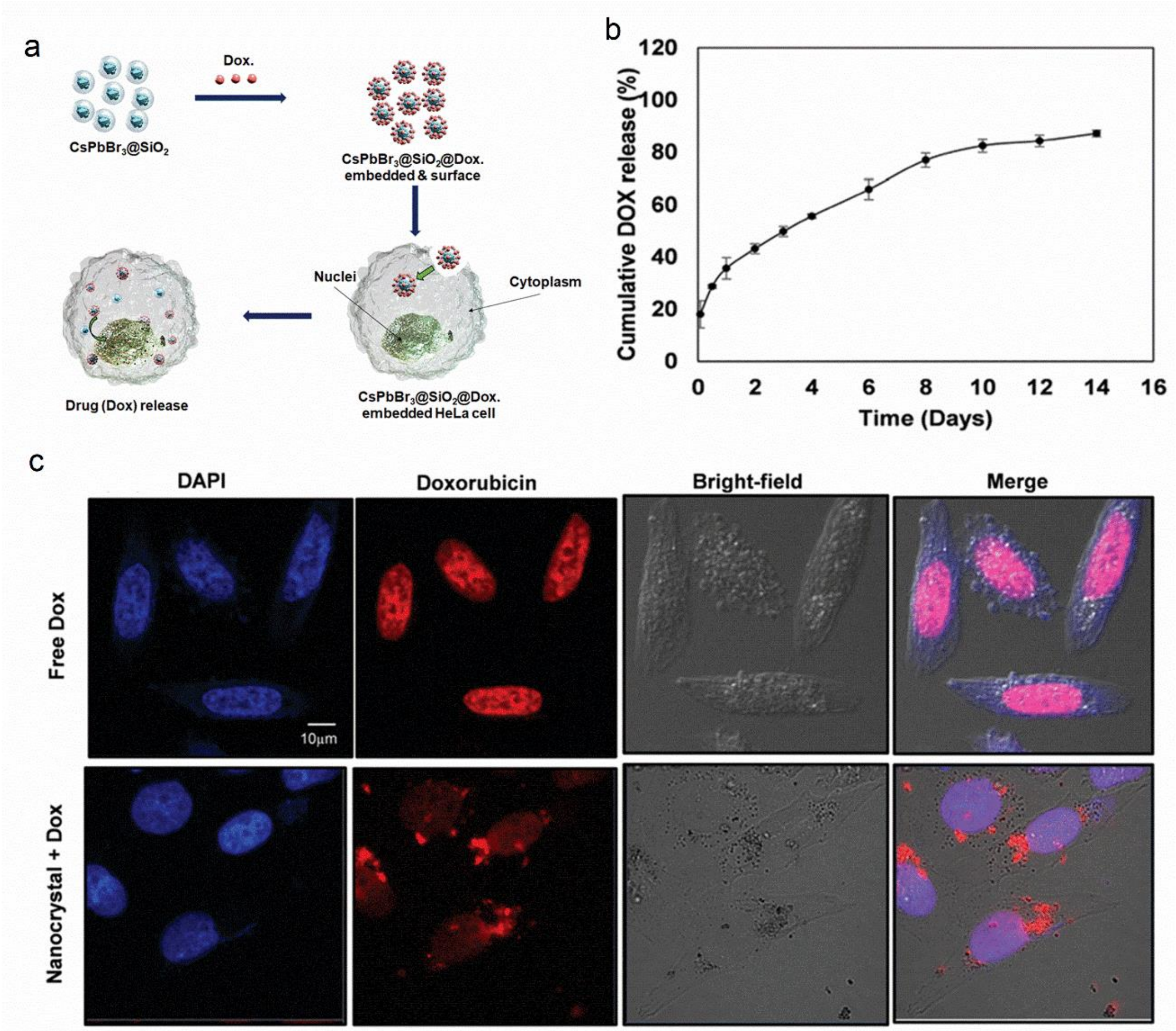

**Figure 22**. a) The proposed mechanism for the drug delivery, b) The cumulative release profile of the drug as the increasing days; c) The confocal imaging of HeLa cells after the drug delivery. a,b,c) Reproduced with permission.[110] Copyright 2020, Royal Society of Chemistry.

## 4.2. MHPs@TiO2 for Photoelectrochemistry

With these advantages of low-cost facile processing, high extinction coefficient, wide light absorption range, tunable band gap, superior charge transport properties, MHPs have obtained impressive promising in the photocatalytic reactions, like HI splitting,[45,241] $CO_2$ reduction,[44,87,242] selective oxidation of benzylic alcohol,[243] photocatalytic degradation of organic compounds,[40] PEC reactions.[88,137,244] Nevertheless, the low efficiency and the poor stability are still far from satisfaction. Therefore, searching for a strategy for both improving the stability and the efficiency is especially urgent. $TiO_2$ is usually selected as an efficient electron transportation layer in the MHPs based solar cells because of the relative band alignment of photogenerated charge carriers within MHPs and $TiO_2$, as well as the good UV-responsive property of $TiO_2$.[245,246] Meanwhile, similar with other $MO_x$, like $AlO_x$, $SiO_2$, the compacted $TiO_2$ shell also has high chemical stability, which can passivate and protect the vulnerable MHPs core, and therefore contributes to the long-term stability.

*4.2.1. $CsPbBr_3$ NCs/amorphous-$TiO_2$ (a-$TiO_2$) for Photocatalytic Reduction of $CO_2$*

In Kuang's work,[87] $CsPbBr_3$ NCs/a-$TiO_2$ was successfully fabricated and applied for the photocatalytic reduction of $CO_2$. As shown in **Figure 23a**, the photoactivity of the $CO_2$ reduction is significantly improved because of the introduced a-$TiO_2$. And during the photocatalytic period of 3 h, the amount of the consumed photoelectrons for $CsPbBr_3$ NC is about 25.72 µmol $g^{-1}$, while $CsPbBr_3$ NC/a-$TiO_2$ samples show 2 ~ 6.5 enhancements, corresponding to about 79.25 ~ 193.36 µmol $g^{-1}$. Moreover, $CsPbBr_3$ NC/a-$TiO_2$ shows higher photocatalytic $CO_2$ reduction selectivity than the pristine $CsPbBr_3$ NC. And the formation of $CH_4$ contributes to most of the improvement in activity, because the formation of $CH_4$ is more thermodynamically favorable than that of CO and $H_2$ as 8 electrons are needed. In addition, the $CsPbBr_3$ NC/a- $TiO_2$ also shows satisfied photocatalytic stability. For the $CsPbBr_3$ NC/a-$TiO_2$ (20), 91.5% of the initial photoactivity can be retained during the five test cycles (**Figure 23b**). These results demonstrate that the deposition of the a-$TiO_2$ contributes to the highly stable performance in the solvent under constant illumination and stirring, as well as the improved photoactivity.

*4.2.2. $CsPbBr_3$/$TiO_2$ for PEC Reaction*

Besides, $CsPbBr_3$/$TiO_2$ is also a good candidate for PEC reaction in real water testing.[84,88] The conduction band (CB) and valence band (VB) edges for the $CsPbBr_3$ are $\approx -1.0$ and $\approx 1.5$ V (vs. normal hydrogen electrode, NHE), respectively.[247] And the CB and VB edges for anatase $TiO_2$ are ≈−0.4 and ≈2.8 V vs. NHE, respectively.[248] Therefore, a type II band alignment lies in the $CsPbBr_3/TiO_2$ heterostructure. And the photogenerated holes will be confined in the VB of the $CsPbBr_3$ core under illumination, while the photogenerated electrons will concentrate in the CB of the $TiO_2$ shell and therefore facilitate the charge transfers.

In Zheng'work,[88] the PEC measurement was performed in a three-electrode setup in $Na_2SO_4$ solution (0.1 M, pH = 6.8), Pt was selected as the counter electrode, Ag/AgCl (3.0 M KCl) as the reference electrode. Ar was introduced prior to and during each measurement. And the light source is a 405 nm LED. As shown in **Figure 23c**, the assembled $CsPbBr_3/TiO_2$ demonstrates about 2 times higher photocurrent than the bare $CsPbBr_3$ NC at −0.1 V vs. NHE. Besides, the long-term measurement shows that the photocurrent of $CsPbBr_3/TiO_2$ composite can maintain at a constant in a period of ca. 8 h test, suggesting outstanding stability in the real water PEC measurement (**Figure 23d**). The minor increase is caused by the dark current effect, proved by the light on/off measurement after 6 h, in which higher dark current is observed compared with the initial dark current.

Then, the dynamics of hot carrier and charge carrier of $CsPbBr_3$ NCs with different $MO_x$, like $TiO_2$, $SnO_2$, $SiO_2$ were also compared in Kuang's work through PEC reaction.[84] The tests were performed in the tetrabutylammonium hexafluorophosphate ($TBAPF_6$)-dichloromethane solution (0.1 M), using a xenon lamp as the light source with a 420 nm filter. As shown in **Figure 23e**, the pristine $CsPbBr_3$ NC exhibits a photocurrent density of 21 μA $cm^{-2}$, while $CsPbBr_3$ NC/$TiO_2$ and $CsPbBr_{3-x}Cl_x$ NC/$SnO_2$ show higher value of 41 and 52 μA $cm^{-2}$, respectively. However, the photocurrent density of $CsPbBr_3$ NC/$SiO_2$ declines to 17 μA $cm^{-2}$. Herein, the enhancement is attributed to the efficient separation of the charge carriers and the high electron mobility in $CsPbBr_3$ NC/$TiO_2$ and $CsPbBr_{3-x}Cl_x$ NC/$SnO_2$. However, coating with insulating $SiO_2$ accelerates radiative recombination and the hot carrier relaxation, and therefore the photocurrent density is inhibited.

#### *4.2.3. $CsPbBr_{1.5}I_{1.5}/TiO_2$ Inverse Opal Electrode for PEC Sensing of Dopamine*

MHPs show great promise in the PEC sensing because of their long effective diffusion lengths, unique carrier separation properties. But their low formation energy limits their application in the aqueous solution. On the other hand, $TiO_2$ has also been recognized as a good candidate in PEC detection. However, $TiO_2$ based PEC detector or sensors can only harvest the UV light (usually no more than 420 nm), which will destroy some biomolecules.[249]

Taking the advantages of both MHPs and $TiO_2$, a visible light triggered $CsPbBr_{1.5}I_{1.5}/TiO_2$ inverse opal electrode was fabricated.[137] The pristine $TiO_2$ electrode shows low and negligible IPCE values below 430 nm and above 430 nm, respectively. But other electrodes all show significantly enhanced IPCE values above 430 nm (**Figure 23f**). In addition, the $CsPbBr_{1.5}I_{1.5}/TiO_2$/Nafion electrode shows excellent stability in the photocurrent under 600 nm irradiation, without decreasing for 200 min (**Figure 23g**). However, the $CsPbBr_{1.5}I_{1.5}$/Nafion electrode shows sharp decline under the same condition, indicating the excellent protection provided by of the $TiO_2$. Meanwhile, the $CsPbBr_{1.5}I_{1.5}/TiO_2$/Nafion electrode also shows high sensitivity and low detection limitation (from 0.1 μM) on the detection of the dopamine, the photocurrent increases linearly as the increasing concentration of the dopamine (**Figure 23h**). A regression equation: $I = 62.17 + 1.41C$ (I: photocurrent, C: the concentration of dopamine) with $R=0.997$ is achieved. More importantly, the $CsPbBr_{1.5}I_{1.5}/TiO_2$/Nafion electrode still shows high selectivity for DA detection as other interfering substances are introduced (**Figure 23i**).

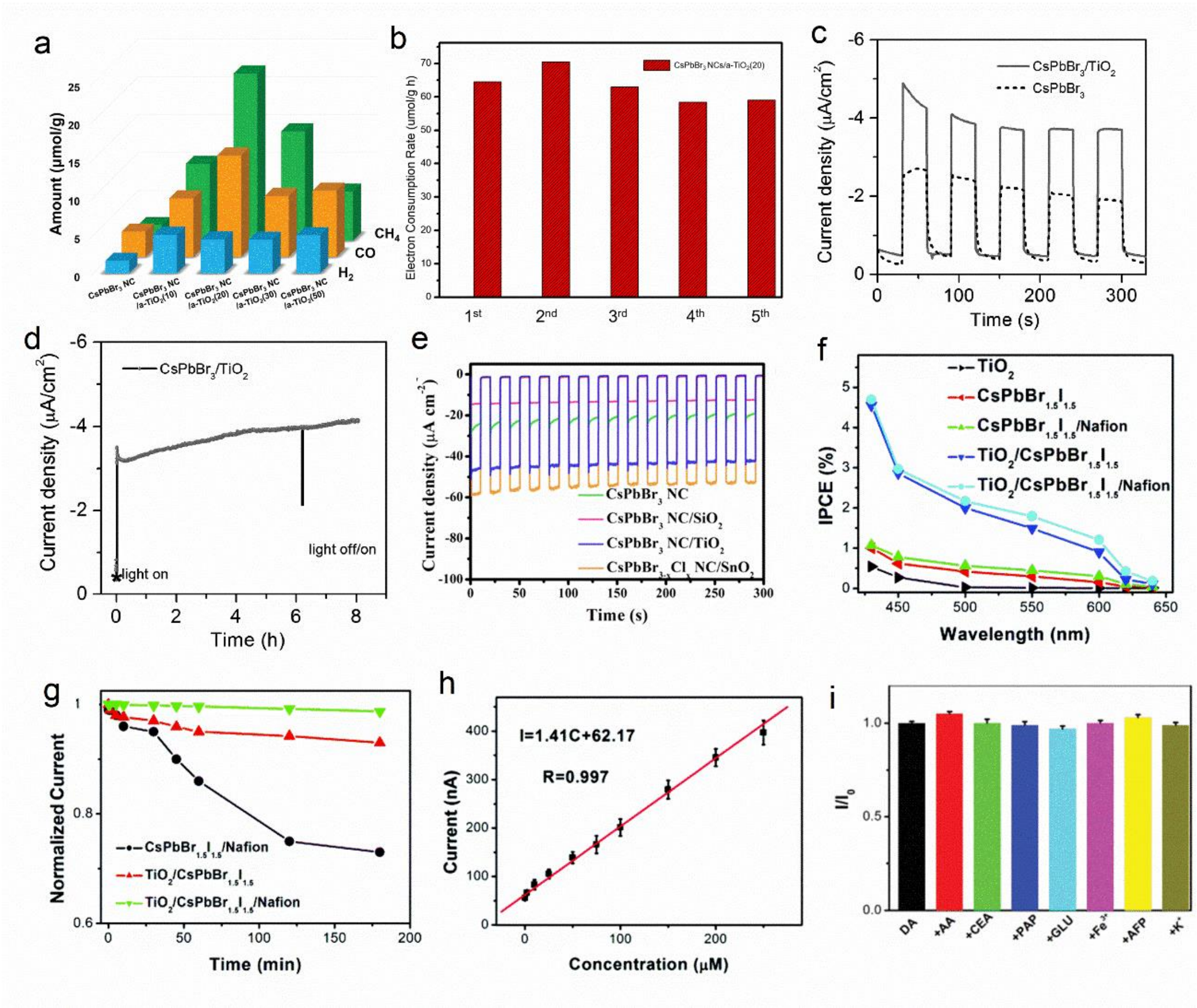

**Figure 23.** a) Photocatalytic $CO_2$ reduction of $CsPbBr_3$ NCs/a-$TiO_2$ and the pristine $CsPbBr_3$ NCs, b) The recycling photocatalytic test for 5 cycles using $CsPbBr_3$ NCs/a-$TiO_2$ (20). a,b) Reproduced with permission.[87] Copyright 2018, Wiley-VCH. c) The on-off photocurrent measurement for the $CsPbBr_3/TiO_2$ and the bare $CsPbBr_3$ at − 0.1 V vs. NHE in 0.1 M $Na_2SO_4$ aqueous solution; d) The long-term photocurrent measurement of $CsPbBr_3/TiO_2$ in neutral water for 8 h. c,d) Reproduced with permission.[88] Copyright 2018, Wiley-VCH. e) The amperometric I−t curves of $CsPbBr_3$ NC, $CsPbBr_3$ NC/$TiO_2$ and $CsPbBr_{3-x}Cl_x$ NC/$SnO_2$ and $CsPbBr_3$ NC/$SiO_2$ electrodes. Reproduced with permission.[84] Copyright 2018, American Chemical Society. f) The IPCE values of different electrodes; g) Time-dependent photocurrent of the electrodes under 600 nm irradiation; h) The curve of the current vs. dopamine concentration; i) The value of $I/I_0$ for the detection of dopamine in the presence of various interfering species. f,g,h,i) Reproduced with permission.[137] Copyright 2018, Royal Society of Chemistry.

### 4.3. $CsPbX_3$@$ZrO_2$

### *4.3.1. $CsPbBr_3/ZrO_2$ as a Green Filter for pc-WLEDs*

The excellent photophysical properties, like high PLQYs, narrow FWHM, highly tunable emission spectra, have made $CsPbBr_3$ as the most popular candidate for LEDs. However, challenges like unsatisfactory efficiency and poor stability are still needed to be addressed. Using semiconductor to fabricate a heterostructure has been proved to be a good way to improve the stability and photophysical properties. This is because the band energy and the charge transfer process can be easily tuned between semiconductors with different CB and VB. The type I heterostructure, in which the CB and VB edges of semiconductor A are higher and lower than the corresponding bands of semiconductor B, is favorable for the carrier confinement in the narrower band gap semiconductor and therefore enhance the radiative recombination.[72,250,251] Meanwhile, the semiconductor can also passivate the surface of MHPs.

Based on the CB and VB location of $CsPbBr_3$ and $ZrO_2$ in **Figure 24a**, a type I $CsPbBr_3/ZrO_2$ composite forms. The photogenerated carriers will confine in the $CsPbBr_3$ because of the energy level difference between $CsPbBr_3$ and $ZrO_2$, and this will favor higher radiative recombination on the $CsPbBr_3$. Taking advantages of these properties, a pc-WLED was fabricated through stacking up layer by layer ~ a blue-emitting GaN chip, the green-emitting $CsPbBr_3/ZrO_2$–PMMA composite, and a red-emitting CdSe–PMMA composite (**Figure 24b**). The corresponding PL spectrum is presented in **Figure 24c**. The $CsPbBr_3/ZrO_2$ based pc-WLED shows a luminous efficiency of 55 lm $W^{-1}$ at 1 mA compared to 30 lm $W^{-1}$ for $CsPbBr_3$ based pc-WLED under the same current. The CIE color coordinate of the $CsPbBr_3/ZrO_2$ based WLED moves to (0.27, 0.30) after operating at 1 mA for 2 h, which is still in the white emission area and close to the standard white emission coordinate (0.33, 0.33) (**Figure 24d)**.

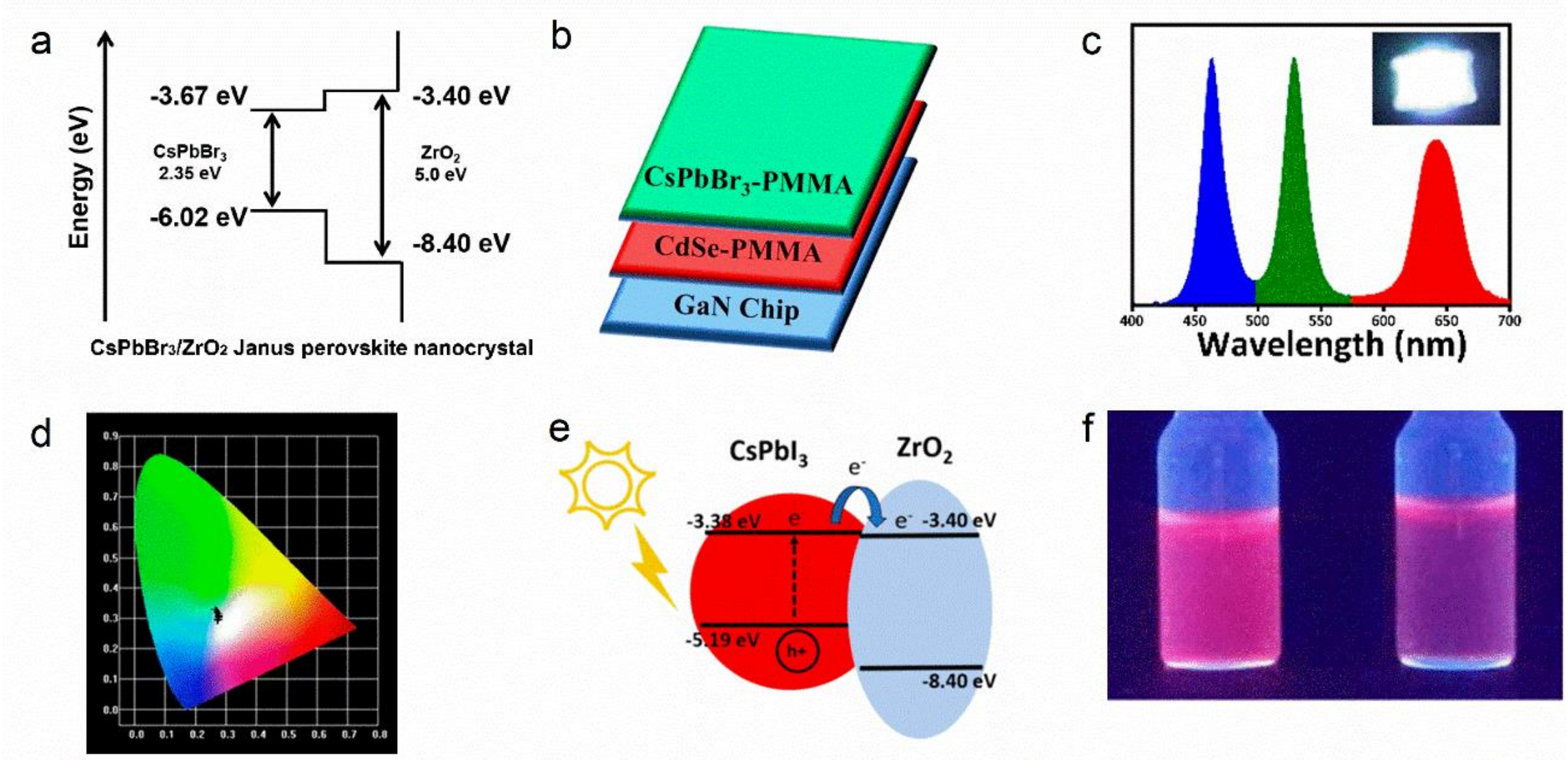

**Figure 24**. a) The energy band alignment of the $CsPbBr_3/ZrO_2$ composite; b) The device structure of the pc-WLEDs; c) The corresponding PL spectra of $CsPbBr_3/ZrO_2$ composite based pc-WLED operated at 1 mA, the inset is a photograph of the working pc-WLED; d) The CIE movement graph of the $CsPbBr_3/ZrO_2$ composite based pc-WLED; e) The energy band alignment of the $CsPbI_3/ZrO_2$ composite; f) The emission of $CsPbI_3/ZrO_2$ solution (right) and the $CsPbI_3$ solution (left) under UV-light. Reproduced with permission.[70] Copyright 2019, American Chemical Society.

*4.3.2. $CsPbI_3/ZrO_2$ as a Potential Photocatalyst for the Photocatalytic $CO_2$ Reduction*

In the band structures of $CsPbI_3/ZrO_2$, $CsPbI_3$ possesses higher CB and lower VB edges than those of $ZrO_2$, therefore a type II heterostructure forms. Photoinduced electrons will accumulate on $ZrO_2$ and holes will transfer to $CsPbI_3$, resulting in the efficient separation of photoinduced carriers (**Figure 24e**). In **Figure 24f**, the $CsPbI_3/ZrO_2$ solution shows lower emission than the $CsPbI_3$ solution. Therefore, the $CsPbI_3/ZrO_2$ composite can be used as a potential photocatalyst for the visible photocatalytic $CO_2$ reduction.

### 4.4. $CsPbBr_3@SiO_2/Al_2O_3$ for Efficient and Stable Green LEDs

The instability of MHPs is still the bottleneck for further practical applications because of the attacks from the surrounding environment, such as the humidity, light, temperature, and oxygen, These attacks will induce the degradation of the MHPs and therefore the decreased PL properties. Normally, single coating is not dense enough and there are still many pinholes, water and oxygen can go inside and cause the degradation. On the other hand, the $SiO_2/Al_2O_3$ binary coating has been proved to be able to provide better protection than individual coating for Austenitic Stainless

Steel[252] and graphite.[253] Because the combination of $SiO_2$ and $Al_2O_3$ can form a much compact surface than the single coating.

Li's group successfully fabricated the $CsPbBr_3$ QDs-SAM and these powders were sealed in PDMS (a mixture of PDMS A and PDMS B).[69] The mixture was dropped on the blue LED chip (peak at 455 nm) after removing the bubbles, and therefore a green LED with the green emissive $CsPbBr_3$ QDs-SAM and a blue LED chip was obtained (**Figure 25a**). The $CsPbBr_3$ QDs-SAM based green LED has a high luminous efficacy of 80.77 lm $W^{-1}$, which is more than 2 times higher than that of the bare $CsPbBr_3$ based green LED (36.15 lm $W^{-1}$). Then, the operational stability was measured under 5 mA and 2.7 V at ambient environment. The $CsPbBr_3$ QDs-SAM shows significantly enhanced stability, 90% of the initial PL emission can be maintained after working for 96 h, while the PL intensities for the $CsPbBr_3/SiO_2$ and the pristine $CsPbBr_3$ drastically decrease to 45 % and 15% after only 12 h, respectively (**Figure 25b**). And the operational stability can be further improved using the matrix with better light resistance, like polyvinylidene fluoride (PVDF).

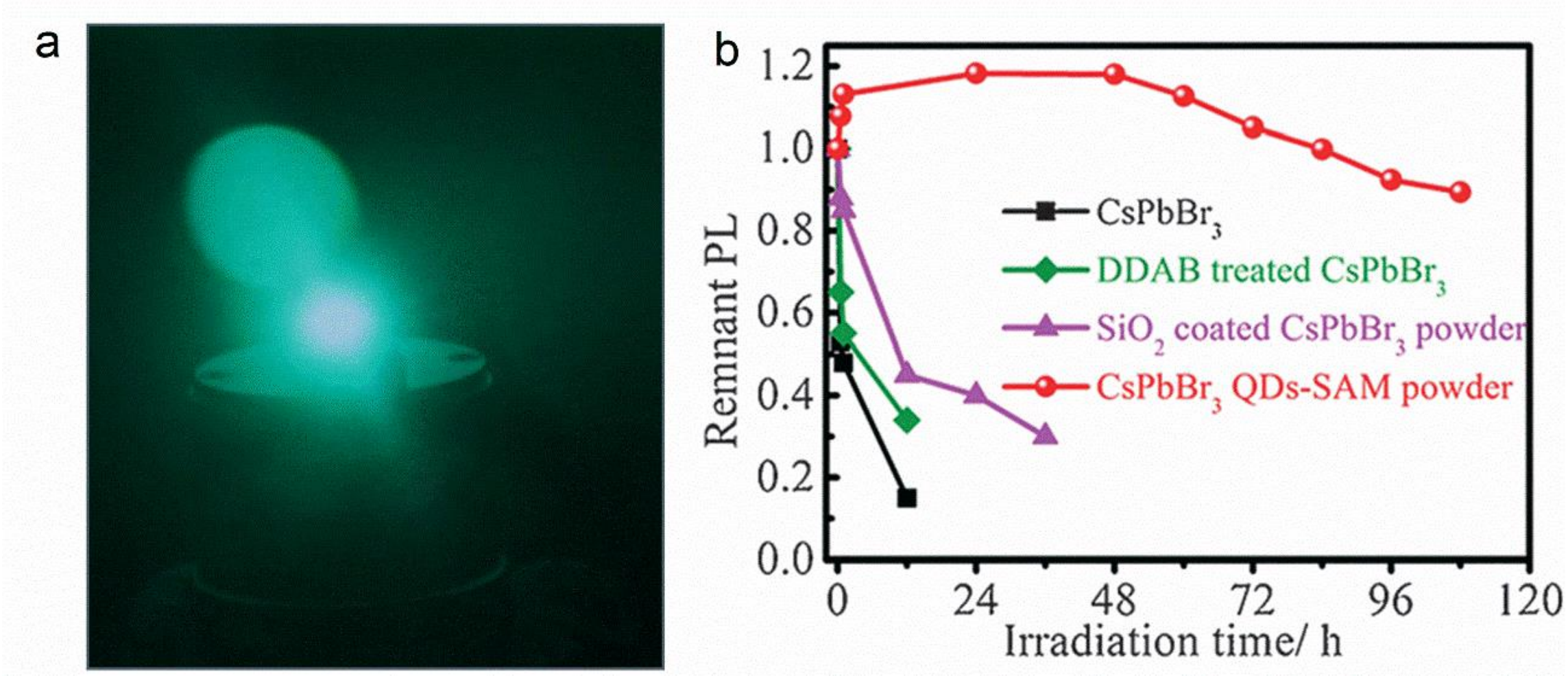

**Figure 25**. a) The working green LED fabricated with the $CsPbBr_3$ QDs-SAM and a blue LED chip; b) The comparison of the stability on a blue LED chip (5 mA, 2.7 V). Reproduced with permission.[69] Copyright 2017, Wiley-VCH.

**4.5. $CsPbBr_3@Al_2O_3$ for Waterproof Laser**

Single-crystalline MHPs nanoplates (NPs) based WGM nanolasers have been widely explored, with the purpose of developing devices with better performances.[254] However, using these devices in the biochemical-compatible environments, like polar solvents, is still limited by the fact that MHPs are vulnerable in these solvents. Therefore, a $Al_2O_3$ layer was deposited through ALD technique to protect $CsPbBr_3$ NPs against the polar solvents.[4] Firstly, the devices are tested in 8 glycerine-$H_2O$ solutions with different refractive indices under the same pump fluence of 49 μJ $cm^{-2}$. Distinct lasing can be detected in all eight solutions (**Figure 26a**). The laser behaviors change from single-mode to dual mode and then to single mode again as the increasing refractive index. This indicates that the laser has a good selectivity towards the refractive index of the solution, favors the application in the communication field. The pump intensity dependent light emission intensity (L−L curves) measurements were further performed in solutions with different refractive indices. Lasing thresholds increase as the increasing refractive indices in both long-wavelength mode and short-wavelength mode (**Figure 26b, c**). Then, the lasing thresholds were extracted from the L-L curves and plotted in **Figure 26d**, and the sensitivities for the long wavelength mode and short wavelength mode are calculated to be 129.7 μJ $cm^{-2}$ $RIU^{-1}$ and 388.2 μJ $cm^{-2}$ $RIU^{-1}$, respectively. Meanwhile, the sensor also shows excellent stability, proved by the repeated measurements (**Figure 26d**).

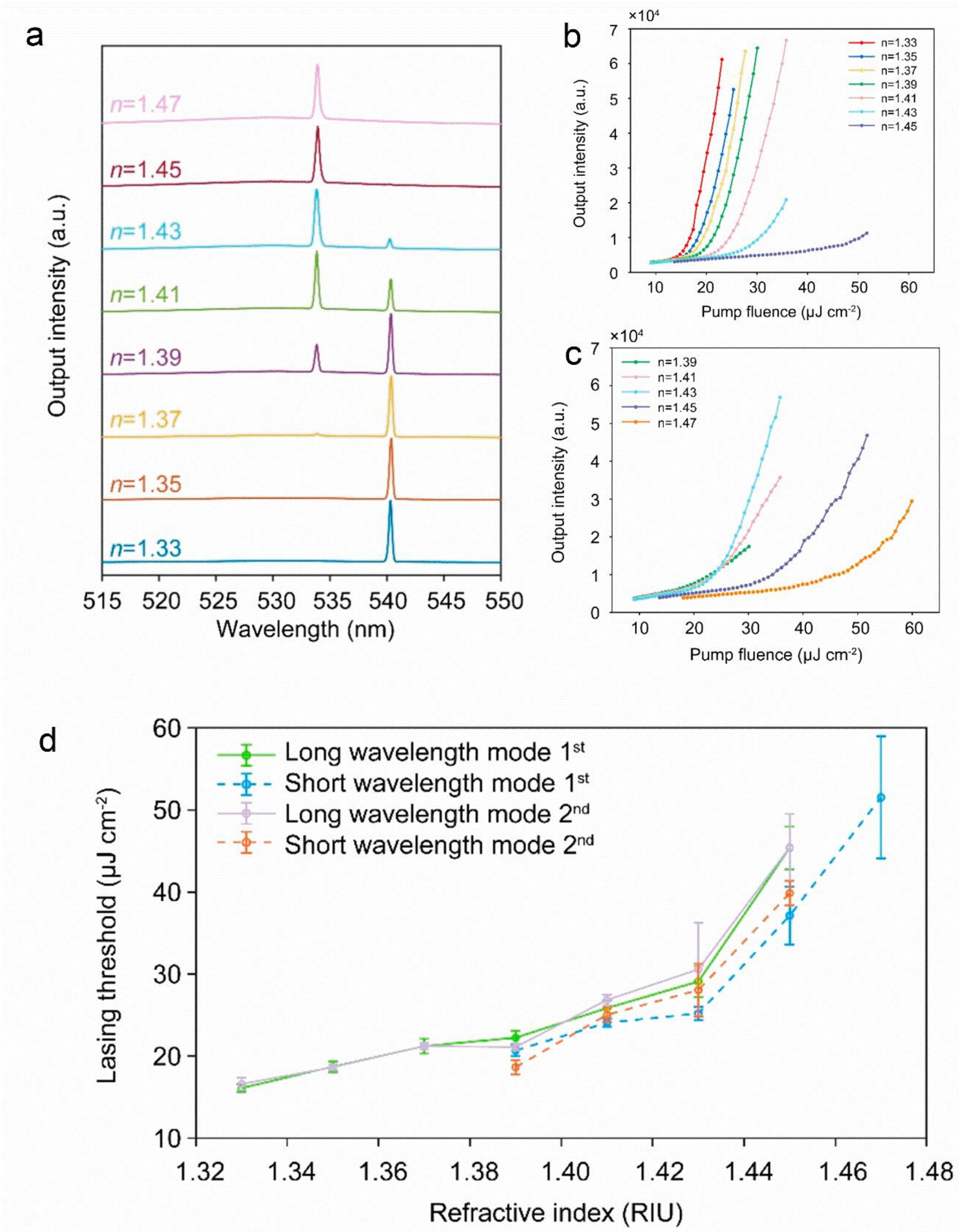

**Figure 26**. a) The normalized lasing spectra in 8 glycerine-$H_2O$ solutions; L−L curves of the b) long-wavelength mode and c) short-wavelength mode; d) Lasing thresholds extracted from the L−L curves in b) and c). Reproduced with permission.[4] Copyright 2020, American Chemical Society.

### 4.6. $CsPbBr_3@SiO_2/ZrO_2$ for Efficient and Stable Green LEDs

Binary $MO_x$ coating shows better performance in enhancing the stability of the MHPs that of the single coating.[69] Meanwhile, most of the current MHPs related works focus on enhancing the stability and efficiency of the MHPs and the MHPs based devices.[37,166,169,255] Up to now, the degradation mechanism of the MHPs based devices is seldom investigated. Therefore, more attention should be attached to understand the deactivation process of the devices.

In our work, the first binary $SiO_2/ZrO_2$ coated $CsPbBr_3$ architecture was successfully fabricated.[89] The green corresponding LED devices were prepared by covering the commercial blue LED chip with the $CsPbBr_3@SiO_2/ZrO_2$/PMMA filters in a remote configuration. The devices operated under 100, 50, 15, and 10 mA show a lifetime (the time to reach 50% of the initial emission intensity) of 188, 289, 535, and 680 h, respectively. At the beginning, all these devices exhibit enhanced brightness because of the photoactivation effect. Interesting, the changes in the brightness during the decay process follow a 3 - step process under the ambient conditions for all these currents (**Figure 27a, b, c, d**). Then, freshly prepared devices were measured in the glovebox at 100 and 10 mA to investigate the deactivation process. However, these devices measured under the inert condition exhibit a quick exponential deactivation behavior (**Figure 27e, f**).

Based on the previous work[123,206,213] and the above results, a deactivation mechanism is proposed. Firstly, the photoinduced desorption of the surface ligands leads to the formation of surface trapping defects and therefore the initial emission deactivation. Then, the presence of moisture under the ambient condition will passivate surface trap - states and promotes the plateau stage. And this stage is the key to realize the excellent operational stability. Finally, severe surface degradation will occur after being continuously exposed to the moisture and constant blue excitation, and this will cause an emission deactivation.

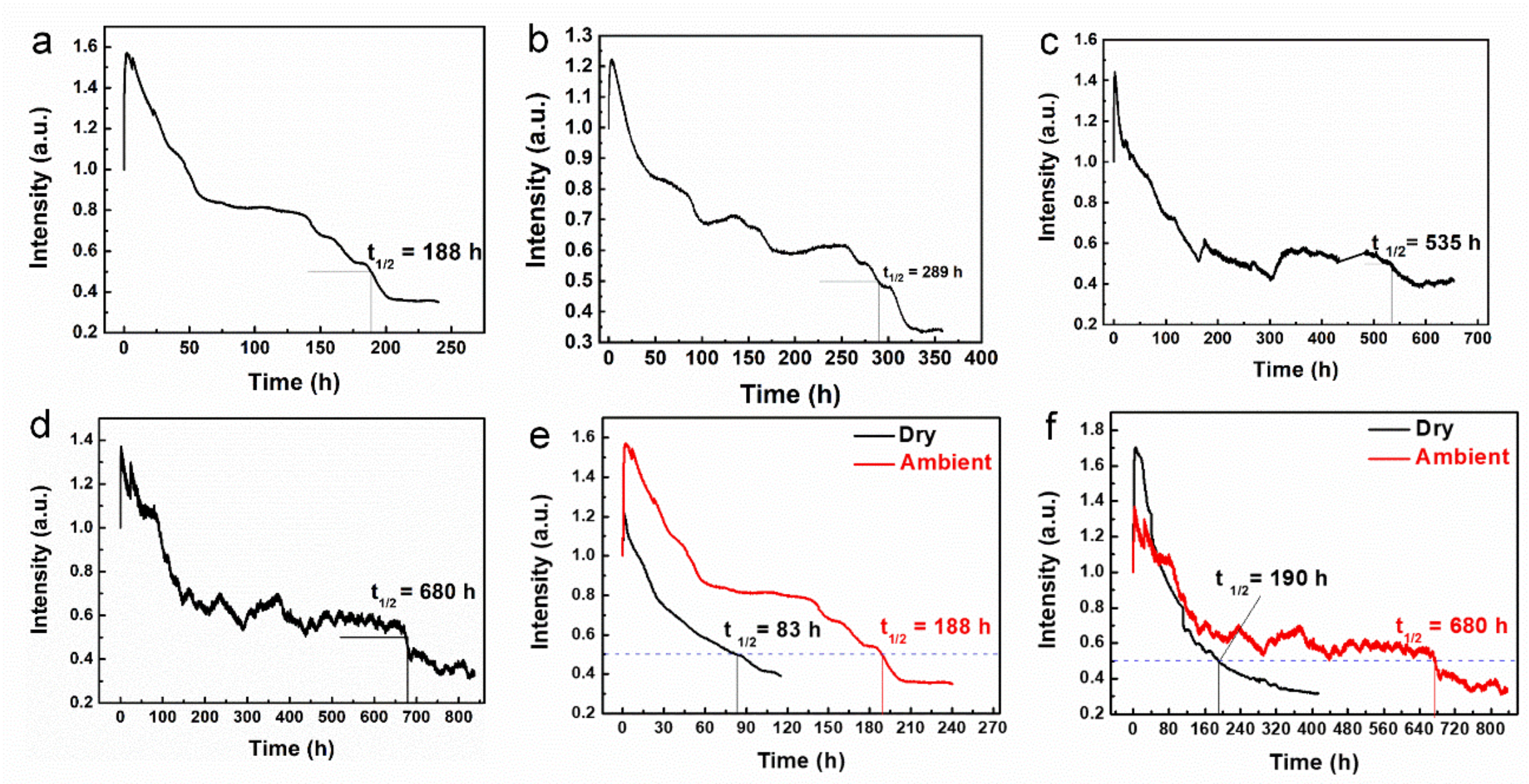

**Figure 27**. The operational stability of devices under ambient condition at a) 100 mA, b) 50 mA, c) 15 mA, d) 10 mA, respectively. Comparison of the devices operated under dry/ambient condition at e) 100 mA, f) 10 mA. Reproduced with permission.[89] Copyright 2020, Wiley-VCH.

### 4.7. Other Applications

In addition to the above wide applications, MHPs/$MO_x$ also show great potential in other areas. For example, the $(CsPbBr_3/Fe_3O_4)$@MPSs@$SiO_2$ composite can be used to capture and target the circulating tumor cells (CTCs) from both breast cancer cell lines and the blood of lung cancer patients.[94] And the sandwiched structured $SiO_2$@$CsPbX_3$@$SiO_2$ is also selected as a fluorescent sensor to detect $Fe^{3+}$ ions in real water samples with high sensitivity.[141] Furthermore, the monodispersed $CsPb_2Br_5$@$SiO_2$ core−shell nanoparticles can also be applied as the luminescent labels for the specific detection of the bovine serum albumin (BSA) protein.[118] Meanwhile, $CsPbX_3$@$SiO_2$, which shows higher cooling efficiency than the conventional white- and metallic-colored surface emitters, are expected to broaden the applications to daytime radiative cooling.[116]

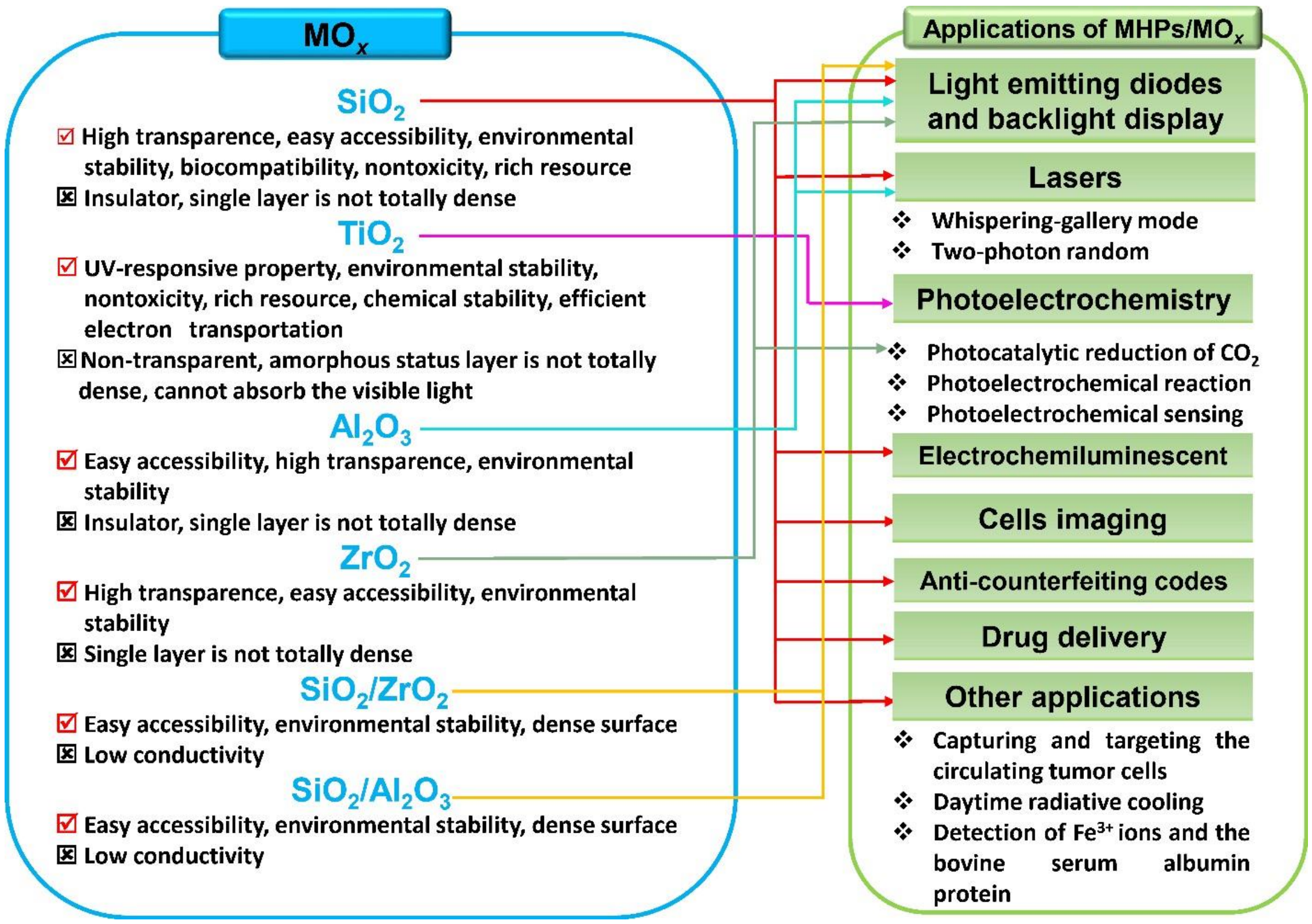

**Figure 28**. Summary of the application of the MHPs@MO$_x$, classified by the type of the MO$_x$ coatings.

### 4.8. Take Home Messages about the Applications of MHP@MO$_x$ with Respect to the Type of MO$_x$ Coating

MHPs@$SiO_2$ is the most attractive composite among the MHP@MO$_x$. It has been extensively used in various areas, ranging from LEDs and backlight display, lasers, cells imaging, to anti-counterfeiting codes, drug delivery, electrochemiluminescent, and even in capturing and targeting the circulating tumor cells as well as the daytime radiative cooling (**Figure 28**). The widespread applications of the MHPs@$SiO_2$ are attributed to the high transparence, easy accessibility, good environmental stability and biocompatibility, and some other properties, like nontoxicity and rich resource. However, the single layer of $SiO_2$ is normally not totally dense, this will limit the further commercialization of the MHPs@$SiO_2$. And the insulativity of $SiO_2$ will accelerate the radiative recombination and the hot carrier relaxation, and therefore detrimental for the PEC reaction. MHPs@$TiO_2$ is a good candidate for the applications in photoelectrochemistry, such as the photocatalytic reduction of $CO_2$, PEC reaction, PEC sensing. Herein, the $TiO_2$ shell can efficiently protect the MHPs from the surrounding environment, and improve the charge carrier transfer because of the relative band alignment of photogenerated charge carriers within MHPs and $TiO_2$ as well as the good UV-responsive property of $TiO_2$. However, the non-transparency of the $TiO_2$ imposes its applications in many areas, like LEDs, lasers, cell imaging, etc. In addition, benefited from the good transparence, easy accessibility, environmental stability of $ZrO_2$, the MHPs@$ZrO_2$ can be a very promising composite for the LEDs. Meanwhile, the $CsPbI_3/ZrO_2$ composite also has great potential in the visible photocatalytic $CO_2$ reduction because of the type II heterojunction forms at the $CsPbI_3/ZrO_2$ interface. Moreover, the robust properties and good transparency of the $Al_2O_3$ also endow the MHPs@$Al_2O_3$ priority to continuously lase in water as well as be applied in LEDs. However, the single layer $MO_x$ is usually not totally dense, which will allow the permeation of the surrounding $H_2O/O_2$. On the other hand, the binary coating, such as $SiO_2/AlO_x$ and $SiO_2/ZrO_2$ can provide more efficient protection for the MHPs because of the dense surface. And the LEDs devices based on the MHPs@binary $MO_x$ all shows excellent stability against surrounding stresses, such as water, heat, oxygen.

## 5. Conclusions and Outlook

MHPs have shown great potential in various areas, including solar cells, LEDs, photodetectors, photocatalysts, cell imaging, lasers, and so on. However, they are usually vulnerable to the surrounding environment, such as the heat, the humidity, the oxygen, and light. Many strategies have been developed to solve these issues.

In this review, we focus on summarizing the recent process about using the $MO_x$ to improve the stability as well as keeping the good optical properties of the pristine MHPs. The literatures cited in this review confirm that different synthetic methods, like sol-gel, ALD, template-assisted method, physical mixture, SILAR, the combination of template-assisted and sol-gel methods, the solvent-free approach are all efficient methods for the fabrication of the MHPs@$MO_x$ structure (**Table 1** and **Table 2**). However, every method has both good side and bad side. For example, the sol-gel method is the most popular strategy to fabricate the MHPs/$MO_x$ structure. But more attention should be paid to control the shell thickness as well as to avoid the byproducts (like water and alcohols) during the hydrolysis process. ALD is a perfect technique to precisely control the thickness of the shell, but it has been restricted to $Al_2O_3$ and the instrumentation requirements are not standard in all the labs. The MHP@$MO_x$ composite fabricated through the ALD method is suitable to be used in the devices which have a high requirement for the film quality, like lasers.

However, the composites prepared through the ALD method tend to have low PLQYs. Further work about should be focused on enhancing the PLQYs of composites through optimizing the parameters, like the temperature, the working cycles, the reaction time, the reactants. The template-assisted method is a facile way to achieve the large-scale synthesis. And no byproducts form and the size of the MHPs can be easily tuned through using templates with different pore sizes for the template-assisted method. But the obtained composites always show low PLQYs because of the lack of the ligands. The physical method is easy to operate but further work should also be done to avoid the aggregation of the MHPs during the process. The SILAR method is a good way with both high PLQYs and homogenously dispersed MHPs. In addition, the composites obtained from the template-assisted, the physical method, and the SILAR method usually show low stability because the lack of a full protection. Therefore, the combination of the high temperature sintering with the template will lead to MHP@$MO_x$ composites with extraordinary dense structure. And the combination the benefits from both the sol-gel method and the template-assisted methods will contribute to get stable and uniform MHPs QDs, while MHP@$MO_x$ composites obtained from molten-salts-based approach under mild conditions also represents an important technological advancement.

Then, we also introduce the widely used techniques, like SEM, TEM, XPS, XRD, PL, UV-Vis, TRPL, and the stability (against water, storage, thermal, light) measurements to characterize the MHP@$MO_x$ composites (**Table 3**). However, one technique is not enough, these methods should be combined with each other to fully understand the change in spectra, morphology, structure before and after encapsulation/the stability measurements.

Next, we provide an overview of the recent exploration of using MHPs/$MO_x$ in pc-LEDs, cells imaging, lasers, photoelectrochemistry, anti-counterfeiting, drug delivery, ECL and other applications (**Figure 28**). Specifically, the MHPs/$SiO_2$ composites have been widespread used in many areas, ranging from pc-LEDs to cell imaging, lasers, anti-counterfeiting, ECL, as well as drug delivery. This is attributed to the accessibility, good environmental stability, excellent biocompatibility, nontoxicity as well as high transparency of the $SiO_2$. And MHPs/$TiO_2$ composites are mainly applied in the photoelectrochemistry, due to the perfect band alignment of photogenerated charge carriers within MHPs and $TiO_2$, high chemical stability, as well as good UV-responsive property of $TiO_2$. In addition, the type I heterostructure fabricated between $CsPbBr_3$ and $ZrO_2$ also enables the efficient radiative recombination for highly efficient and stable pc-LEDs. And the type II heterostructure between $CsPbI_3$ and $ZrO_2$ makes $CsPbI_3/ZrO_2$ composite a potential photocatalyst for the visible photocatalytic $CO_2$ reduction. And the high diffusion barrier for water provided by the $AlO_x$ coating also provide great opportunity for the $CsPbBr_3@Al_2O_3$ to be excellent waterproof laser. Importantly, the $SiO_2/AlO_x$ and $SiO_2/ZrO_2$ binary coating will provide MHPs dense protection, and therefore highly stable and efficient pc-LEDs based on $CsPbBr_3@SiO_2/Al_2O_3$ and $CsPbBr_3@SiO_2/ZrO_2$ can be achieved.

However, many important issues still need to be solved for the further application of the MHPs/$MO_x$. Firstly, there should be a standard to evaluate the stability of the MHPs/$MO_x$ composite. For example, the power intensity and wavelength of the lamp (used for the photostability measurement) should be standardized. And the condition (the relative humidity and the temperature) for the storage stability should be the same for each paper. Moreover, binary coating should be encouraged to be widely used to achieve better protection. And it is highly urgent to enlarge the preparation process from the lab to the industrial level for the commercialization of the MHPs/$MO_x$ composites. In addition, MS is the most widely used template used for the fabrication of the MHPs/$MO_x$ currently. More templates, like mesoporous alumina (MA) and mesoporous $TiO_2$ can also be good candidates and should be encouraged to be widely used for different applications. Meanwhile, the number of works about using ALD are still small and only focused on $AlO_x$ coatings currently, more works should be expanded to other $MO_x$, like $TiO_2$, $SiO_2$, ZnO. Most importantly, new strategies, like the combination of different methods, the solvent free approach should be highlighted by more future work.

Besides, the current MHPs/$MO_x$ structure are usually at the microscale level. However, optoelectronic devices like the PeLEDs, the solar cells, the transistors, all have a strict requirement for the thickness (should be in nanoscale) and the quality (should have a full coverage and highly homogeneous) of the MHPs layers. Therefore, it is urgent to seek a method which is able to provide a totally compacted protective $MO_x$ layer for the MHPs at low temperature or even room temperature at a single particle level precisely. In addition, there are limited $MO_x$ are applied as the coating at this moment; and these widely used $MO_x$, like $SiO_2$ and $Al_2O_3$, are transparent and can provide efficient protecting for MHPs, but they are insulators. In future, more kinds of $MO_x$ with good conductivity should be explored to expand the encapsulation materials, such as NiO, ZnO, CuO, and $V_2O_5$, $Cr_2O_3$. Meanwhile, $CeO_2$, which is a good photocatalyst, can also be used to fabricate the MHPs/$CeO_2$ structure to improve the charge separation and the stability of the pristine MHPs in the application of the photoelectrochemistry. We hope researchers can gain a deep insight about the rent process about the MHPs/$MO_x$; and this review can provide a guideline for researchers to develop highly stable and efficient MHPs/$MO_x$.

**List of Acronyms and Abbreviations**

AAO: Anodized aluminum oxide

AHFS: Ammonium hexafluorosilicate

ALD: Atomic layer deposition

APTES: (3-Aminopropyl)triethoxysilane)

ASB: Aluminum sec-butoxide

ASE: Amplified spontaneous emission

a-$TiO_2$: Amorphous-$TiO_2$

BSA: Bovine serum albumin

C-ALD: Colloidal ALD

CB: Conduction band

CCT: Correlated color temperature

CIE: Commission internationale de l'éclairage

CoR: Coreactants

CPB: $CsPbBr_3$

CRI: Color rendering index

CTCs: Circulating tumor cells

CVD: Chemical vapor deposition

DA: Decanoic acid

DAm: Dodecylamine

DAPI: 4',6-diamidino-2-phenylindole

DBATES: Di-sec-butoxyaluminoxytriethoxysilane

DDAB: Didodecyl dimethyl ammonium bromide

DBAE: (Dibutylamino)ethanol

DFT: Density functional theory

DI: Deionized

DMF: N,N-Dimethylformamide

DMSO: Dimethyl sulfoxide

Dox: Doxorubicin

DRS: Diffuse reflectance spectra

ECL: Electrochemiluminescence

EDS: Energy dispersive spectroscopy

EELS: Electron energy loss spectroscopy

EL: Electroluminescent

EQE: External quantum efficiencies

FA: Formamidinium

$FSiO_2$: Fluorinated silica

FTIR: Fourier transform infrared spectroscopy

FTO: Fluorine doped tin oxide

FWHM: Full width at half maximum

GC: Gas chromatography

GPC: Growth per cycle

HAADF: High-angle annular dark-field

h-$Al_2O_3$: Hierarchical-$Al_2O_3$

HRTEM: High resolution transmission electron microscopy

HSA: Human serum albumin

HSNSs: Hollow siliceous nanospheres

IOPCs: Inverse opal photonic crystals

LARP: Ligand-assisted reprecipitation

LEDs: Light-emitting diodes

MA: Methylammonium

MFA: Methylformamide

MHPs: Metal halide perovskites

MOF: Metal organic framework

$MO_x$: Metal oxide

MPSs: Mesoporous polystyrene microspheres

MPTMS: (3-mercaptopropyl) trimethoxy silane

MS: Mesoporous silica

MSNs: Mesoporous silica nanoparticles

MTT: 3-(4, 5-dimethylthiazol-2-yl)-2, 5-diphenyltetrazolium bromide

NCs: Nanocrystals

NHE: Normal hydrogen electrode

NMR: Nuclear magnetic resonance

NPs: Nanoplates

NSs: Nanosheets

NTSC: National Television System Committee

OA: Oleic acid

ODE: 1-octadecene

PBS: Phosphatebuffered saline

PCE: Power conversion efficiencies

pc-LEDs: Phosphor-converted light-emitting diodes

pc-WLEDs: Phosphor-converted white light-emitting diodes

PDMS: Polydimethylsiloxane

PEC: Photoelectrochemical

PeLEDs: Perovskite LEDs

PHPS: Perhydropolysilazane

PLQYs: Photoluminescence quantum yields

PS: Polystyrene

PVDF: Polyvinylidene fluoride

QDs: Quantum dots

RATQ: Resistibility against thermal quenching

RH: Relative humidity

RP: Ruddlesden-Popper

RT: Room temperature

SEM: Scanning electron microscope

SILAR: Successive ionic layer adsorption and reaction

STEM: Scanning transmission electron microscopy

TBOT: Titanium butoxide

TBT: Tetrabutyl titanate

TEM: Transmission electron microscopy

TEOS: Tetraethyl orthosilicate

TGA: Thermogravimetic analysis

TMA: Trimethylaluminum

TMOS: Tetramethoxysilane

TOAB: Tetraoctylammonium bromide

TON: Turnover number

TOPO: Trioctylphosphine oxide

TPA: Tripropylamine

TPOS: Tetrapropoxysilane

TRPL: Time-resolved photoluminescence

TTEO: Tantalum(V) ethoxide

VB: Valence band

WGM: Whispering-gallery mode

WLEDs: White light-emitting diodes

XPS: X-ray photoelectron spectroscopy

XRD: X-ray diffraction

ZTB: Zirconium(IV) tert-butoxide

**Acknowledgements**

Y.Y.D. thanks the financial support from China Scholarship Council (CSC, No. 201808440326). R. D. C. acknowledges the European Union's Horizon 2020 research and innovation FET-OPEN under grant agreement ARTIBLED No. 863170, the ERC-Co InOutBioLight No. 816856, and the MSCA-ITN STiBNite No. 956923.

**Conflict of Interest**

The authors declare no conflict of interest.

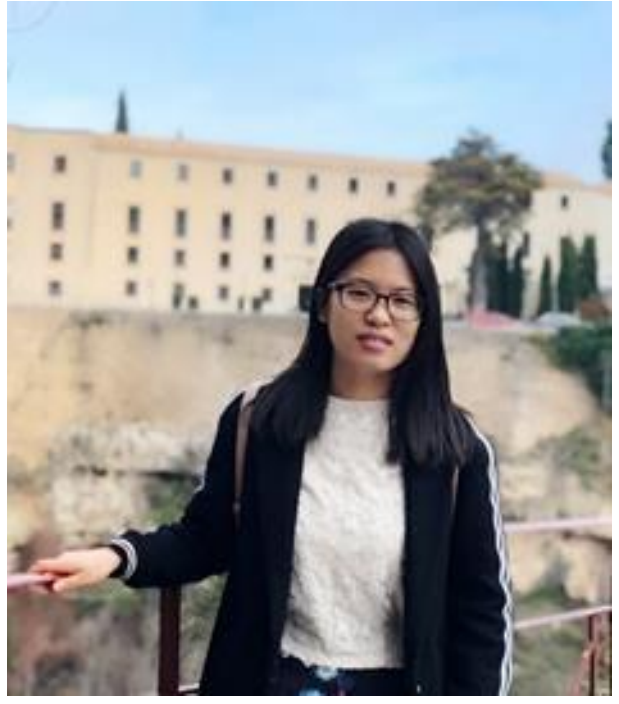

Yanyan Duan received her master degree (2017) in Materials Engineering from University of Chinese Academy of Science (CAS). She worked as a Research Assistant in Shenyang Institute of Automation, Guangzhou, CAS from 2017 to 2018. Then, she got the financial support from China Scholarship Council and started her PhD from October of 2018 in IMDEA Materials Institute/Universidad Politécnica de Madrid (Madrid, Spain) under the supervision of Prof. Dr. Rubén D. Costa . Currently, her research focuses on the preparation, characterization, and application of the perovskites for hybrid light emitting devices.

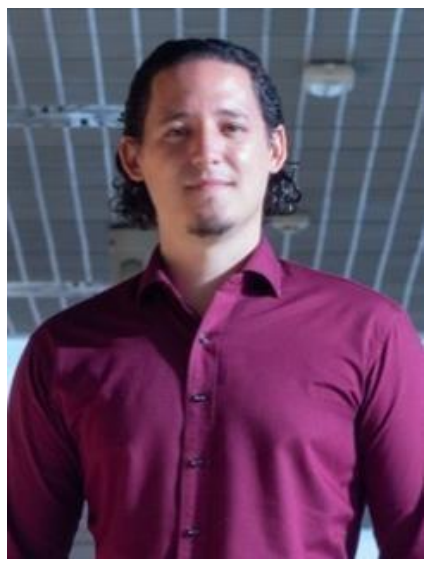

Prof. Dr. Rubén D. Costa received his PhD from the U. Valencia (Spain) in 2010 and was Humboldt post-doc at the U. Nürnberg -Erlangen (Germany) from 2011-2013. In 2014 he started the Hybrid Optoelectronic Materials and Devices Lab as Liebig group leader (2014-2017). In 2017, he moved his group to IMDEA Materials (Spain) and expanded to U. Waseda (Tokyo) as associate Professor in 2018. Since 2020 he leads the chair for Biogenic Functional Materials at the Technical University of Munich. He has reported >150 scientific publications/books/patents and has been recipient of >35 awards/mentions/fellowships.

Table of Contents Entry

This review summaries the latest research related to the strategies to fabricate the metal halide perovskites@metal oxide (MHPs@MO$_x$) composites, techniques to systematically evaluate the performances and structures of the composites. And recent achievements in various applications are also presented based on different MO$_x$. Finally, conclusions and future research prospects are also outlined to ensure a bright future of MHPs@MO$_x$.

Y.Y. Duan, D.Y. Wang, and R.D. Costa*

Recent Progress on Synthesis, Characterization, and Applications of Metal Halide Perovskites@Metal Oxide

ToC figure

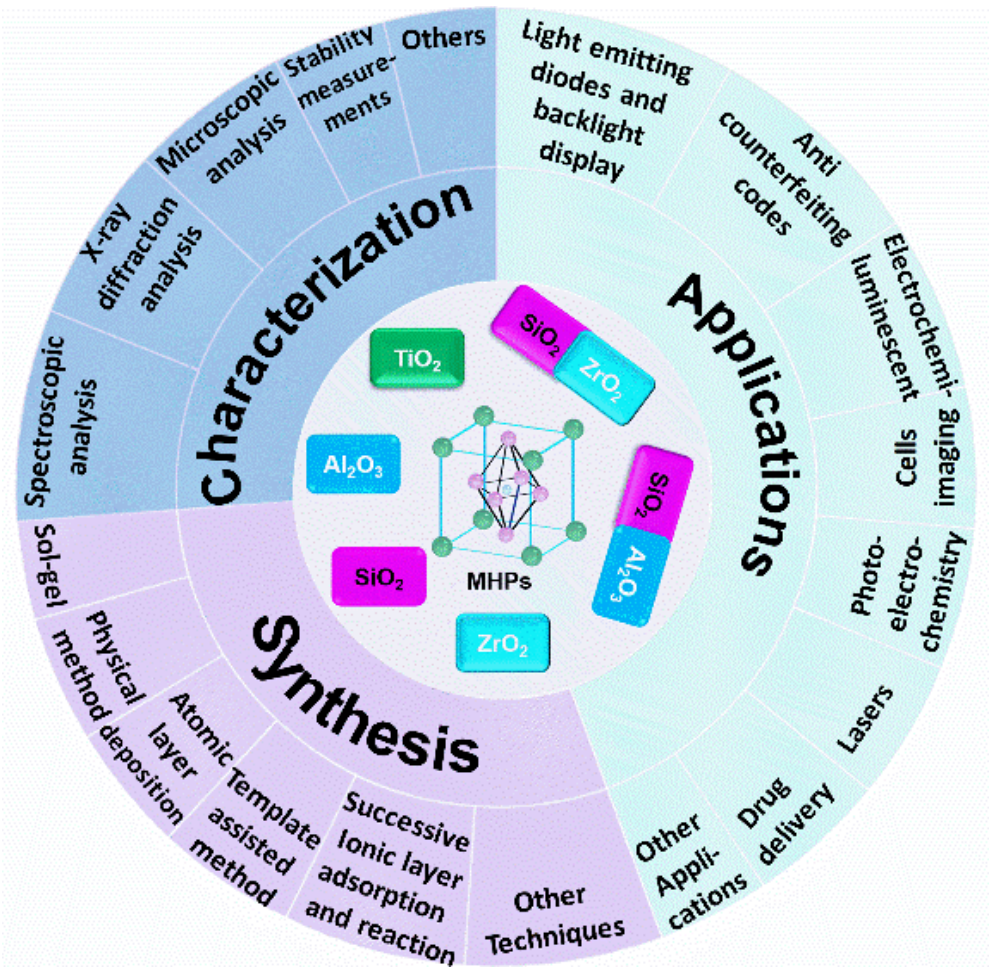